 \documentclass[twocolumn,trackchanges,usenames,dvipsnames]{aastex631}
\usepackage[english]{babel}

\usepackage{amsmath,amssymb}
\usepackage{graphicx}
\usepackage{xcolor}
\usepackage{latexsym}
\usepackage{dcolumn}
\usepackage{amsmath}
\usepackage{epsf}
\usepackage{float}
\usepackage{enumerate}
\usepackage{enumitem}
\usepackage{graphicx}
\usepackage{dcolumn}
\usepackage{bm}
\definecolor{mynavy}{RGB}{18,18,155}
\usepackage{soul}  
\setstcolor{red}

\usepackage{wasysym}
\usepackage{rotating}
\usepackage{multirow}
\usepackage{chemfig}
\usepackage{enumitem}
\usepackage{ulem,xcolor}

\usepackage{natbib}
\usepackage{chngcntr}
\usepackage{tikz}
\usetikzlibrary{arrows.meta}
\usepackage{scalerel}
\definecolor{plum}{rgb}{0.36078, 0.20784, 0.4}
\definecolor{chameleon}{rgb}{0.30588, 0.60392, 0.023529}
\definecolor{cornflower}{rgb}{0.12549, 0.29020, 0.52941}
\definecolor{scarlet}{rgb}{0.8, 0, 0}
\definecolor{brick}{rgb}{0.64314, 0, 0}
\definecolor{sunrise}{rgb}{0.80784, 0.36078, 0}
\definecolor{black}{rgb}{0.15,0.35,0.75}
\definecolor{carolina}{RGB}{153, 186, 221}
\definecolor{darkcyan}{RGB}{57, 144, 145}

\begin{document}
\hypersetup{linkcolor=plum, citecolor=plum}

\title{Unveiling the central engine of core-collapse supernovae in the Local Universe: NS or BH?}


\author{Maurice H.P.M. van Putten}
\altaffiliation{Corresponding author, mvp@sejong.ac.kr}
\affiliation{Department of Physics and Astronomy, Sejong University, 98 Gunja-Dong, Gwangjin-gu, Seoul 143-747, South Korea}
\affiliation{INAF-OAS Bologna via P. Gobetti 101 I-40129 Bologna Italy, Italy}

\author{Maryam Aghaei Abchouyeh}
\affiliation{Department of Physics and Astronomy, Sejong University, 98 Gunja-Dong, Gwangjin-gu, Seoul 143-747, South Korea}

\author{Massimo Della Valle}
\affiliation{INAF – Osservatorio Astronomico di Capodimonte, Salita Moiariello 16, 80131 Napoli, Italy}



\date{\today}

\begin{abstract}
The physical trigger powering supernovae following the core collapse of massive stars is believed to involve a neutron star (NS) or a black hole (BH), depending largely on progenitor mass. 
A potentially distinct signature  
is a long-duration gravitational wave (GW) burst from BH central engines by their ample energy reservoir $E_J$ in angular momentum, far more so than an NS can provide. 
A natural catalyst for this radiation is surrounding high-density matter in the form of a non-axisymmetric disk or torus.
Here, we derive a detailed outlook on LVK probes of core-collapse supernovae CC-SNe during the present observational run O4 based on their event rate, an association with normal long GRBs and mass-scaling of GW170817B/GRB170817A. 
For BH central engines of mass $M$, 
GW170817B predicts a descending GW-chirp of energy 
${\cal E}_{GW}\simeq 3.5\% M_\odot c^2 \left(M/M_0\right)$ 
at frequency 
$f_{GW}\lesssim 700\,{\rm Hz}\left(M_0/M\right)$, 
where $M_0\simeq 2.8\,M_\odot$.
For a few tens of events per year well into the Local Universe within 50-100Mpc, probes at the detector-limited sensitivity are expected to break the degeneracy between their NS or BH central engines by GW calorimetry.
\end{abstract}



\section{Introduction}
\label{Intro}

The operation of LIGO-Virgo-KAGRA (LVK) is presently in transition from detector to observatory, and it is opportune to develop new observational strategies to exploit this radically new window to broadband gravitational radiation fully. 
Of particular interest to {gravitational wave (GW) observations} are core-collapse supernovae (CC-SNe) signaling the endpoint of the evolution of massive stars at the tail of the Salpeter Initial Mass Function \citep{sal55,sma09,par2022}. {While most CC-SNe are expected to produce an NS \citep[e.g.][]{par2022}, some must be producing stellar-mas BHs from the more massive progenitors. This introduces the astrophysical conundrum of ``NS or BH" central engine for individual events, notably challenging to solve based on electromagnetic (EM) and neutrino ($\nu$) observations alone. Generally, the angular momentum of these central engines determines the energy reservoir $E_J$ in angular momentum powering the event \citep{bis70,bis71}. 
{If accompanied with a GRB, exhausting $E_J$ may be indirectly observable in the GRB light curve {\em on average} \citep{van12}.}
A hallmark signature would be a direct detection in a descending chirp in gravitational radiation. 
A signature of this kind is notably distinct from the ascending chirps produced in compact binary coalescence.
} 

The MeV-neutrino burst accompanying the iconic type II SN1987A with progenitor Sanduleak -69 202 \citep{gil87,wes87} in the LMC at a distance of 50\,kpc \citep{pie19} revealed the formation of matter at supra-nuclear densities \citep{hir87,arn89}. However, its EM$\nu$ light curves have been found inadequate to rigorously identify the physical trigger of the explosion - a rotating NS or BH. For 37 years, this left the final remnant unknown \citep{bou93,alp18,gal18} 
until the recent JWST possible identification of a NS \citep{fra24}. This outcome is consistent with the estimated mass of about $20\,M_\odot$ of the progenitor Sk -69 202 \citep{alp18,sma09,son87,suk18}. 

For the hundreds of CC-SNe in the Local Universe per year, there is a clear need for a different approach. To this end, we propose to identify the physical trigger of the {aforementioned} explosion mechanism by directed probes in gravitational radiation, expected to provide a tight constraint on the potential {central engine, the remnant and, by implication, the mass of the progenitor.}


\begin{figure}[ht]
    \center{\includegraphics[scale=0.4]{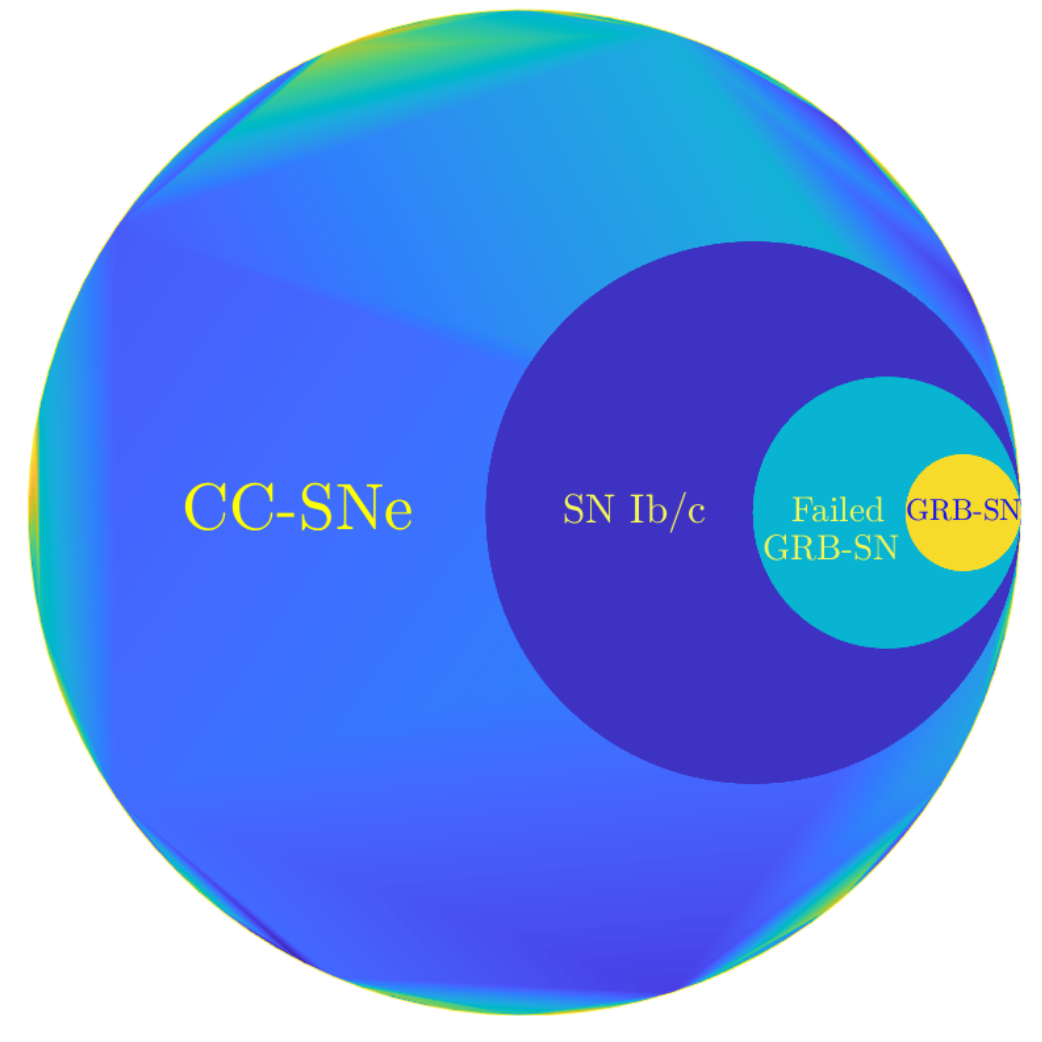}}
    \vspace{-1cm}
    \center{\includegraphics[scale=0.23]{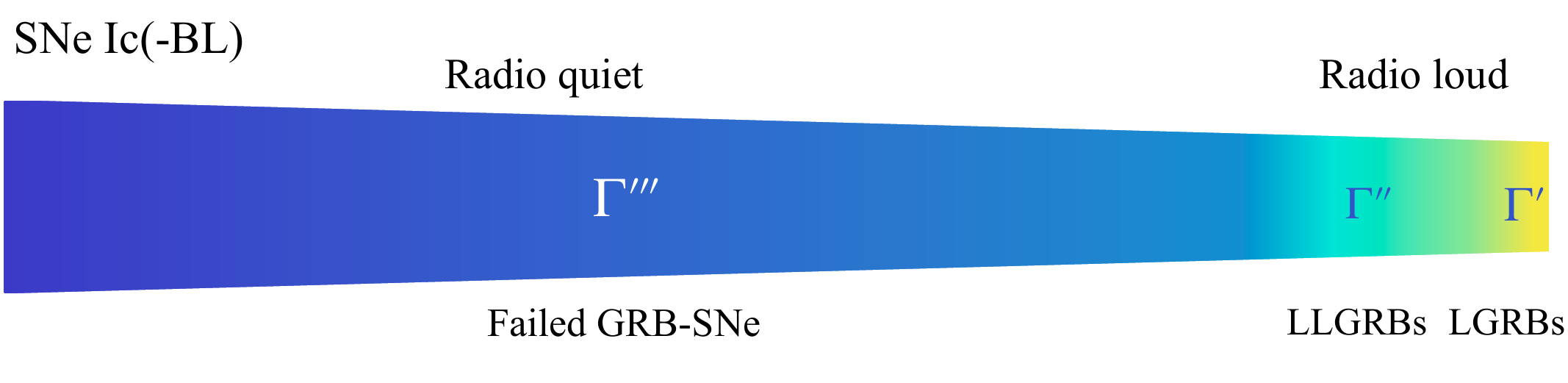}}
    \caption{
    Venn diagram of CC-SNe (not to scale). The energetic CC-SN Ib/c have a small fraction $\Gamma^{\prime}$ producing GRB-SNe ({\em yellow}) with an additional fraction $\Gamma^{\prime \prime}$ of failed GRB-SN ({\em turquoise}). 
    Both may have a dominant output in gravitational radiation (GW-SN) when powered by a rotating black hole \citep{van03b}. 
    The progenitor and remnant of the remainder ({\em dark blue}) is unknown, largely by ambiguity in their central engine (NS or BH). Similarly for the larger group of SN II ({\em gradient fill}). The large stellar radius of their progenitors appears prohibitive for any central engine to be sufficiently long-lived to successfully produce a GRB, even as some may be GW-SN. {The gradient-filled bar zooms in on SNe of type Ic, highlighting a diversity in narrow (NL) and broad line (BL) events \citep{mau10} and branching ratios $\Gamma$ (Appendix \ref{Apc}) of the latter into (L)LGRBs following \cite{cor16,cor23}.}}
    \label{fig1b}
\end{figure}

{Among all CC-SNe, type Ib/c events are relatively more energetic than SNe of type II but less numerous with an event rate of about 100 per year within a distance of 100\,Mpc (Fig. \ref{fig1b}).}
They are believed to be produced by relatively more massive and envelope-stripped stars in the brightest regions of star-formation \citep{kel08,ras08} and their progenitors are more likely massive stellar binaries rather than single stars \citep{woo93,fry14,rue21}. As the parent population of normal long gamma-ray bursts (GRBs) {and low-luminosity GRBs (LLGRB)} \citep{gal98}, the central engines and remnants of these energetic events are possibly BHs rather than NSs. 
{Accordingly, these events hold promise to be leading candidates for long-duration GW bursts.}

{To date, SN Ic-BL are the only ones with { normal LGRBs and LLGRBs} \citep{cor16}, though SNe Ib are not excluded as potential {progenitors as well}. The main difference between these two classes might be the low-mass envelope for the {LLGRB associated events} that prevent the jet from being ultra-relativistic, opposite to the case of LGRBs \citep{nak15}, or the difference in their central engines \citep{van04,van12,nak12,dec22}. LGRBs events have small {true event rate of $\sim 1$ per year in 100\,Mpc (with an observed event rate $\sim 10^{-2}$ per year)}, while the event rate of their low-luminosity counterparts { may be} ten times higher (Fig. \ref{fig1b}) \citep{fra01,van03,pod04,van04,del06,woo06,gue07,hjo12,van19,lev20,cor23}.}

 \begin{figure*}
    \center{\includegraphics[scale=0.35]{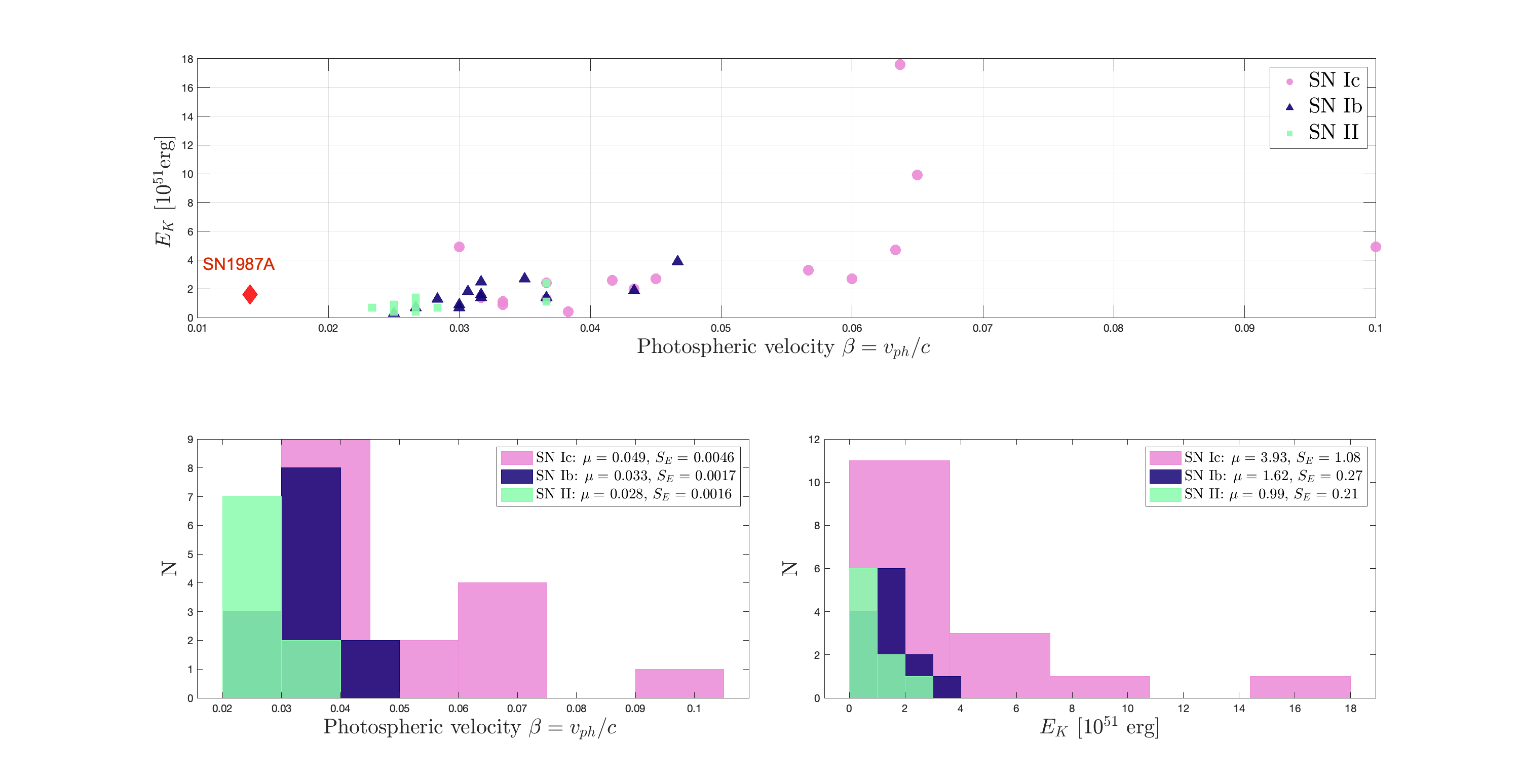}}
    \caption{{Distributions of CC-SNe} of type II and Ib/c by their photospheric velocity $\beta=v_{ph}/c$ and kinetic energy derived from EM observations \citep{ash87,pas05,utr11,lym16}. 
    $\beta$ is derived from photospheric velocities about 1-2 months after the shock breakout time $t_b$.
    (Top panel) SNe of type Ic stand out by their kinetic energy and photospheric velocity, especially compared to SN 1987A at the far left of the distribution. (lower panels) Histogram of $\beta$ and kinetic energy of SNe Ib/c and II indicating a clear distinction between the three in EM observations. }
    \label{fig1a}
\end{figure*}


SNe II are less energetic but relatively more numerous than SN Ib/c by a factor of a few \citep{cap99,wei11,cap15}. Their progenitors are sufficiently massive to produce a NS or BH, largely depending on their mass \citep{sal55,par2022}. In fact, they appear to have a bi-modal distribution in normal and broad-line (BL) events associated with non-relativistic ejecta and, respectively, the more energetic mildly-relativistic ejecta \citep{mau10,van11}.
It is not known how this relates to their central engines and remnants, if at all.
With the envelope retained and the progenitor stellar radius about two orders of magnitude larger than SN Ib/c,  conceivably no central engine has sufficient lifetime for a successful jet breakout of the progenitor stellar envelope. Though uncertain, SN II events producing BHs may hereby be loud in GWs despite not being associated {with a GRB}. 

Fig. \ref{fig1a} shows the kinematic {properties of CC-SNe} complementing Fig. \ref{fig1b}, by kinetic energy and photospheric velocity \citep{lym16}.

\begin{figure*}[t]
   \vskip -0.1in
    \center{\includegraphics[scale=0.6]{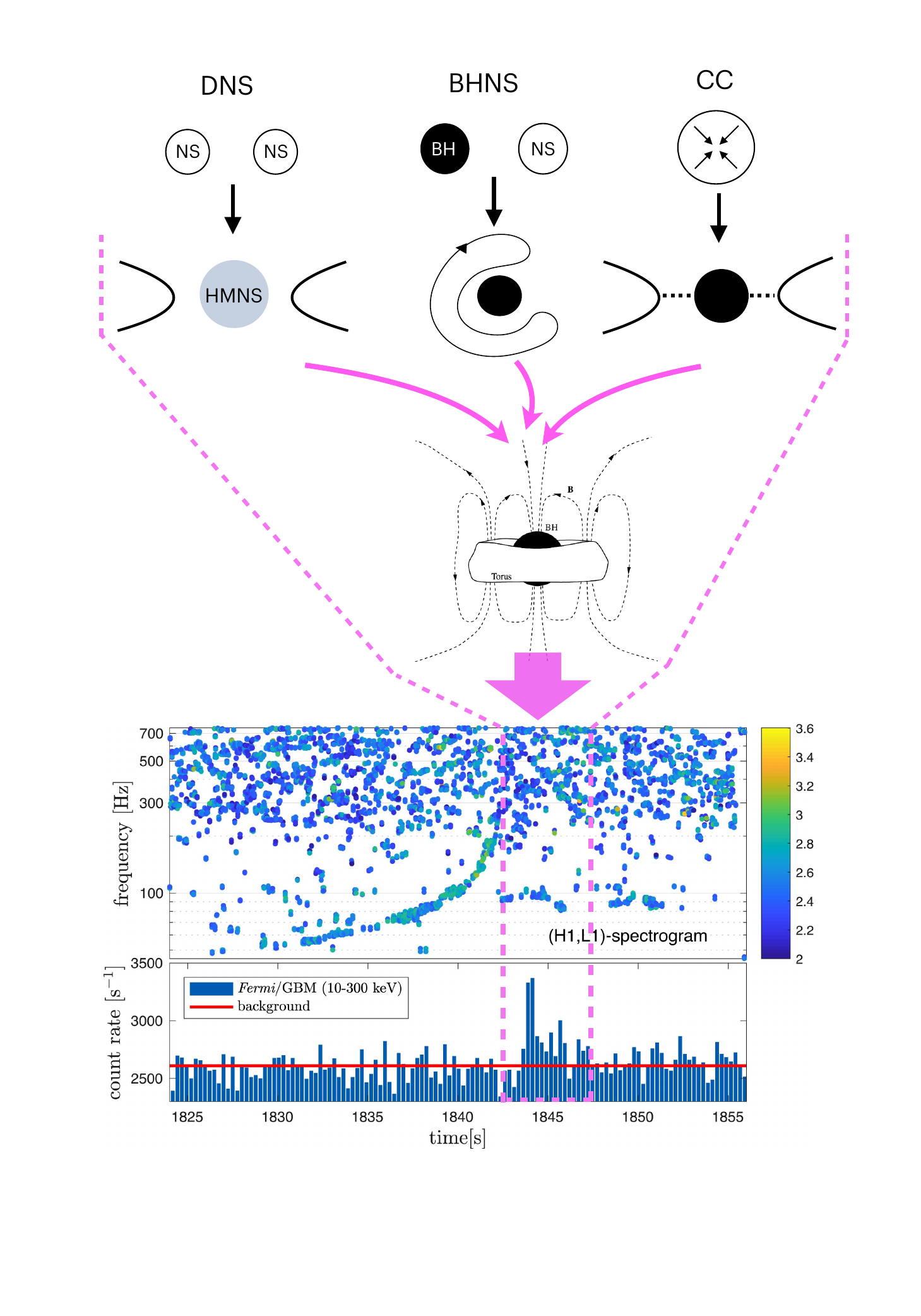}}
    \vskip -1in
    \caption{{{(Top)} BH central engines are a universal outcome of mergers (DNS, NSBH; left, middle) and core-collapse (CC) of sufficiently massive stars (right). DNS produces rapidly rotating BHs in prompt or delayed gravitational collapse of an initial HMNS (left). NS-BH and CC are expected to produce BHs with diversity in spin, the latter following a surge of hyper-accretion {(See Fig. \ref{fig5-5})}. High-density matter surrounding the BH provides a catalyst to convert its ample energy reservoir into multi-messenger emission. A descending GW-chirp signals an initially rapidly rotating BH. Accompanying sub-dominant emissions in MeV-neutrinos, magnetic winds and a baryon-poor jet emanating from the BH (not shown) accounts, from mergers, for SGRBEEs or LGRBs with no SN (LGRBN) sharing the Amati-correlation and, from CC, CC-SNe [Reprinted from van Putten 2015.] (Lower panels.) Observation of the descending GW-chip GW170817B/GRB170817 following the ascending GW-chirp GW170817 of a DNS, signaling spin-down of a Kerr BH formed in delayed gravitational collapse of the initial HMNS. [After \cite{van01,van03b,van15,van19b}.]}}
    \label{fig0817}
\end{figure*}

In light of the above, and by the relatively large event rate of CC-SNe in the Local Universe, we here consider SN-triggered directed searches for their potential output in gravitational radiation. Our focus is on the tail of CC-SNe that are the most energetic. {These are most likely powered by a rotating BH for their ample energy reservoir (see \eqref{EQN_Erot} in \S \ref{LT} below)
\begin{eqnarray}
E_J\lesssim 0.29Mc^2
\label{EQN_EJ}
\end{eqnarray}
in angular momentum, far beyond what an NS can provide with $E_J^{NS}\lesssim 0.03 M_\odot c^2$}. This can potentially drive an energetic multi-messenger disk emission sustained for the lifetime of rapid spin of the BH. Such will satisfy rigorous mass scaling based on the Kerr metric as an exact solution to the theory of general relativity. 

{At low energy radiation, the power of BH central engines is evident in {launching extragalactic radio jets},} e.g., in M87 and AT2020ocn \citep{cui23,pas24}. In the case of stellar mass rotating BHs surrounded by high-density matter in the form of a disk or torus, this can include gravitational radiation \citep{van01}. This is demonstrated for the first time by the descending GW signal GW170817B powered by a rapidly rotating BH of mass $M\simeq2.8M_\odot$ \citep{van19a}.

{Crucially, by the universality of BHs (\S2), this may carry over to produce GW-SN from some of the energetic CC-SNe, prompting the present study.}
In \S\ref{BH} the BH formation in transient events is discussed motivated by GW170817B/GRB170817A. We then revisit the physics of BH central engines with their potential radiation channels and the energy budget of each (\S \ref{discussiona}). Next, we estimate the LVK horizon distance for a GW-SN from the population of energetic CC-SNe powered by BH central engines by { mass-scaling} of GW170817B. This includes three gain factors originating from detector improvement in O4 over O2, the mass of the BH, and the observed frequency. \S\ref{Outlook} discusses the technical limitations and observational opportunities for detecting such a signal, providing an estimate of the search window { given the time of first observation of the CC-SNe}. In \S\ref{discussion} we discuss three possible outcomes of LVK searches for GW-SNe.
 
\section{Universality of Black holes}
\label{UBH}
{According to the theory of general relativity, BHs are described by the three parameters of mass ($M$), angular momentum ($J$) and electric charge ($Q$) \citep{tho86}. This is known as the no-hair theorem. For astrophysical BHs, it is well known that the gravitational contribution of $Q$ can be safely neglected even when non-zero. As a result BHs have no memory of their formation history or progenitor except for $M$ and $J$. This shows a high degree of universality for BH as central engines and the associated observational consequences arising from mergers involving an NS or core-collapse of massive stars (Fig. \ref{fig0817}).}

{Characteristic for rotating BHs is their ample energy reservoir {in $E_J$ \eqref{EQN_EJ}}. Releasing $E_J$ to power energetic astrophysical processes requires a catalyst, without which the BH remains passive. The primary candidate for a catalyst considered in the present work is surrounding high-density matter in the form of a disk or torus, expected from catastrophic events shown in Fig. \ref{fig0817}. As a result, such disk or torus will be the primary radiator across various emission channels. }

{For the catalyst to operate, required is an interaction between the BH and the surrounding matter, the result of which will follow the first law of thermodynamics. In the original proposal for this catalyst in \cite{van99}, this interaction is mediated by the inner torus magnetosphere through exchange of Maxwell stresses between the BH and surrounding matter. The result is described by a dominant fraction of BH luminosity incident on the inner face of a surrounding torus and a subdominant emission in a baryon-poor jet. This is in contrast with a subsequent alternative idea of magnetically arrested disk (MAD) which appeals to magnetic pressure - the scalar part of the magnetic stress-energy tensor sans Maxwell stresses - leaving $E_J$ untapped \citep{nar03}.}

{Given a high-angular momentum BH-torus system {satisfying $\Omega_H>\Omega_T$ (\S \ref{LT} and Fig. \ref{fig5-4} below), the BH will be losing angular momentum against the torus by exchange of Maxwell stresses with 
the torus in accord with the first law of thermodynamics (Appendix B). In this process, the lifetime of BH spin scales inversely with the strength of this interaction mediated by a torus magnetosphere.
Detailed stability considerations show that this is bounded by about 10\% of the torus kinetic energy $e_{k}$ \citep{van99,van03b}. This bound puts a lower limit on the lifetime of spin of the BH}. This is a {\em secular time scale} extending over seconds to tens of seconds depending on the inverse of the mass ratio $\sigma$ of the torus to the BH (Appendix \ref{AppA}).}

{During this process one may expect prodigious emission in GWs from the surrounding matter, expected to comprise quasi-monochromatic radiation from the innermost regions where density and temperature are highest, and broadband emission from the extended accretion flow beyond, if present \citep{van16,van19}. The first is expected to be the dominant channel and mostly so in quadruple GW-emission, based on the anticipated power-law spectrum \citep{van02,van03b,bro06}. Some confirmations for this anticipation can be found in theoretical studies and numerical simulations \citep{kiu11,tos19,tos20,wes21}. These results are potentially of interest also to LISA probes of tidal disruption events around SMBHs \citep{ama23}.} 

{Furthermore, GW-emission from an inner disk or torus may develop side-bands by modulation due to Lense-Thirring precession (see \S \ref{discussiona}). These emissions may be accompanied by broadband emissions from the extended accretion flow. When sufficiently cold, emission may derive from clumpiness in the outer parts of accretion flows formed by gravitational instabilities \citep{pir07}, bar-mode instabilities \citep{kob03} and/or wave-structures in accretion flows \citep{lev15}.}

{The first (and so far only) direct observation of BH spin-down is GW170817B: a descending GW chirp lasting $\sim 3.7$\,s following the ascending merger chirp GW170817 (Fig. \ref{fig0817}). An earlier indirect observation may be found in GRB060614 \citep{del06a,fyn06,gal06,geh06}. With no supernova detected \citep{del06,fyn06}, the progenitor is possibly a merger. Its anomalously long duration of $\sim 102$\,s, therefore, represents a secular time-scale, orders of magnitude larger than any dynamical time-scale in such systems, that is naturally accounted for by the lifetime of rapid spin of a BH. If so, GRB060614 indirectly points to GW radiation of the same duration \citep{van08} realized by aforementioned $\sigma$ considerably smaller than that of GW170817B. }

\section{BH formation in energetic transients}
\label{BH}

 The normal long GRBs associated with SN Ib/c satisfy the Amati-relation \citep{ama02,ama06,ama09}. This includes the Extended Emission to short GRBs, pointing to a common central engine \citep{van14}. SN1998bw was the first detection of a CC-SNe associated with a GRB at a distance of about 40Mpc ($z=0.0085 \pm 0.0002$) \citep{gal98,fol06,tin98}, though with no observational channel other than EM \citep{woo99}.
Short GRBs are believed to be produced by mergers of a NS with another NS (DNS) or stellar-mass BH (NSBH) \citep{eic89,van14,dav15,ghi16,fro19,ama21,bia24}. 
The DNS merger GW170817 detected by LIGO-Virgo during observational run O2 is the first such progenitor of GRB170817A, detected at a distance of 40\,Mpc \citep{abb17}.

Interestingly, GRB170817A is accompanied by a coincident long-duration descending gravitational wave chirp GW170817B with energy output and frequency sweep
\begin{eqnarray}
{\cal E}=\left(3.5\pm1\right)\%M_\odot c^2,~~
{\Delta F}\simeq \left(700-200\right)\,\mbox{Hz}.
\label{EQN_ES}
\end{eqnarray}
This detection (\ref{EQN_ES}) is at 5.5\,$\sigma$ significance based on two statistically independent time-symmetric analyses
following a unitary transformation of LIGO H1 and L1 data \citep{van23}, where $c$ denotes the velocity of light. 
With ${\cal E} \simeq 3.5\%M_\odot c^2$ exceeding the rotational energy of a hypermassive neutron star (HMNS) by a factor of a few, this signature 
reveals the central engine of GRB170817A: a Kerr black hole of mass $M_0\simeq 2.8M_\odot$ spinning down over $\sim 3.7\,$s consistent with $T_{90}^{8-70\mbox{keV}}=\left( 2.9\pm 0.3\right)$\,s of GRB170817A \citep{poz18}.

The present estimate for $M_0$ is an updated value based on the data in \cite{van19a}. Since this derives from GW170817B, independent from all else, it is remarkable for its one $\sigma$ consistency with the mass estimate $2.74^{+0.04}_{-0.01}M_{\odot}$ from the merger GW170817 \citep{abb17a}. This consistency strengthens the association of GW170817 with GRB17081A, given the simultaneity of GW170817B and the latter.

Given the merger time $t_m$ of GW170817, the trigger time $(0.92\pm 0.08)$\,s of GW170817B reveals the {\em delayed gravitational collapse} \citep{van23} of the initial HMNS to a Kerr BH.
This outcome shows consistent GW-EM event timing, given the estimated lifetime $\left(0.98\pm 0.3\right)$\,s of the initial HMNS inferred from the kilonova AT2017gfo \citep{gil19}.

Taking advantage of the universality of BHs, in the next section we discuss the physics of BH central engines for the above mentioned stellar mass BHs alongside super-massive BHs in active galactic nuclei (AGN).

\section{Physics of BH central engines}
\label{discussiona}

\begin{figure*}[t]
    \center{\includegraphics[scale=0.20]{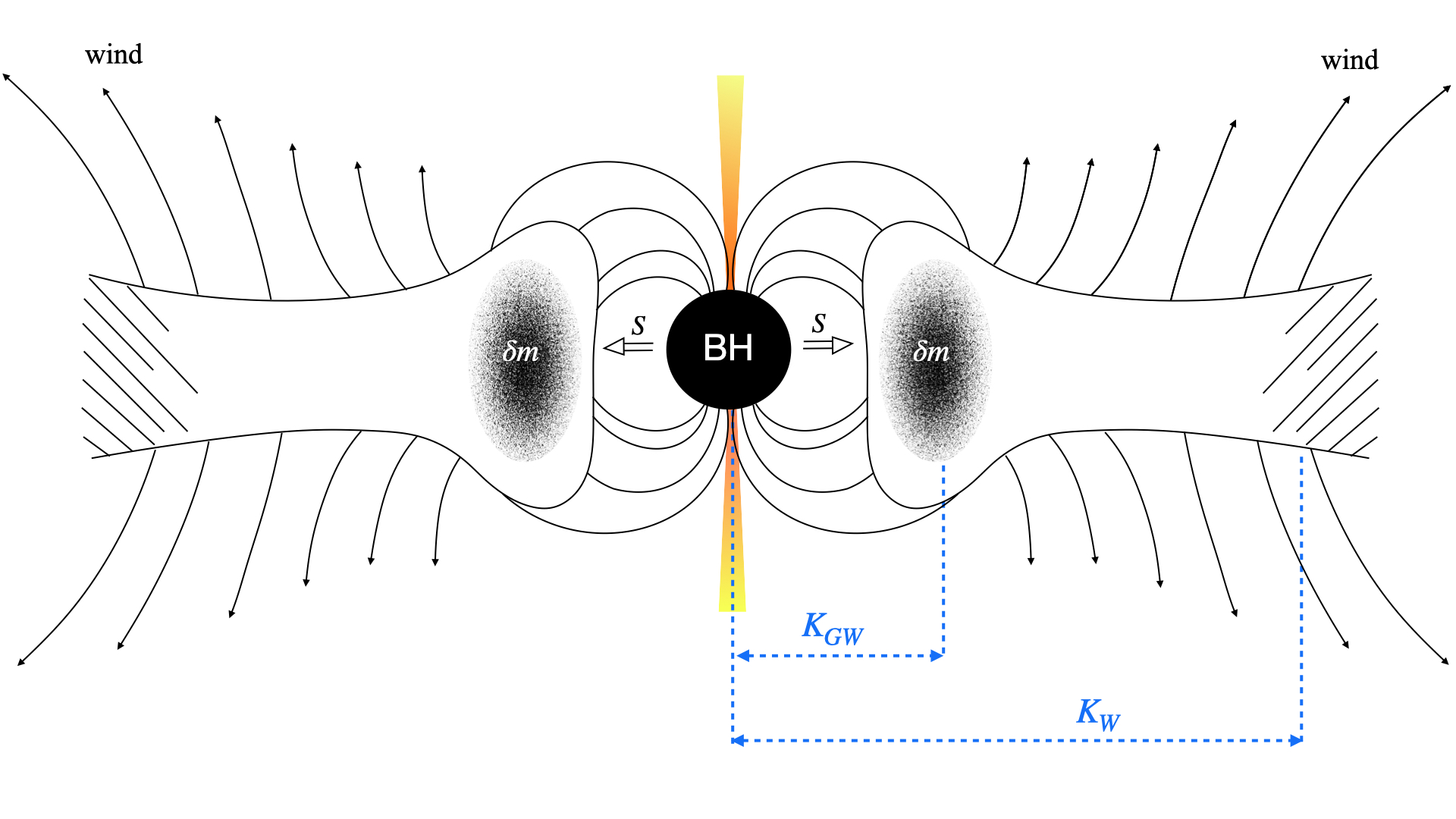}}
    \caption{Schematic of the ISCO-normalized BH central engine, in which radii $R$ are normalized to $R_{{\scaleto{ISCO}{3.5pt}}}= zR_g$.  GW-emission and magnetic winds are radiated at the effective normalized radii $K_{{\scaleto{GW}{3pt}}}$ and, respectively $K_{{\scaleto{W}{3pt}}}$. These are expected to satisfy the inequality $K_{{\scaleto{GW}{3pt}}}<K_{{\scaleto{W}{3pt}}}$, provided the system is in the state of a suspended accretion sustained by a poynting flux $S$ emanating from the BH through Alfv\'en waves mediated by an inner torus magnetosphere surrounding the horizon in its lowest energy state \citep{van01}. A relatively subdominant radiation is released in to an open outflow powering a jet. This process lasts while the BH is effectively spinning faster than the angular velocity of the ISCO. [After \cite{van99,van01,van03b}].
    }
    \label{fig4}
\end{figure*}

{Spinning BHs} (Fig. \ref{fig4}) hold unique prospects for multi-messenger radiation for two reasons. First, the exact Kerr solution shows an energy reservoir in angular momentum that, per unit mass, can far exceed what is offered by baryonic objects i.e. neutron stars and white dwarfs. Secondly, rotating BHs readily spin faster than matter orbiting at the innermost stable circular orbit (ISCO), while the ISCO around rapidly spinning BHs shrinks down to the event horizon (Fig. \ref{fig5-3}). Rapidly spinning BHs therefore, can open an essentially direct interaction between the BH and disk, powering up the surrounding high density matter for a multi-messenger radiation output. This outlook is particularly attractive for the case of BHs surrounded by high density matter in CC-SNe events or BH-NS and DNS mergers. In this case the event is expected to be luminous in GWs from non-axisymmetric mass motions. 

A newly formed rapidly spinning BH during a core collapse is expected to be surrounded by high density matter in a torus (disk). This torus may develop a state of suspended accretion at the ISCO, schematically indicated in Fig. \ref{fig4}. This state is sustained by continuous input from the BH mediated by Alfv\'en waves propagating over closed magnetic fields over the torus magnetosphere supported by the surrounding matter.  The suspended accretion state around this BH appears across a forbidden region enclosed by closed magnetic field lines that collectively form a torus magnetosphere (Appendix \ref{AppA}). This state lasts for the lifetime of rapid spin of the BH during which the torus receives an excess angular momentum from the BH (Fig. \ref{fig5-3}) available for heating of the surrounding high-density matter. This leads to the formation of multiple mass-moments with an inevitable cooling in gravitational radiation and associated MeV-neutrino emission  \citep{van02,van12}.

\subsection{Multi-messenger energy budgets}
\label{ME}

It has long since been proposed that CC-SNe are powered by a rapidly spinning central engine, horboring a compact object \citep{bis70, van99,van01,van11}. A CC-SNe has various radiation channels including baryon poor and baryon rich jets (produced by magnetic winds), dissipation, gravitational radiation, and neutrinos, each of them carrying a different fraction of energy \citep{van03b}.  A calorimetry on the energy radiated in each channel in an energetic CC-SNe powered by a BH, shows $E_{{\scaleto{GW}{3.5pt}}}/E_{J}\simeq 10\%$ where $E_{J}\equiv E_{rot}$ and $E_{{\scaleto{GW}{3.5pt}}}$ are the total spin energy of the BH, and respectively the GW energy. Therefore $E_{{\scaleto{GW}{3.5pt}}}\simeq 3\times 10^{53} \mbox{erg} \simeq 0.1 M_{\odot}c^2$ for a BH of mass $5M_{\odot}$ \citep{van03b}. The energy radiated in magnetic winds is an order of magnitude lower giving $E_{{\scaleto{W}{3.5pt}}}\simeq 3\times 10^{52} \mbox{erg} \simeq 0.01 M_{\odot}c^2$. $E_{{\scaleto{W}{3.5pt}}}$ is to account for both $E_K$ powering the SN explosion and internal photon radiation irradiating the remnant stellar envelope from within. The efficiency of a non-relativistic wind ($\beta_{{\scaleto{W}{3.5pt}}}<<1$) in making a successful SN turns out to be higher than that of an ultra-relativistic wind ($\beta_{{\scaleto{W}{3.5pt}}}\simeq 1$) \citep{van11}. The dissipated energy in thermal or MeV-neutrino emissions is estimated to be in between the other two with $E_{\nu,th}\simeq 5\times 10^{52} \mbox{erg}\simeq 0.03 M_{\odot}c^2$ \citep{van03b}. Beside these channels, a successful GRB carries a small fraction of energy $E_j$ in a baryon-poor jet, amenating from the BH over a moderate half-opening angle on its event horizon (see Fig. \ref{fig4} and \cite{van15}). Therefore, we expect the following ranking of energies
\begin{equation}
E_j<E_{{\scaleto{W}{3.5pt}}}<E_{\nu,th}<E_{{\scaleto{GW}{3.5pt}}}<E_d,
\label{EQN_E}
\end{equation}
where $E_d$ is the energy dissipated unseen in the BH event horizon. Accordingly, the gravitational radiation $E_{{\scaleto{GW}{3.5pt}}}$ is the most opportune observational channel if the event falls within the horizon distance of the GW observatories. In \eqref{EQN_E}, $E_{{\scaleto{GW}{3.5pt}}}$ being maximal formalizes our definition of being loud in \S \ref{Intro}.

\begin{figure*}[t]
    \center{\includegraphics[scale=0.3]{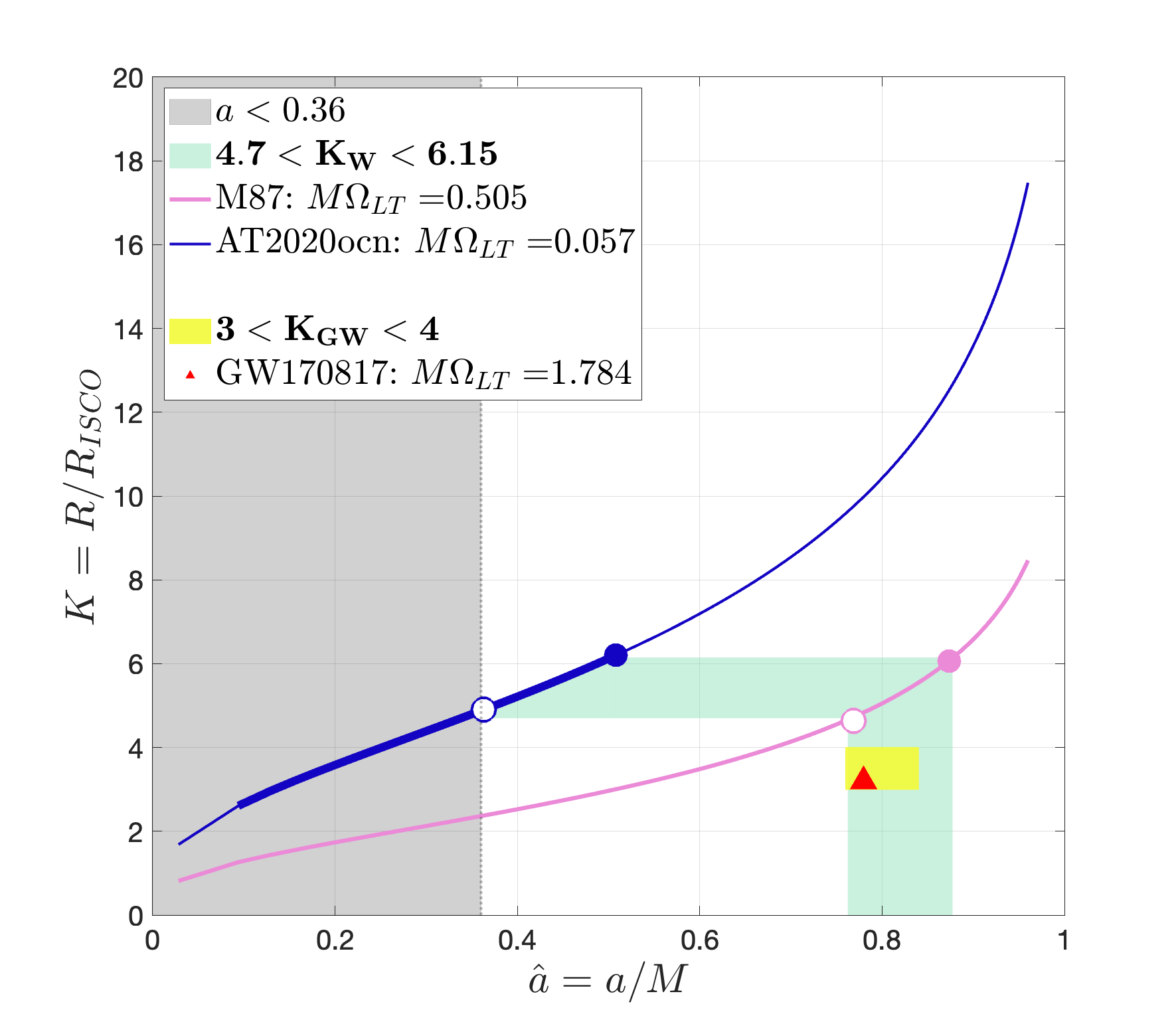}\includegraphics[scale=0.3]{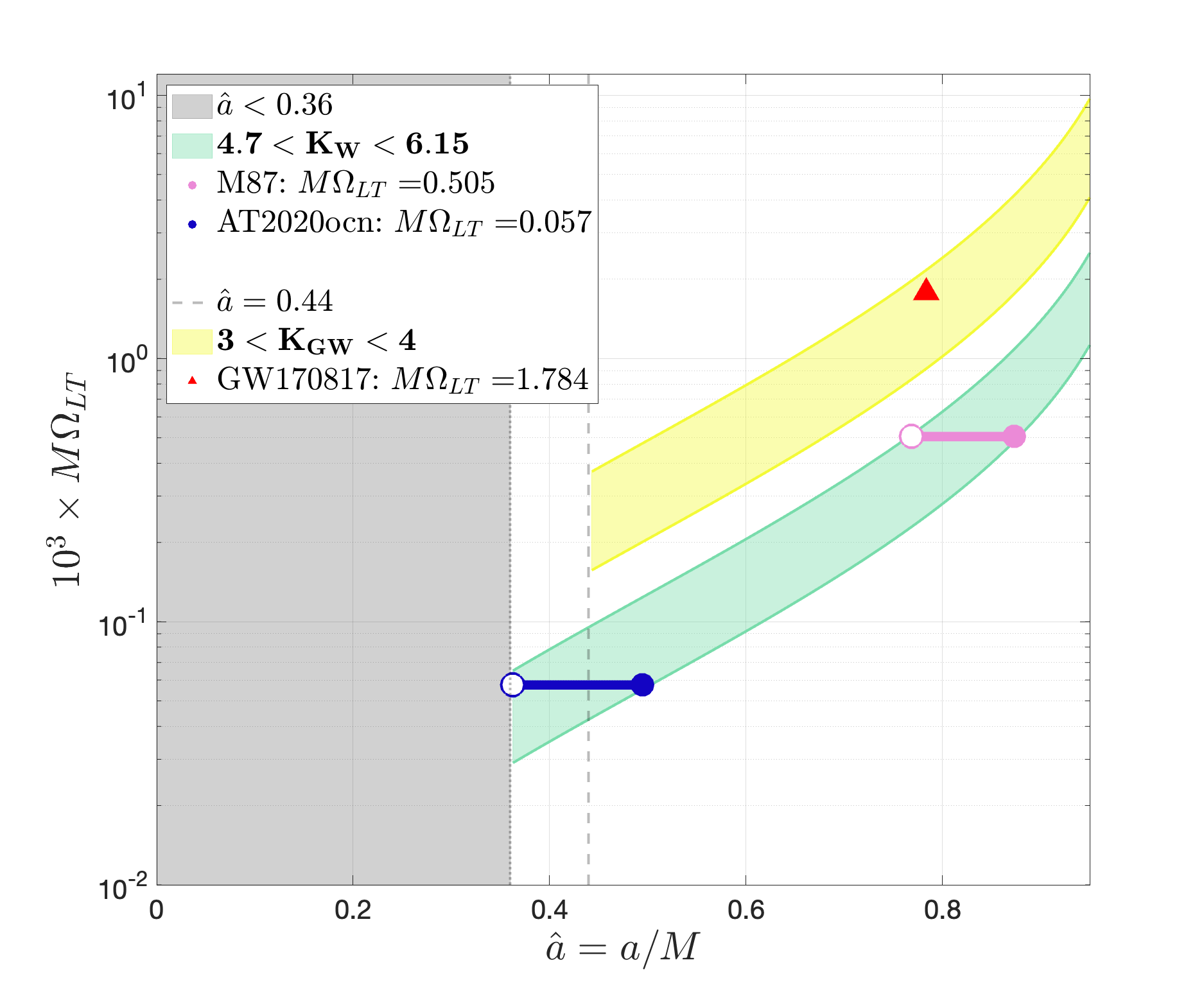}}
    \caption{({\it Left panel}.) Multi-messenger overview of Lense-Thirring precession about BH central engines for M87 and AT2020ocn with an outlook on their stellar mass counterparts such as GW170817B/GRB170817A. Navy and pink curves show $K_{_{\scaleto{W}{3.5pt}}}$ values for a running $\hat{a}=a/M$ based on the the observational values for wind precession frequency for the central SMBH of M87 and AT2020ocn \citep{cui23, pas24}. The $\hat{a}$ values reported  for AT2020ocn multi-wavelength observations are marked by thicker lines. The gray dashed line shows $\hat{a}=0.36$ which is the boundary to stop accretion from the disk in BH-torus system. By the universality of $K_{_{\scaleto{W}{3.5pt}}}$ for BHs, $\hat{a}$ interval for M87 is estimated by horizontal mapping from AT2020ocn (green band) giving $0.76<\hat{a}<0.86$. $K_{{\scaleto{GW}{3pt}}}$ for GW170817B is shown by the red triangle, and the uncertainties marked by the yellow box, illustrative for \eqref{EQN_K}. 
   ({\it Right panel}.) The product of Lense-Thirring angular velocity and mass of the BH ($M\Omega_{{\scaleto{LT}{3.5pt}}}$) for $K_{_{\scaleto{W}{3.5pt}}}$ and $K_{{\scaleto{GW}{3pt}}}$ derived from \eqref{EQN_OLT}, within the uncertainty range of $\hat{a}$ borrowed from the observational results of M87 and AT2020ocn for $K_{_{\scaleto{W}{3.5pt}}}$ ({\it green}), and GW170817B for $K_{{\scaleto{GW}{3pt}}}$ ({\it yellow}) (see Fig. \ref{fig5-1}). Yellow and green bands are observational support for \eqref{EQN_K}. The red triangle shows the central value of $M\Omega_{{\scaleto{LT}{3.5pt}}}$ for GW170817B using \eqref{EQN_OLT}, based on updated results from \cite{van19a}. Recent observational result for M87 and AT2020ocn jet precessions are shown by the pink and respectively navy dots. {Navy line corresponds to the thick navy band for AT2020ocn in the left panel. Likewise, pink line corresponds to the inferred range of $\hat{a}$ for M87 in the left panel.  }}
    \label{fig5-2}  \label{fig5-1}
\end{figure*}

\subsection{ISCO-normalized Lense-Thirring precession}
\label{LT}

Notably, emissions from a BH-torus (disk) system may {show effects of Lense-Thirring precession} \citep{len18} due to the misalignment between the disk and the equatorial plane of the rotating BH. For a BH with angular momentum $J=aM$, this precession has the angular velocity \citep{wil72,ste98}
\begin{equation}
    \Omega_{{\scaleto{LT}{3.5pt}}}=\frac{2J}{R^3}
    \label{EQN_OLT}
\end{equation}
at orbital radius $R$, where $a$ is the specific angular momentum. Here, we use geometric units $c=G=1$, further in what follows unless otherwise stated. 

{We introduce the ISCO-normalized radius $R$,}
\begin{equation}
K=\frac{R}{R_{{\scaleto{ISCO}{3.5pt}}}},
    \label{EQN_isco}
\end{equation}
 where $R_{{\scaleto{ISCO}{3.5pt}}}=zR_g$ is the radius of the ISCO in terms of gravitational radius $R_g=GM/c^2$.
 In applying \eqref{EQN_isco} to multi-messenger emission, different radiation channels may originate from different values of $K$.
For GWs and magnetic winds in particular, we denote these geometric radiation parameters by $K_{{\scaleto{GW}{3pt}}}$ and $K_{_{\scaleto{W}{3.5pt}}}$, respectively (Fig. \ref{fig4}), proposing them to be universal.

 Interestingly, the quadruple GW emission in BH spin-down is expected to be dominant over all other frequencies and radiation channels \eqref{EQN_E}, essentially set by twice the Keplerian orbital frequency determined by the orbital radius $K_{{\scaleto{GW}{3.5pt}}}>1$ in \eqref{EQN_isco}. Accordingly, the forbidden region may well extend beyond the ISCO (Fig. \ref{fig4}). 

\begin{table*}[]
\centering
\footnotesize
\begin{tabular}{|l|l|l|l|l|l|}
\hline
{ Parameter}              & { Definition }                      & { M87}                 & { AT2020ocn}                                                           & { GW170817B}              & { Comments}  \\ \hline
$M$ {[}$M_{\odot}${]}  & \footnotesize{BH mass}                          & $5.7\times 10^9$    & $10^{6.4\pm 0.6}$                                                   & 2.8                    & \footnotesize{[1], [2], [3]  }                                                                                                    \\ \hline 
$\hat{a}$              & \footnotesize{Spin parameter $a/M$}             & $0.76-0.87$       & $0.05-0.5$  & $0.78^{+0.06}_{-0.02}$ & \footnotesize{Eq. \eqref{EQN_OLT}, Figs. \ref{fig5-1}-\ref{fig5-2}, [3]} \\ 
            &            &       &  $0.36-0.5$ &  &  \\ \hline \hline
$E_{J}/E_{J}^{max} [\%]$                  & \footnotesize{Percentage of rotational energy}  & $32-47$       & $0.5-9$                                                       & $32-42$    & \footnotesize{Fig. \ref{fig5-3}  }                                                                    \\ 
         &            &       &  $6-9$ &  &  \\ \hline
$\Omega_{LT}$ {[}rad/s{]} & \footnotesize{Lense-Thirring angular velocity}  & $1.8\times 10^{-8}$ & $4.57\times 10^{-6}$                                                & 170 [27 Hz]                     & \footnotesize{[1], [2], Eq. \eqref{EQN_OLT}  }                                                                        \\ \hline
$M\Omega_{LT}$         & \footnotesize{Mass-normalized angular velocity} & $0.505$             & $0.057$                                                             & 1.784                  & \footnotesize{Eq. \eqref{EQN_OLT}, Fig. \ref{fig5-2}}                                                                  \\ \hline

$K$                  & \footnotesize{ISCO normalized radial distance}  &  \multicolumn{2}{c|}{ $K_{{\scaleto{W}{3pt}}}= 5.42\pm 0.73$ \hspace{-0.5cm}}                                                         & $K_{{\scaleto{GW}{3pt}}}= 3.2^{+0.8}_{-0.2}$ & \footnotesize{Fig. \ref{fig5-1}, [3]   }                                                                    \\ \hline
\end{tabular}
\caption{Some physical properties of rotating BHs in M87, AT2020ocn and their stellar-mass counterpart GW170817B/GRB170817A. Following normalization by mass $M$ and ISCO radius $R_{ISCO}=zR_g$ (Fig. \ref{fig4}), the dimensionless values shown highlight the universality of astrophysical BHs interacting with their surrounding disk or torus across the super-massive and stellar-mass spectrum of transient sources. In particular, the inequality $K_{_{\scaleto{W}{3.5pt}}}>K_{_{\scaleto{GW}{3.5pt}}}$ \eqref{EQN_K} appears consistent. This table summarizes the output presented in \S \ref{discussiona}
 and observational results reported in $^{[1]}$\cite{cui23},$^{[2]}$\cite{pas24} and  $^{[3]}$updated from \cite{van19a}.}
\label{T1}
\end{table*}

{As mentioned above,} GWs associated with energetic CC-SNe powered by rapidly spinning BHs are expected to be quadruple emissions at leading order (\S \ref{UBH}). By Lense-Thiring precession \eqref{EQN_OLT} this will {have accompanying side-bands at frequencies}
\begin{equation}
f_{{\scaleto{GW}{3.5pt}}}\pm f_{{\scaleto{LT}{3.5pt}}},
\label{EQN_sb}
\end{equation}
above and below the central frequency $f_{{\scaleto{GW}{3.5pt}}}$ \citep{arm99,van04b,van04c,ban19}. A detected GW from a BH central engine must therefore, include at least one of these three. 

The Lense-Thirring effect produces both GWs and magnetic wind/jet precessions with the latter at an effective radius larger than that of the former {(Table \ref{T1} and Fig. \ref{fig4})}
\begin{equation}
    K_{{\scaleto{GW}{3pt}}} < K_{_{\scaleto{W}{3.5pt}}},
    \label{EQN_K}
\end{equation}
because the open magnetic field lines are effectively at larger distances from the BH compared to multiple mass-moments, not being directly interacting with the BH {(see Fig. \ref{fig4})}. 

{Lense-Thirring precession effect on GWs (i.e. \eqref{EQN_sb})} is still to be reported but its effect on
magnetic winds is reported recently in the precession of the jet of M87  \citep{cui23}, and AT2020ocn \citep{pas24}: 

\begin{itemize}
\item The mass-normalized jet precession angular velocity of SMBH of M87 is $M\Omega_{{\scaleto{LT}{3.5pt}}}\simeq 0.505$ (Fig. \ref{fig5-1}). The spin of this BH is uncertain \citep{rey21}. Numerical simulations speculate $\hat{a}=a/M\simeq 0.9375$ \citep{cui23}. On the other hand, EM multi-wavelength observation are ambiguous, pointing to a central engine which is low-power by FR I morphology (in the radio), yet high-power by morphology in the optical (as if FR II)  \citep{fan74,bir93,bir95,luc19,cui23}. 
\item The SMBH of AT2020ocn has the mass-normalized angular velocity of $M\Omega_{{\scaleto{LT}{3.5pt}}}\simeq 0.057$ (Fig. \ref{fig5-1}) with reported spin parameter to be $0.05<\hat{a}<0.5$. Accordingly the AT2020ocn BH is slowly rotating falling in to the left hand side of Fig. \ref{fig5-3}.
\end{itemize}

The suspended accretion state requires the BH to be spinning more rapidly than the surrounding matter, $\Omega_{BH}>\Omega_{T}$, i.e. a minimum $\hat{a}>0.36$ and more realistically $\hat{a}>0.44$ (Figs. \ref{fig5-3}-\ref{fig5-4}; \cite{van99,van12}). Putting $\hat{a}=sin\lambda$, the BH rotational energy satisfies $E_{J}=2Mc^2\, \sin^2(\lambda/4)$, i.e. 
\begin{equation}
    E_{J}=4\times 10^{54}\, \sin^2(\lambda/4)\left(\frac{M}{M_{\odot}} \right)\mbox{erg}\lesssim 29\%Mc^2.
    \label{EQN_Erot}
\end{equation}
$E_{J}$ at $\hat{a}=0.36$ is a mere $6\%$ of the maximum rotational energy $E_{J}^{max}=29\%Mc^2$ (Fig. \ref{fig5-3}). Below this, $\hat{a}<0.36$, accretion is inevitable. Prospects for observational consequences of \eqref{EQN_Erot} therefore, coincides with the conditions for the suspended accretion state. In DNS mergers, if the BH is formed shortly after the merger, the BH should be rapidly spinning satisfying $\hat{a}\sim 0.8$ \citep{bai08,van13}, making \eqref{EQN_Erot} accessible. In fact GW170817B/GRB170817A demonstrates just such case at $5.5\sigma$ significance \citep{van23}. Accordingly, we proceed our discussion for BHs satisfying $\hat{a}>0.36$.

With the observational results on $\Omega_{{\scaleto{LT}{3.5pt}}}$ for M87 and AT2020ocn, and considering the uncertainties in BH spin, we leave $\hat{a}=a/M$ to be a running parameter to get $K_{{\scaleto{W}{3pt}}}$ by (\ref{EQN_OLT}-\ref{EQN_isco}) as shown in Fig. \ref{fig5-1}. By the proposed universality of $K_{{\scaleto{W}{3.5pt}}}$ (\S \ref{LT}) and the reported spin parameter range for AT2020ocn, we estimate M87 to horbor a rapidly rotating BH by $0.76<\hat{a}<0.87$ (Fig. \ref{fig5-1}). This is consistent with the FR II morphology in the optical. 

To be more specific, we infer $4.7\lesssim K_{{\scaleto{W}{3pt}}}\lesssim 6.15$. This is remarkably consistent with the constraint \eqref{EQN_K}, given $K_{{\scaleto{GW}{3pt}}}\simeq 3.2^{+0.8}_{-0.2}$ for GW170817B/GRB170818A. {\it This geometrical inequality provides observational support for the proposed universality of these radiation parameters $K_{{\scaleto{W}{3pt}}}$ and $K_{{\scaleto{GW}{3pt}}}$}.

Considering $K_{{\scaleto{GW}{3pt}}}\simeq 3.2_{-0.2}^{+0.8}$, $z\simeq 2.9$, $\hat{a}\simeq 0.78^{+0.06}_{-0.02}$ and $M=2.8M_{\odot}$ for GW170817B (updated from \cite{van19a}), the associated Lense-Thirring frequency inferred from \eqref{EQN_OLT} is estimated to be
\begin{equation}
f_{{\scaleto{LT}{3.5pt}}}\simeq 27\left(\frac{2.8M_{\odot}}{M} \right)\,\mbox{Hz}.
\label{fLT}
\end{equation}
This makes the side-bands \eqref{EQN_sb} to similar events, though challenging, technically observable by LVK.

Table. \ref{T1} summarizes this section.

\section{Horizon distance by scaling GW170817B}
\label{Horizon}

As mentioned in \S \ref{BH}, GW170817B shows the power of GW-observations to confirm its DNS progenitor {\em and} the central engine of GRB170817A.
In fact, this event dramatically demonstrates the GW-emission to be the {\em dominant} channel in multi-messenger radiation, far exceeding what is observed in EM \citep{van03b,moo17,moo18,van23}. 
This opens a new window to probing CC-SNe, the most energetic of which may be likewise dominant in GW-emission, powered by a BH.

Notable recent CC-SNe are the type II SN2023ixf in M101 at 6.4\,Mpc during an engineering run ER15 just prior to LVK O4, SN2024ggi in NGC3621 at about 7\,Mpc and SN2018cow (AT2018cow), 
a {\em Fast Blue Optical Transient} (FBOT) at 60\,Mpc, more recently classified as SN Ic-BL (Ib) \citep{kil23,xia21,gon24,sri24}. 
The estimated mass of the SN2018cow progenitor points to a BH remnant \citep{sun22,sun23,leu20}, distinct from SN1987A, though LVK was not in observing mode during this time. 
SN2023ixf and SN2024ggi are evidently of interest by their close proximity. SN2023ixf and SN2024ggi occurred during LVK observing time, though not in observing mode during the first. Regardless, these diverse CC-SNe events post-GW170817 are illustrative for ample new opportunities beyond mergers to advance above-mentioned transition from detector to observatory. 

To be specific, we will discuss the feasibility of the above in terms of the anticipated horizon distance in un-modeled GW-probes of the physical trigger powering CC-SNe when powered by BH.

\subsection{LVK sensitivity improvement O2-O4}
\label{LVK}

Sensitivity of LVK O4 is improved by a factor $k_D\simeq 1.8$ over O2 \citep{LVK21,LVK24}. For DNS mergers, the horizon distance hereby extends to $\simeq 160$ \citep{LVK24} over the LVK bandwidth of maximal sensitivity (Fig. \ref{fig1})
\begin{eqnarray}
B\simeq 100-250\,\mbox{Hz}.
\label{EQN_B}
\end{eqnarray}

The LVK horizon distance to CC-SNe producing neutron stars is conventionally believed to be limited to galactic events \citep{sri19,var23} with some potential to go beyond by emission from non-axisymmetric tori around (proto-)NS or BHs \citep{pir07,gos16}. A different prospect holds for the above-mentioned energetic SN II and SN Ib/c, that may be powered by rapidly rotating BH as in GRB170817A \citep{van19}. 
Their horizon distance may then extend out to the Local Universe \citep{van01} by virtue of the ample energy reservoir in angular momentum according to the Kerr solution, that, in interaction with surrounding high-density matter, may power multi-messenger emission including GWs similar to (\ref{EQN_ES}) over the course of the lifetime of BH-spin.

\subsection{Mass-scaling of GW170817B/GRB170817A}
\label{mass-scaling}

Specific to BH central engines of CC-SNe, the output (\ref{EQN_ES}) generalizes by mass-scaling $M/M_0$, satisfying
\begin{eqnarray}
E&\simeq& 3.5\%\left(\frac{M}{M_0}\right)M_\odot c^2,\nonumber\\
\Delta F &\simeq& \left( 700 - 200 \right)\left(\frac{M_0}{M}\right)\mbox{Hz}.
\label{EQN_SM}
\end{eqnarray}
Here, we assume a dimensionless Kerr parameter $\hat{a}\simeq 0.8$ similar to that of GW170817B/GRB170817A \citep{van15,van23}. 

For instance, the distant DNS merger GW190425 \citep{abb20a} during O3 points to a remnant of $M=3.4\,M_\odot$, suggesting considerable overlap with above-mentioned $\Delta F$ and $B$. 
On the other hand, the remnant of the candidate NS-BH merger GW190814 with $M \simeq 25.8 M_\odot$ \citep{abb20b} suggests $\Delta F$ to fall below $B$, though still within the LVK bandwidth of sensitivity. 

Energetic CC-SNe are expected to produce BHs with masses broadly correlated to their pre-collapse He-core mass. Their envelope-stripped and rotating progenitors have a mass-estimate of $M_{He}$ \citep{koc14}. 
Combined with an estimated mass-fraction thereof collapsing to a BH, this shows the formation of rapidly rotating BH with $M\gtrsim 5M_\odot$ (\cite{van04} and references therein). 
Scaling (\ref{EQN_SM}) hereby predicts
\begin{eqnarray}
{E} & \simeq 7 \% M_\odot c^2,~~{\Delta F} & \simeq [100, 380]\,\mbox{Hz}. 
\label{EQN_S}
\end{eqnarray}
Fortuitously, $\Delta F$ in (\ref{EQN_S}) and $B$ in (\ref{EQN_B}) overlap while $E$ is substantially larger than the canonical output $2.5\%M_\odot c^2$ of GW170817 \citep{abb17} giving a second gain factor of $k_M\simeq 2$. This can also be seen from scaling of $h_{char}$ in $E_{{\scaleto{GW}{3.5pt}}}$ of (\ref{EQN_ES}) according to \citep{fla98,van01,cut02}
\begin{eqnarray}
h_{char} = \frac{\sqrt{2}}{\pi D} \sqrt{\frac{E}{\left| \Delta F \right|}}
\sim \frac{M}{D}\sqrt{\frac{{\cal E}_{GW}}{M}} \propto \frac{M}{M_0}
\label{EQN_h}
\end{eqnarray}
for a source at distance $D$, where proportionality follows from (\ref{EQN_SM}). This, along with aforementioned improvement in detector sensitivity $k_D$ over O2 presents a compelling outlook to probing the central engines of energetic CC-SN in the Local Universe during O4.

The horizon distance to (\ref{EQN_S}) derives from a signal-to-noise ratio (SNR) in characteristic strain amplitude $h_{char}$ of the source relative to the strain noise-amplitude of the detector.
By numerical evaluation, Fig. \ref{fig2} shows the gain in signal-to-noise by $\Delta F$ in (\ref{EQN_ES}) scaled by $M/M_0$. 
Energetic CC-SNe hereby receive a gain $k_n\gtrsim1.27$ in $h_{char}$ for $M\gtrsim 5M_\odot$, shifting $\Delta F$ downwards to increasingly overlap with $B$ in (\ref{EQN_B}). 

Combining the above, a total gain in horizon distance to energetic CC-SNe during O4 satisfies
\begin{eqnarray}
k=k_{n}\, k_M\, k_D\simeq 4.4.
\label{EQN_k}
\end{eqnarray}

\begin{figure}[t]
    \center{\includegraphics[scale=0.215]{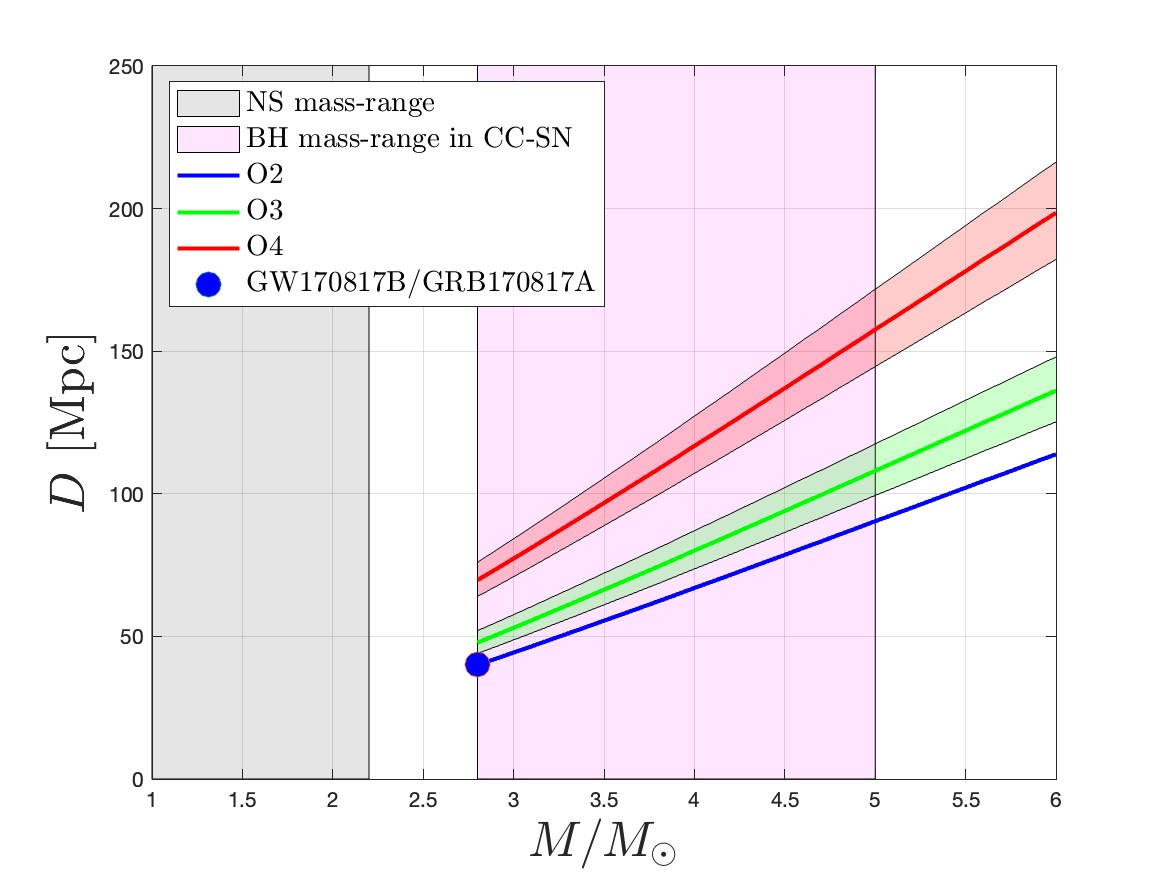}}
    \caption{Horizon distances as a function of mass of initially rapidly spinning black holes by mass-scaling of 
    GW170817B/GRB170817A ({\em blue dot}) and detector 
    improvement in O4 over O2.
    Highlighted is the expected mass range of black holes produced in
   energetic CC-SNe ({\em magenta}). The upper limit of the excluded zone, representative for neutron stars ({\em gray}), is uncertain, indicated by a mass-gap of a few tenths of a solar mass ({\em white}). Likewise, the upper limit
    of black hole masses produced in energetic CC-SNe is uncertain, awaiting detailed probes by gravitational-wave observations. 
    }
    \label{fig3}
\end{figure}

Fig. \ref{fig3} shows the estimated horizon distance for CC-SNe 
as a function of a generic BH mass-range $3M_\odot<M<5M_\odot$ \citep[e.g.][]{bou93}. 
Including the estimate $M_0\simeq 2.8M_\odot$ of GW170817B/GRB170817A, 
this extends to $2.8M_\odot \lesssim M < 5M_\odot$ with
horizon distances $70\lesssim D/\mbox{Mpc} \lesssim 160$.

\section{Observational outlook}
\label{Outlook}

The probability of observing GW-transients depends critically on local cosmic volume, $V$, proportional to the cube of the horizon distance, the total duration $T$ of an observational run, duty cycle of the gravitational-wave detectors adjusted for data-quality (below) and any trend in source event rate with redshift \citep{abc23}. 
SNe are generally more frequent with redshift, raising to a maximum around the epoch of maximal cosmic star formation \citep{mel12,mad14,str15}. 
The probability of a detection hereby scales with at least $VT$, normalized to the local event rate at redshift zero. 

Corrected for beaming (\S1), 
the true event rates of long and short GRBs are  similar though not identical with different redshift distributions \citep{van14}.  
The former is indicative of the above-mentioned relatively small branching ratio 
(\ref{EQN_G1a}), leaving a true event rate of LGRBs of about 1\,yr$^{-1}$ within 100\,Mpc. 

Observing one GW170817-GRB170817A event during O2 and none during O3 is statistically consistent considering the detectors duty cycle and yield factors \citep{abc23}.
By the above-mentioned similar event rates of normal long and short GRBs, it is likewise consistent to have no GW-detection of a SN Ib/c during the runs O2-3.

{The ground rule of a GW detection is a correlated response in two or more well-separated detectors. In more practical terms,  we infer a GW detection by confirmation in at least two detectors operating at comparable performance. We quantify observational epochs accordingly by the {\em detector yield} $y$, the fraction of time when, for instance, both H1 and L1 are successfully operating simultaneously at nominal sensitivity 
\citep{abc23}. }
The detector yield factor $y$ for such joint operation varies with observational runs, reducing the effective observational time to $Ty$. A detailed study shows $VTy$ for O2-3 to be comparable; the sensitivity of detectors in O3 improved over O2 but at the cost of a lower $y$ for long-duration data segments lasting seconds \citep{abc23}, relevant to our GW-signal of interest (\ref{EQN_ES}).

The observational run O4, recently starting its second half O4b, promises to be different, provided that the sensitivity is higher than that of O3 and the detector yield factor $y$ of O4 is at least on par with O2, when $y$ was relatively high by consistent joint H1L1-operations over time.

Probing CC-SNe for their central engines is to be carried out by un-modeled GW searches at {\em detector noise-limited sensitivity} by time-symmetric search methods with equal sensitivity to ascending and descending GW-chirps \citep{van23,abc23}. 
Ultra-long duration GW-emission may be produced by superluminous supernovae \citep{van17}.

Probing CC-SNe for (\ref{EQN_SM}-\ref{EQN_S}) may be pursued in blind all-sky searches or in directed searches to specific events \citep{van16}. 
Notably, TOOs are provided by modern automated surveys such as the {\em Zwicky Transient Facility} \citep{gra19} and accompanying high-energy EM emission by existing and planned satellite missions such as THESEUS \citep{ama18}. For the latter, only a small fraction of events may be producing a detectable GW-emission. While (\ref{EQN_G2}) may be larger than (\ref{EQN_G1a}), most SN Ib/c may be unsuccessful in producing jets and breakout {producing a (L)LGRBs}. Indeed, most may be producing NS rather than BH. Even when producing a BH, they might be slowly rotating with insufficient spin-energy to match (\ref{EQN_B}-\ref{EQN_SM}), or central engines with insufficient lifetime to allow jet breakout and a successful GRB, yet making a GW signal at (\ref{EQN_G2}).
Taking into account finite detector yield factors $y$, an EM-survey of $\sim 10^3$ SN Ib/c is required for a few hundred SN Ib/c to be probed by LVK for their GW-emission similar to (\ref{EQN_ES}). 

\begin{table*}[]
\begin{tabular}{c|c|c|c|c||c|}
\cline{2-6}
                             & $N$ (within 100\,Mpc)         & LGRB (true) & LLGRB (ture)  & Failed GRB        & GW-SN       \\ \hline
\multicolumn{1}{|c||}{SN Ibc} & $\sim 100$ & 1-3                                                   & $\sim$10                                               & few$\times$ LLGRB & few tens \\ \hline
\multicolumn{1}{|c||}{SN II}  & $\gtrsim 200$ & not observed                                         & not observed                                          & $N/A$              & $\gtrsim 0$     \\ \hline
\end{tabular}
\caption{{Number of CC-SNe Ibc and II within 100Mpc per year and their associated (L)LGRBs versus expected number of events that are loud in gravitational radiation. Crucially, there is window for a substantial fraction of GW-SNe amongst the failed GRB-SNe.}}
\label{T2}
\end{table*}

\subsection{Gravitational wave search window}
GW-emission of a CC-SN happens during the active lifetime of its central engine about the explosion time ($t_0$). SNe are serendipitously discovered ($t_O$) through their light curves hours to weeks after the shock breakout ($t_b$).  This introduces
\begin{eqnarray}
    \Delta_0&=&t_b-t_0,\nonumber\\
    \Delta_1&=&t_O-t_b,\\
    \Delta_2&=&\Delta_0+\Delta_1=t_O-t_0.\nonumber
    \label{EQN_delta}
\end{eqnarray}

$\Delta_0$ can be estimated for different type of CC-SNe based on the radius of the progenitor star and the expansion velocity giving \citep{mau09,van11,lym16,sah09,van11,tad19,ham06,san19,gut17}
\begin{equation}
\mbox{few hours} \lesssim \Delta_0\lesssim \mbox{several days}.
\end{equation}

A more precise estimation of $\Delta_0$ can be done for individual events. $\Delta_1$ is particularly uncertain, as it inevitably derives from extrapolating the SN light curve. 
In practice
\begin{equation}
    \Delta_0 << \Delta_1.
\end{equation}
If we are lucky to detect the SN early enough, $\Delta_1 \lesssim 1$d.

SN2011hd, a type IIb supernova in M51 \citep{arc11}, stands out as it was detected in X-rays. It signals a moment of discovery close to the shock breakout, whence $\Delta_1\sim 12$\,hr (Fig. 5 of \cite{sod12}).
{If an event is observed well before the peak in the SN light curve and at high energy, $\Delta_2$ is estimated to be at least several hours, 
as the temperature of the shock front at $t_O$ generally reduces after the peak. }
For instance, SN2023ixf and SN2024ggi light curves both point to similar $\Delta_2\sim 1\,$d \citep{hos23,shr24}. 
Such timing reduces to (tens of) seconds only for directed searches to GRB-SNe, but these TOOs are exceptionally rare (\S1). 

\section{Discussion and conclusion}
\label{discussion}
CC-SNe are more promising sources for GW observation compared to DNS mergers because of their larger event rates \S \ref{Intro}. Even core-collapse events that are not successful in GRB-emission, can potentially be loud in GW radiation (\S\ref{BH}). Accordingly, their event rate may well be a few times more numerous than that of DNS mergers. For BH central engines, calorimetry on the energy budget of different radiation channels shows that GW-emission may be dominant \eqref{EQN_E}. 

The above appears opportune for rotating BHs when spinning faster than the surrounding matter (Fig. \ref{fig5-4}). It should be emphasized that the duration of the output radiation can be long as it is determined by the lifetime of BH spin. This timescale readily covers seconds to tens of seconds \citep{van99,van03b}.

In surveying energetic CC-SNe in the Local Universe, some may reveal central engines of this kind. The recent observation of the wind precession of M87 and AT2020ocn lend credence to the exciting possibility of herein finding {Lense-Thirring precession} counterparts in gravitational radiation (\S \ref{discussiona}).

\subsection{Horizon distance to GW from CC-SNe}

For the present LVK observational runs, we estimate a horizon distance of $\sim 160$\,Mpc to BH central engines of energetic CC-SNe (Fig. \ref{fig3}), based on the ample energy reservoir in angular momentum of rotating black holes of {mass $\lesssim 5M_\odot$}. For the observer, we introduce, an effective horizon distance corrected for the detector yield factor \citep{abc23} by
\begin{equation}
D_{eff}=y^{\frac{1}3}D\simeq [45-100]\, \mbox{Mpc},
    \label{EQN_HD}
\end{equation}
where $D=[70-160]$\,Mpc shown in Fig. \ref{fig3} (\S \ref{Horizon}) is the horizon distance. {This is based on O4 detector sensitivity, and detector yield factor $y\simeq 0.28$ for long duration burst searches (the case of $W=8$\,s in \cite{abc23}) if similar to that of O2 and O3.} 
This outlook may be contrasted with the
galactic horizon distance ($<1$\, Mpc) in conventional LVK searches for long duration GW-bursts \citep{abb16,abb20c,szc21}. It accounts for  the undesired bias in detecting compact binary coalescence (CBC) \citep{abb23} rather than observing extreme transient events more generally. \eqref{EQN_HD} removes this bias through the open window offered by the broadband sensitivity of the LVK observatories in searches at detector-{limited sensitivity also for un-modeled signals}. 

{{SNe Ibc and II event rates} \cite{cap99,wei11,cap15} show that there are about 100 and, respectively, $\gtrsim 200$ events within 100\,Mpc per year. Considering \eqref{EQN_G1a}-\eqref{EQN_fG} and by \eqref{EQN_HD}, there are about 0.1-few LGRBs associated with these events (maybe unobserved due to beaming) within the horizon distance of $D_{eff}$. Including LLGRB-SNe in \eqref{EQN_fGL}, this fraction may increase by a factor of $\sim 10$, {increasing (L)LGRB events} within $D_{eff}$ that are potentially loud in GW. Accordingly by \eqref{EQN_G3} (which includes LGRBs, LLGRBs and failed GRBs), we expect $>10$ of SNe Ibc and II within the 100\,Mpc per year to be targets of opportunity for aforementioned GW searches (GW-SN). These results are summarized in Table \ref{T2}.  The estimated rate of events interesting for gravitational radiation within the lower bound of 50\,Mpc in \eqref{EQN_HD} is therefore $\dot{N}_{\rm{GW-SN}}\times (50/100)^{(1/3)} \gtrsim$ few per year.}

This outlook derives from \eqref{EQN_SM} and \eqref{EQN_k}, upon scaling of GW170817B (\ref{EQN_ES}) by application of time-symmetric searches for un-modeled GW-signals at detector-limited sensitivity on par with CBC \citep{van23}. 

Our estimated horizon distance is considerably beyond
what may be expected in GW-emission in process of core-collapse, in the formation and evolution of rotating (proto-)NS \citep[e.g.][]{lin23}. By their limited spin energy, the horizon distance to NS central engines is considerably smaller, especially at GW-frequencies about the LVK bandwidth of maximal sensitivity (\ref{EQN_B}).
Nevertheless, spin-down of proto-NS might be detectable at extra-galactic distances.

\subsection{Possible long duration burst GW search results}

In light of the above, a probe of a directed search from a CC-SN within the 160Mpc, using a method with detector-noise limited sensitivity in the Local Universe may reveal the following for the central engine and remnant:

\begin{itemize}
\item 
\textbf{Case A.}
No detection at low frequency (0-900Hz). Based on \eqref{EQN_SM}, this outcome rules out a BH central engine and points to an NS central engine, likely leaving an NS remnant and, accordingly, a progenitor mass below $20M_\odot$. 
Note that a BH remnant can not be ruled out in the exceptional case of a belated collapse to a BH.
Regardless, {a rapidly rotating NS central engine} is expected to have its GW signature above 1kHz, well above the LVK bandwidth of maximal sensitivity (\ref{EQN_B}).
\vspace{0.2cm}
\item 
\textbf{Case B.}
The detection of a GW-signal below 1\,kHz but energetically below what is expected from (\ref{EQN_SM}). This may, instead, point to the spin-down of a slowly rotating proto-NS, pointing to an NS remnant and a progenitor mass below 20\,$M_\odot$. Alternatively, the signal is produced from a distant BH central engine which is not associated with the CC-SN at hand.
\vspace{0.2cm}
\item 
\textbf{Case C.} The detection of a GW-signal below 1\,kHz consistent with \eqref{EQN_SM}, evidencing a BH central engine, a BH remnant and hence a progenitor mass greater than $20M_\odot$. 
Possibly, this outcome would require corroborating evidence from the kinetic energy in supernova ejecta (Fig. \ref{fig1a}) \citep{van11}. 
Notably, these signals may appear with characteristic side-bands \eqref{EQN_sb} from disk precession (\S\ref{discussiona}).
\end{itemize}

Subject to the availability of simultaneous H1L1-data at nominal sensitivity, Case A-C provide radically novel observational constraints on the physical nature of the central engine, remnant, and the mass of the progenitor, otherwise notoriously challenging to determine by EM$\nu$-observations alone. 

Current targets of opportunities include SN II events SN2023ixf and SN2024ggi at $\sim 7$Mpc, and the SN Ic events SN2018cow at 60\,Mpc, SN2020bvc at 120\,Mpc and SN2020oi at 23\,Mpc \citep{kil23,xia21,gon24,sri24}. These events can be probed at LVK detector-limited sensitivity for their un-modeled output in gravitational radiation by application of Butterfly Matched Filtering \citep{van23}. According to Case A-C, such is expected to conclusively break the degeneracy between an NS or BH central engine.

\section*{acknowledgments}
The authors thank the anonymous reviewer for detailed reading and constructive comments. This research is supported, in part, by NRF RS-2024-00334550. 

\appendix

\section{Branching ratios of SN Ib/c}
\label{Apc}

{By their association to LGRBs, SN Ic-BL stand out as candidates for GW-SNe. 
However, the observed branching ratio of {SN Ib/c to LGRBs}, is quite small}. We have
\begin{equation}
\Gamma_{obs}=\frac{\dot{N_{obs}}\left[\mbox{\rm GRB-SN}\right]}{\dot{N_{obs}}[\rm SN Ib/c]}\simeq 10^{-4}
    \label{EQN_G1a}
\end{equation}
{This notoriously small ratio is due} to the beaming factor $f_b^{-1}\sim 100$ in GRB-emission \citep{fra01,van03,pod04,van04,del06,gue07,van19,lev20}. Therefore, we have a true-to-observed event rate of GRB-SNe, satisfying \citep{woo06,hjo12,cor16,ann20,cor23}
\begin{eqnarray}
\Gamma^{\prime}=\frac{\dot{N}\left[\mbox{\rm GRB-SN}\right]}{\dot{N}\left[{\rm SN Ib/c}\right]}\simeq f_b^{-1}10^{-4}\simeq 1\%-3\%.
\label{EQN_fG}
\end{eqnarray}
 {This outlook is relevant to directed searches of gravitational-wave emission from the parent population of energetic SNIb/c. Consistently observational results show that SN Ic-BL are about 5\% of the total population of SN explosions. In comparison, a tiny fraction of about 1\% of these SNe are associated with LGRBs, which is about an order of magnitude less than LLGRBs \citep{cor23}, {though because of uncertainties these fractions are not well constrained} \citep{cor16,cor23}. Still, it can introduce an additional branching ratio}
 \begin{equation}
\Gamma^{\prime \prime}=\frac{\dot{N}\left[\mbox{\rm LLGRB-SN}\right]}{\dot{N}\left[{\rm SN Ib/c}\right]}\simeq f_{bL}^{-1}10^{-4}\simeq 10\%.
     \label{EQN_fGL}
 \end{equation}
 
{SN1998bw/GRB980425 is the first observation of a GRB associated with an SN, followed by more observations afterward. Recent estimations for SN1998bw like SN predicts their population to be 19\% of total SN Ic-BL \citep{cor16,cor23}.}

{Crucially, a successful GRB-SN requires the central engine to be long-lived for about ten seconds \citep{van04,van12,nak12,dec22}.  
Shorter lifetimes are expected to produce LLGRBs or failed GRB-SNe {with less-energetic jet or no jet} breakout of the progenitor stellar envelope, the latter with unknown branching ratio}
\begin{equation}
  \Gamma^{\prime \prime \prime}=
  \frac{\dot{N}\left[\mbox{failed \rm GRB-SN}\right]}{\dot{N}\left[{\rm SN Ib/c}\right]}.
\label{EQN_G4}  
\end{equation}
In defining \eqref{EQN_G4}, we consider the failed GRB-SN events to be the continuation of successful (LL)GRB-SN originating from the same type of central engine, whereby $\Gamma^{\prime}+\Gamma^{\prime \prime}+\Gamma^{\prime \prime \prime}\le 1$. 

In the core collapse of massive stars, {inevitably, the angular momentum of the central engine} momentum will be broadly distributed with a positive correlation to lifetime and energetic output subject to efficiency. The more extreme cases of large angular momentum may support a successful GRB-SN, while less extreme may produce a SN sans GRB by insufficient lifetime for the above-mentioned successful jet breakout. Even such failed GRB-SNe still holds promise to be loud in GW-emission. Conceivably this extends to the case of negligible angular momentum failing to produce a central engine altogether, and hence no SN.
 
Illustrative for successful central engines is the rapidly spinning BH powering GW170817B/GRB170817A for the duration of 3.7\,s $(<10\,$s) \citep{van19a,van23}. While $\Gamma^{\prime \prime}$ is not well-constrained, conceivably
\begin{eqnarray}
\frac{\Gamma^{\prime \prime \prime}}{\Gamma^{\prime} +\Gamma^{\prime \prime}}=\frac{\dot{N}\left[\mbox{failed \rm GRB-SN}\right]}{\dot{N}\left[\mbox{(L)LGRB-SN}\right]} \simeq \mbox{few}.
\label{EQN_G2}
\end{eqnarray}
Considering the diversity in the central engine of SN Ib/c, the fraction of all SNe Ib/c that are loud in GWs (GW-SN) hereby satisfies
\begin{equation}
    \Gamma_{GW}\gtrsim \Gamma^{\prime}+\Gamma^{\prime \prime}+\Gamma^{\prime \prime \prime}\gg \Gamma_{obs}.
    \label{EQN_G3}
\end{equation}
By the event rate of SN Ib/c in the Local Universe, 
(\ref{EQN_G3}) points to an event {rate of $>10$} GW-SN within 100\,Mpc yr$^{-1}$. 
In directed searches, SN Ib/c may hereby be significantly more attractive than (normal) long GRBs.

\section{Formation and evolution of BH central engines in CC-SNe}
\label{AppA}

\begin{figure*}[h!] 
    \center{\includegraphics[scale=0.3]{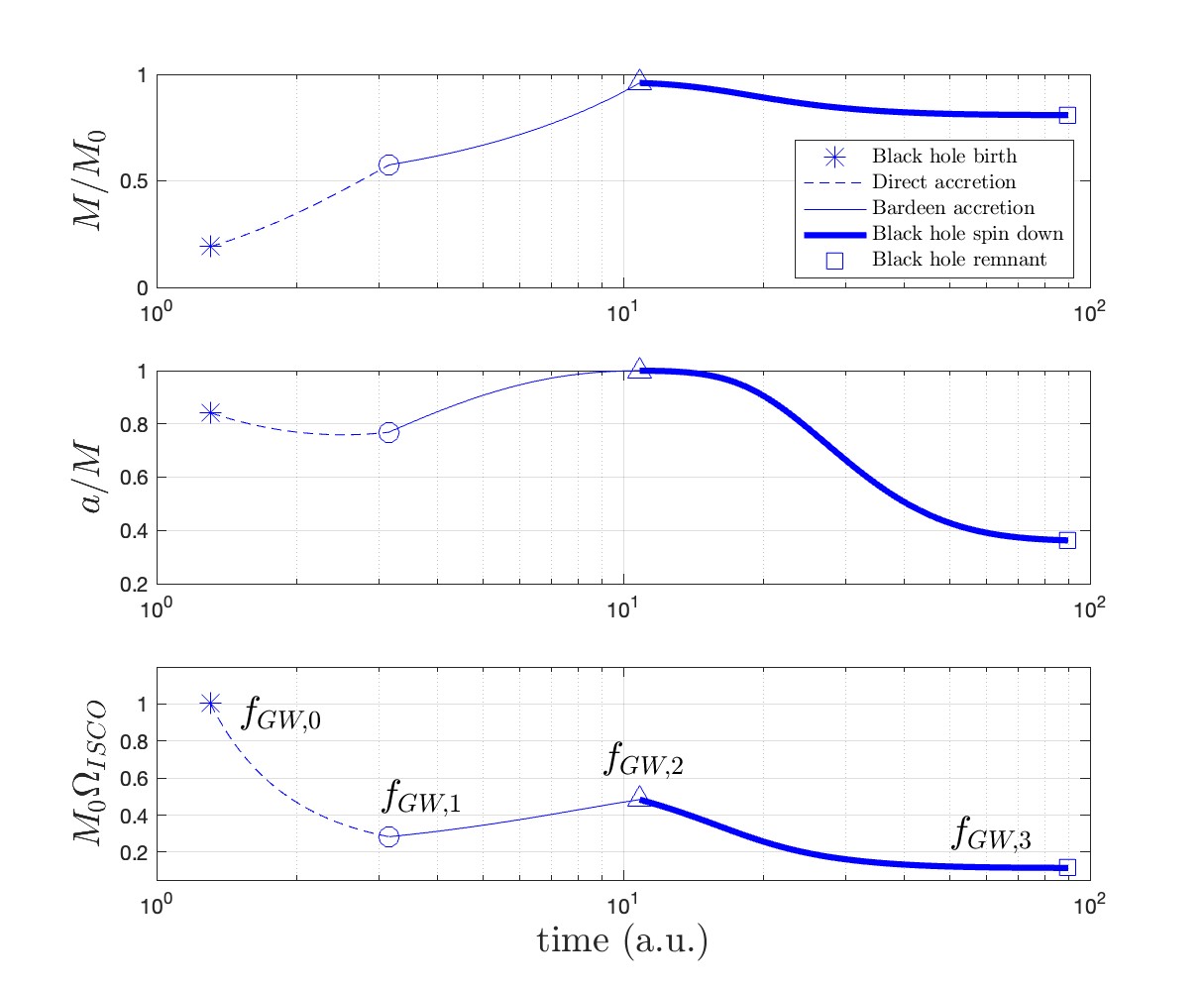}}
    \caption{Following the birth of the BH shown is the opening (dashed), middlegame (slide line) and endgame (solid thick line) of its evolution, referring to, respectively, direct accretion of initially low specific angular momentum, Bardeen accretion and BH spin-down against a surrounding high-density inner accretion disk. Included are the various gravitational-wave frequencies $f_{\scaleto{GW,i}{4.5pt}}$ at the various junctions. Here $M_0$ is the progenitor Helium core mass. {Not all three phases are necessarily passed through in all events. For instance, a sufficiently low angular momentum progenitor may enter the middle game but not beyond by premature accretion of all of $M_0$ into the BH. [Reprinted from \cite{van17}.]}}
    \label{fig5-5}
\end{figure*}

Following birth in core collapse of a massive star, a BH experiences three potential phases of evolution (Fig. \ref{fig5-5}; \cite{van17}):
\begin{itemize}
\item {\it Opening:} Initial surge of BH mass by in-fall of low specific angular momentum from the inner regions of the progenitor stellar envelope. This direct in-fall creates a rapid increase in mass without dramatic change in $\hat{a}$. This specific angular momentum gradually increases as accretion continues, derived from the outer layers of the remnant stellar envelope. This surge, hereby, ceases when the critical value of the specific angular momentum defined by the ISCO of the BH is reached.
\item {\it Middlegame:} When a disk is formed, the BH may still continue to grow by Bardeen accretion of matter dripping across the ISCO on the viscous time scale of the newly formed accretion disk. This process may take the BH to a near-extremal state.  
\item{\it Endgame:} A rapidly spinning BH with $\Omega_{H}>\Omega_{T}$ may produce a suspended accretion state by a transfer of energy and angular momentum to the accretion disk via inner torus magnetosphere. This heats up the torus giving rise to multiple mass moments, allowing for a balance by cooling through the accompanying prodigious output in gravitational radiation. This state lasts for the lifetime of rapid spin of the BH.
\end{itemize}

The process of BH spin-down during the {\it Endgame} is effectively described by \citep{van99}
\begin{eqnarray}
\dot{M}=-\Omega_T\dot{J},\, \dot{J}=-\kappa e_k (\Omega_H-\Omega_T),
    \label{EQN_PDE}
\end{eqnarray}
where $J$ is the BH angular momentum, $\kappa$ parametrizes the interaction strength between the BH and surrounding matter with specific kinetic energy $e_k$ \citep{van03b,van15,van16,van17}. {The associated BH luminosity transferred onto the surrounding high-density matter is available for catalytic conversion into various radiation channels, described by the equations of suspended accretion \citep{van01}.}

The BH luminosity is released predominantly in the form of gravitational radiation during the ensuing state of suspended accretion. {This state lasts for the lifetime of BH spin, which scales inversely with $e_{k}$ in \eqref{EQN_PDE}}. The output is accompanied by minor emissions in an ultra-relativistic baryon-poor jet emanating from the BH in its lowest energy state, and MeV-neutrinos and magnetic winds from the disk \eqref{EQN_E}. GW luminosity of BH during the spin-down follows from the quadruple gravitational radiation formula \citep{pet63,van2000,van19}
\begin{equation}
L_{GW}=\frac{32}{5}\left(\frac{\delta m}{M} \right)\left(\frac{1}{zK_{GW}} \right)^5L_0=2\times 10^{51}\left(\frac{\xi}{0.1} \right)^2\left(\frac{\sigma}{0.01} \right)^2\left(\frac{4}{zK_{{\scaleto{GW}{3pt}}}} \right)^5 \mbox{erg\, s}^{-1},
    \label{EQN_L}
\end{equation}
where, $L_0=c^5/G$, $\xi=\delta m/M_T$ and $\delta m$ is the mass inhomogeneity in the torus created due to the heating. \eqref{EQN_PDE} and \eqref{EQN_L} present a 2+2 parameter space for Kerr BH evolution and its GW output, respectively defined by ($M$,\, $\hat{a}=a/M$) and ($\sigma$,\,$\xi$).

\begin{figure*}[h!] 
    \center{\includegraphics[scale=0.24]{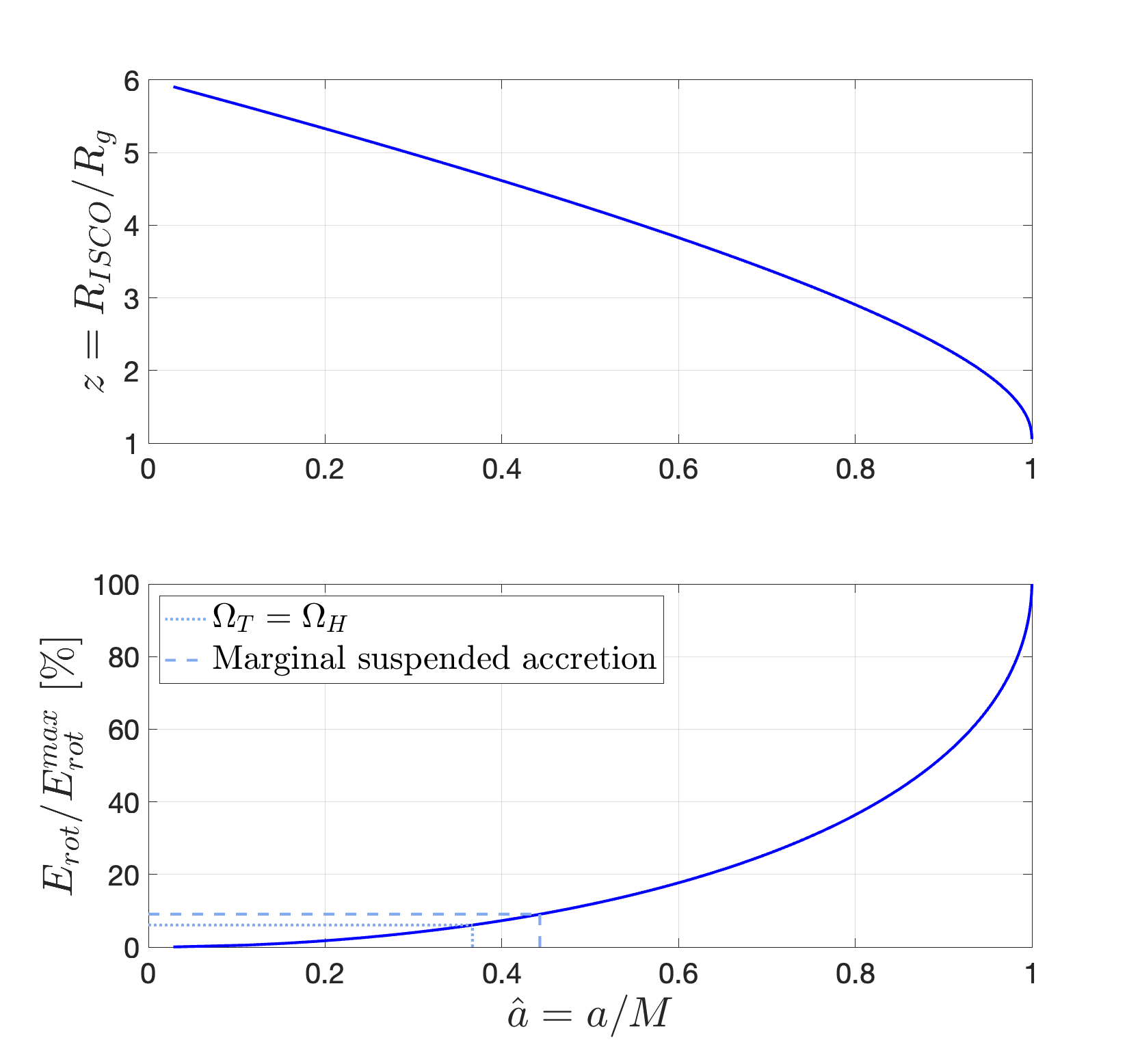}
    \includegraphics[scale=0.20]{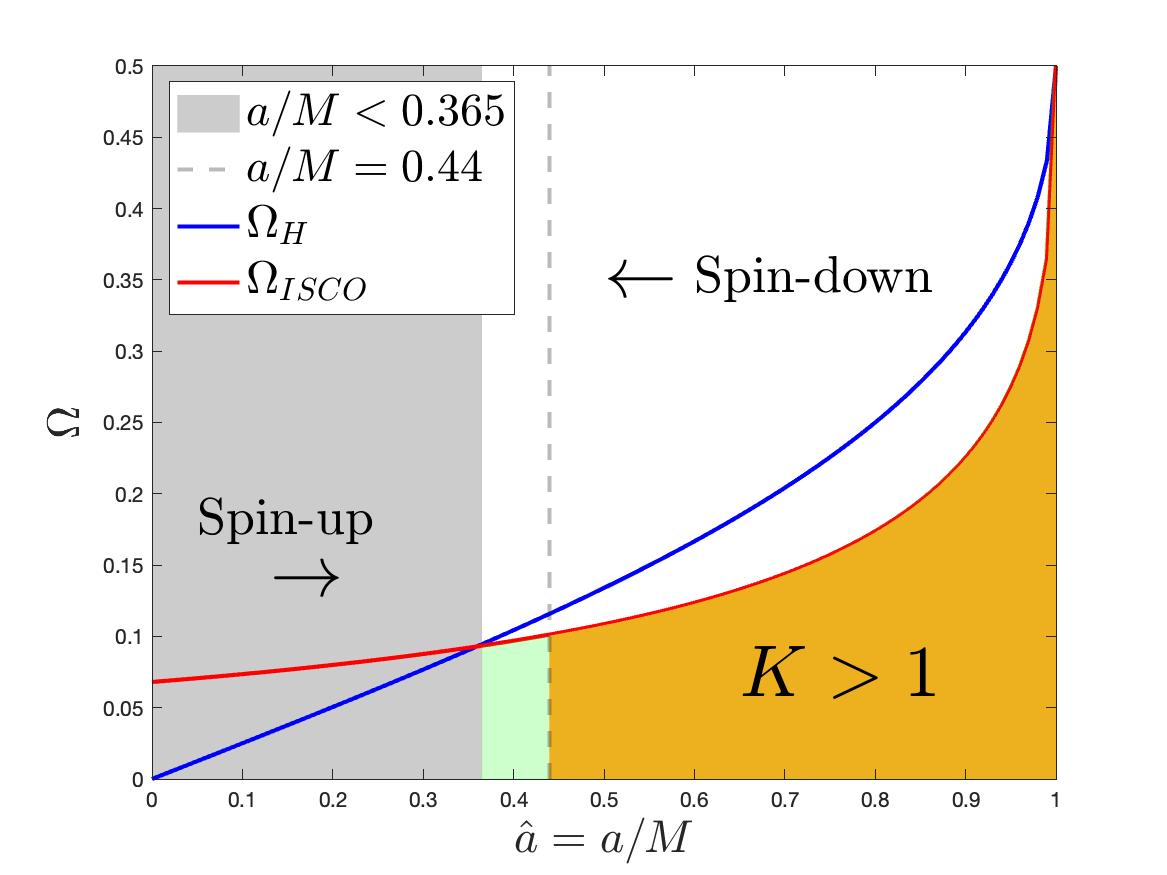}}
    \caption{({\it Left top panel.}) Radius of ISCO, $R_{{\scaleto{ISCO}{3.5pt}}}$, as a function of specific angular momentum $a=J/M$ here expressed as $\hat{a}=a/M$ - the dimensionless spin parameter of a Kerr BH. Rapidly rotating BHs have smaller $R_{{\scaleto{ISCO}{3.5pt}}}$. In its lowest energy state \citep{van01}, it
    facilitates strong interactions with surrounding matter via a torus magnetosphere. 
    We use $R_{{\scaleto{ISCO}{3.5pt}}}$ in a radial normalization of the interaction between the BH and the surrounding high-density matter to parameterize the ensuing multi-messenger radiation \eqref{EQN_isco}. 
    ({\it Right bottom panel.}) Spin energy $E_{J}$ as a function of $\hat{a}$, plotted relative to the maximal spin energy of an extremal Kerr BH ($\hat{a}=1$). 
    The threshold $\hat{a}\simeq0.44$ with $E_{J}/E_{J}^{max}\simeq 9\%$ is the marginal suspended accretion state with no excess energy for radiation in other channels (dashed line; \citep{van12}). Higher spin parameters are required for additional radiation in GWs and MeV-neutrinos. The transition to accretion from the disk is at $\hat{a}\simeq0.36$ (dotted line), where the rotational energy of the BH is further reduced to $E_{J}/E_{J}^{max}\simeq 6\%$. Lower $\hat{a}$ results in accretion with possible black hole spin-up. The intermediate region of $0.36<\hat{a}<0.44$ is a gray area in which $E_{J}$ does not allow disk accretion, but accretion from disk winds is present.
    {\it Right panel.} Schematic overview of the BH spin-down/up when the BH is rotating faster/slower than the surrounding matter with no-slip boundary conditions to its torus magnetosphere and slip boundary conditions on the BH event horizon (Fig. \ref{fig4}). The angular velocities of the BH $\Omega_{H}$ and surrounding matter $\Omega_{{\scaleto{ISCO}{3.5pt}}}$ at the limit of $R_{T}=R_{{\scaleto{ISCO}{3.5pt}}}$ are shown respectively in blue and red curves (Fig. \ref{fig5-3}). The former case, $\Omega_{H}>\Omega_{{\scaleto{ISCO}{3.5pt}}}$ is promising for its potential to energize the surrounding matter, reprocessing this input in different radiation channels.  These various radiations originate at their respective ISCO-normalized radii $K$ (orange). The transition between the two cases is theoretically indicated by $\Omega_{H}=\Omega_{{\scaleto{ISCO}{3.5pt}}}$ equivalent to $\hat{a}\simeq0.36$ (gray), and more realistically at $\hat{a}\simeq0.44$ (green)\citep{van12}. }
    \label{fig5-3}  \label{fig5-4}
\end{figure*}

In this process, the spin of the BH decreases, the ISCO expands and, consequently, the radius of the torus increases (Fig. \ref{fig5-3}). The duration of this spin-down is determined by the lifetime of rapid spin of the BH \citep{van17} that scales with the inverse mass-ratio of torus-to-BH, $\sigma=M_T/M$,
\begin{equation}
T_s\simeq 15\,s\left (\frac{0.01}{\sigma}\right )\left (\frac{M}{7M_\odot}\right ).
    \label{EQN_duration}
\end{equation}
  Eventually, $\Omega_H$ reaches $\Omega_T$, completing the {\it Endgame} with a slowly rotating remnant $(\hat{a}\simeq 0.36$, 
  Figs. \ref{fig5-5}-\ref{fig5-4}).

GW170817B/GRB170817A offers the first observation of a descending chirp powered by the spin-down of a Kerr black hole of $M_0\simeq 2.8M_\odot$ against high-density matter with an initial dimensionless spin $\hat{a}\simeq 0.8$ (\S \ref{BH}, \S \ref{discussiona}). The mass of BHs formed in CC-SNe are expected to be larger, e.g., covering a fiducial range $3M_\odot \lesssim M\lesssim 5M_\odot$ given their diversity (Figs. \ref{fig1b}-\ref{fig1a}). By universality of BHs with no memory of their progenitors except for mass and angular momentum (\S \ref{BH}), they are expected to have similar behavior subject to mass scaling.

\section{Mass-scaling of LVK sensitivity to descending GW-chirps}
\label{APb}
\begin{figure}
    \center{
    \includegraphics[scale=0.20]{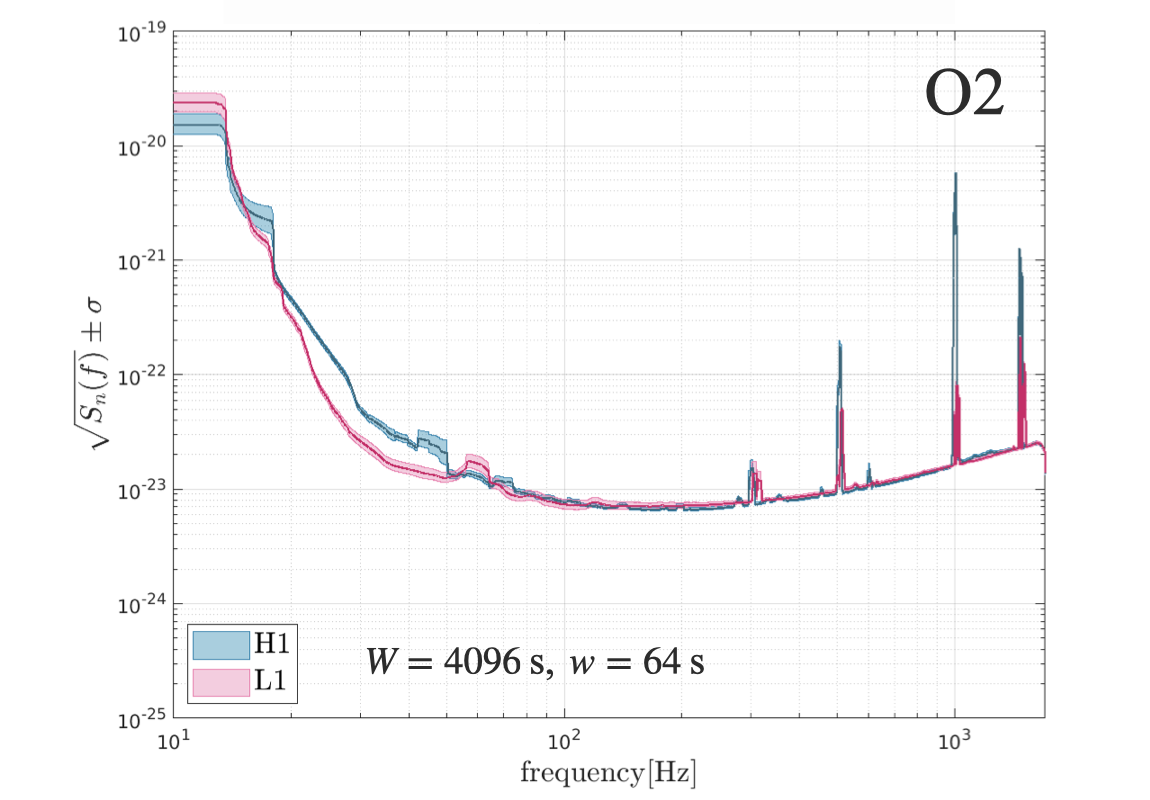}
    \includegraphics[scale=0.19]{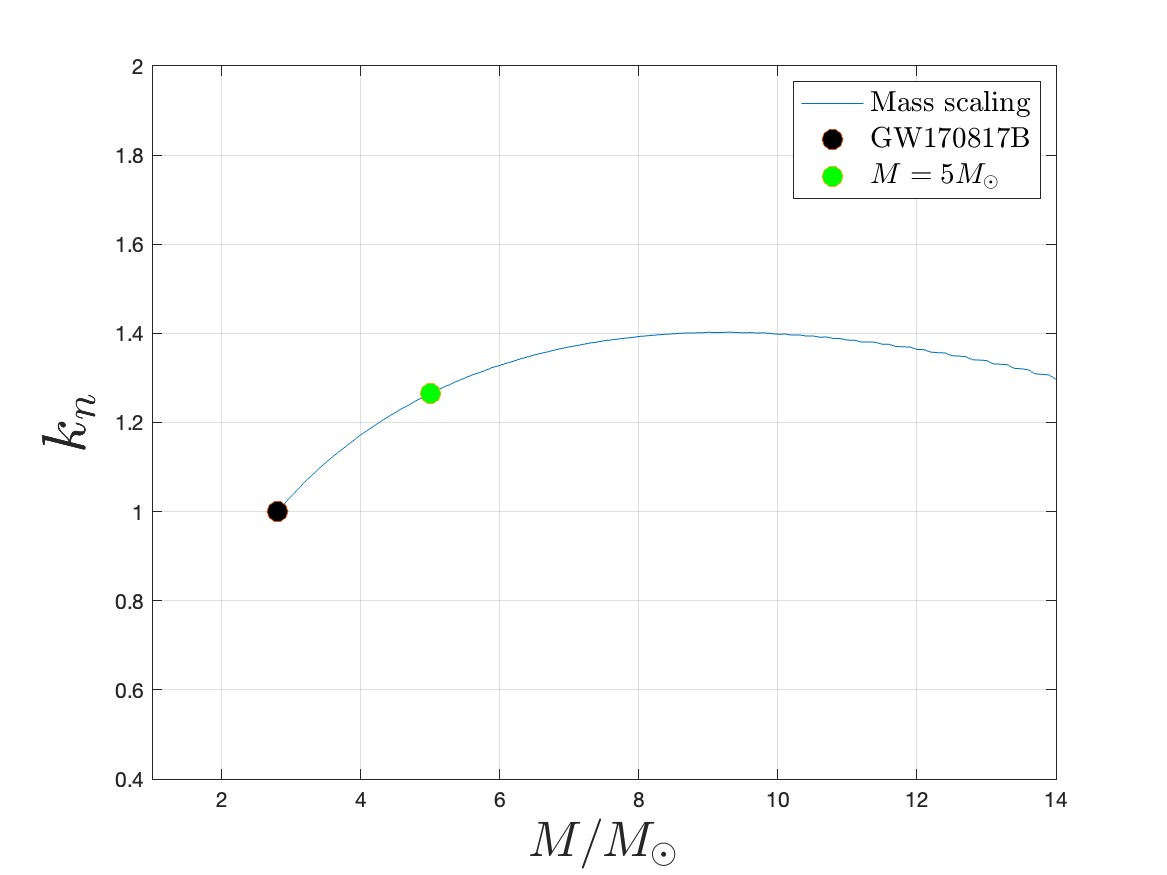}
    }
    \caption{
    ({\it Left panel.}) Detector strain-noise amplitude in O2, smoothed by a moving average in frequency. Shown is the mean with $1\sigma$ variation of strain H1 (blue) and L1 (red). The bandwidth of maximal sensitivity $B\simeq 100-250\,$Hz corresponds to the minimum of these curves. O4 realizes an improvement by a factor $k_D\simeq1.8$ compared to O2 with a commensurate increase in horizon distances. (Reprinted from \citep{abc23}.)
    ({\it Right panel.}) Signal-to-noise ratios in observing BH spin-down improves as the frequency sweep $\Delta f_{{\scaleto{GW}{3.5pt}}}=\Delta f_{{\scaleto{GW}{3.5pt}}}(M_0/M)$ moves 
    closer to the bandwidth of maximal sensitivity $B$ (Fig. \ref{fig1}) when $M\ge M_0$, 
    where $M_0\simeq2.8M_\odot$ is the mass of the black hole in GW170817B/GRB170817A.
    Highlighted is the gain $k_n\simeq1.27$ for $M=5M_\odot$ expected to be produced in 
    CC-SNe. Additional gain $k_M$ derives from the scaling $E_{{\scaleto{GW}{3.5pt}}}={\cal E}_{GW}(M/M_0)$, where ${\cal E}_{GW}$ denotes the energy output (\ref{EQN_ES}).}
    \label{fig1} \label{fig2}
\end{figure}

A descending GW-chirp powered by an initially rapidly rotating Kerr BH
energizing high-density matter beyond the ISCO (Fig. \ref{fig4}) satisfies a canonical GW-frequency $f_{GW}$ scaled by $1/M$ and GW-energy $E_{GW}$ scaled by $M$ for a given initial spin parameter $\hat{a}$ \eqref{EQN_SM}. 
In light of the LVK detector strain-noise amplitude (Fig. \ref{fig1}), in case of an event with BH mass $M\gtrsim 3M_\odot$, scaling
of $f_{GW}\sim 1/M$ moves the output closer to or into the bandwidth $B=100-250\,$Hz of maximal sensitivity. We express this by the gain $k_n$ obtained by projecting the output frequency on the noise spectrum of detectors, with B to be the most sensitive band Fig. \ref{fig1}.
Total gain in horizon distance results from $k_n$, together with two other gain factors \eqref{EQN_k}. Scaling of $E_{GW}\sim M$ provides an additional gain $k_M$, and the gain $k_D$ realized by improvement of the detector, now O4 over O2 during GW170817B. We summarize the combined gain in (\ref{EQN_k}). 

\clearpage
\bibliographystyle{aasjournal}
\bibliography{mybibfile}

\begin{thebibliography}{}
\expandafter\ifx\csname natexlab\endcsname\relax\def\natexlab#1{#1}\fi
\providecommand{\url}[1]{\href{#1}{#1}}
\providecommand{\dodoi}[1]{doi:~\href{http://doi.org/#1}{\nolinkurl{#1}}}
\providecommand{\doeprint}[1]{\href{http://ascl.net/#1}{\nolinkurl{http://ascl.net/#1}}}
\providecommand{\doarXiv}[1]{\href{https://arxiv.org/abs/#1}{\nolinkurl{https://arxiv.org/abs/#1}}}

\bibitem[{A.(1993)}]{bir93}
A., B.~J. 1993, in Space Telescope Sci. Inst. Symp. 6. Astrophysical Jets, ed. C.~O. D.~Burgarella, M.~Livio (Cambridge University Press), 263--304

\bibitem[{{Abbott} {et~al.}(2016){Abbott}, {Abbott}, {Abbott}, {Abernathy}, {Acernese}, {Ackley}, {Adams}, {Adams}, {Addesso}, {Adhikari}, {Adya}, {Affeldt}, {Agathos}, {Agatsuma}, {Aggarwal}, {Aguiar}, {Aiello}, {Ain}, {Ajith}, {Allen}, {Allocca}, {Altin}, {Anderson}, {Anderson}, {Arai}, {Araya}, {Arceneaux}, {Areeda}, {Arnaud}, {Arun}, {Ascenzi}, {Ashton}, {Ast}, {Aston}, {Astone}, {Aufmuth}, {Aulbert}, {Babak}, {Bacon}, {Bader}, {Baker}, {Baldaccini}, {Ballardin}, {Ballmer}, {Barayoga}, {Barclay}, {Barish}, {Barker}, {Barone}, {Barr}, {Barsotti}, {Barsuglia}, {Barta}, {Bartlett}, {Bartos}, {Bassiri}, {Basti}, {Batch}, {Baune}, {Bavigadda}, {Bazzan}, {Behnke}, {Bejger}, {Bell}, {Bell}, {Berger}, {Bergman}, {Bergmann}, {Berry}, {Bersanetti}, {Bertolini}, {Betzwieser}, {Bhagwat}, {Bhandare}, {Bilenko}, {Billingsley}, {Birch}, {Birney}, {Biscans}, {Bisht}, {Bitossi}, {Biwer}, {Bizouard}, {Blackburn}, {Blair}, {Blair}, {Blair}, {Bloemen}, {Bock}, {Bodiya}, {Boer}, {Bogaert}, {Bogan}, {Bohe}, {Bojtos}, {Bond},
  {Bondu}, {Bonnand}, {Boom}, {Bork}, {Boschi}, {Bose}, {Bouffanais}, {Bozzi}, {Bradaschia}, {Brady}, {Braginsky}, {Branchesi}, {Brau}, {Briant}, {Brillet}, {Brinkmann}, {Brisson}, {Brockill}, {Brooks}, {Brown}, {Brown}, {Brown}, {Buchanan}, {Buikema}, {Bulik}, {Bulten}, {Buonanno}, {Buskulic}, {Buy}, {Byer}, {Cadonati}, {Cagnoli}, {Cahillane}, {Calder{\'o}n Bustillo}, {Callister}, {Calloni}, {Camp}, {Cannon}, {Cao}, {Capano}, {Capocasa}, {Carbognani}, {Caride}, {Casanueva Diaz}, {Casentini}, {Caudill}, {Cavagli{\`a}}, {Cavalier}, {Cavalieri}, {Cella}, {Cepeda}, {Cerboni Baiardi}, {Cerretani}, {Cesarini}, {Chakraborty}, {Chalermsongsak}, {Chamberlin}, {Chan}, {Chao}, {Charlton}, {Chassande-Mottin}, {Chen}, {Chen}, {Cheng}, {Chincarini}, {Chiummo}, {Cho}, {Cho}, {Chow}, {Christensen}, {Chu}, {Chua}, {Chung}, {Ciani}, {Clara}, {Clark}, {Cleva}, {Coccia}, {Cohadon}, {Colla}, {Collette}, {Cominsky}, {Constancio}, {Conte}, {Conti}, {Cook}, {Corbitt}, {Cornish}, {Corpuz}, {Corsi}, {Cortese}, {Costa}, {Coughlin},
  {Coughlin}, {Coulon}, {Countryman}, {Couvares}, {Coward}, {Cowart}, {Coyne}, {Coyne}, {Craig}, {Creighton}, {Cripe}, {Crowder}, {Cumming}, {Cunningham}, {Cuoco}, {Dal Canton}, {Danilishin}, {D'Antonio}, {Danzmann}, {Darman}, {Dattilo}, {Dave}, {Daveloza}, {Davier}, {Davies}, {Daw}, {Day}, {DeBra}, {Debreczeni}, {Degallaix}, {De Laurentis}, {Del{\'e}glise}, {Del Pozzo}, {Denker}, {Dent}, {Dergachev}, {De Rosa}, {DeRosa}, {DeSalvo}, {Dhurandhar}, {D{\'\i}az}, {Di Fiore}, {Di Giovanni}, {Di Girolamo}, {Di Lieto}, {Di Pace}, {Di Palma}, {Di Virgilio}, {Dojcinoski}, {Dolique}, {Donovan}, {Dooley}, {Doravari}, {Douglas}, {Downes}, {Drago}, {Drever}, {Driggers}, {Du}, {Ducrot}, {Dwyer}, {Edo}, {Edwards}, {Effler}, {Eggenstein}, {Ehrens}, {Eichholz}, {Eikenberry}, {Engels}, {Essick}, {Etzel}, {Evans}, {Evans}, {Everett}, {Factourovich}, {Fafone}, {Fair}, {Fairhurst}, {Fan}, {Fang}, {Farinon}, {Farr}, {Farr}, {Favata}, {Fays}, {Fehrmann}, {Fejer}, {Ferrante}, {Ferreira}, {Ferrini}, {Fidecaro}, {Fiori}, {Fiorucci},
  {Fisher}, {Flaminio}, {Fletcher}, {Fournier}, {Frasca}, {Frasconi}, {Frei}, {Freise}, {Frey}, {Frey}, {Fricke}, {Fritschel}, {Frolov}, {Fulda}, {Fyffe}, {Gabbard}, {Gair}, {Gammaitoni}, {Gaonkar}, {Garufi}, {Gaur}, {Gehrels}, {Gemme}, {Genin}, {Gennai}, {George}, {Gergely}, {Germain}, {Ghosh}, {Ghosh}, {Giaime}, {Giardina}, {Giazotto}, {Gill}, {Glaefke}, {Goetz}, {Goetz}, {Gondan}, {Gonz{\'a}lez}, {Gonzalez Castro}, {Gopakumar}, {Gordon}, {Gorodetsky}, {Gossan}, {Gosselin}, {Gouaty}, {Grado}, {Graef}, {Graff}, {Granata}, {Grant}, {Gras}, {Gray}, {Greco}, {Green}, {Groot}, {Grote}, {Grunewald}, {Guidi}, {Guo}, {Gupta}, {Gupta}, {Gushwa}, {Gustafson}, {Gustafson}, {Hacker}, {Hall}, {Hall}, {Hammond}, {Haney}, {Hanke}, {Hanks}, {Hanna}, {Hannam}, {Hanson}, {Hardwick}, {Harms}, {Harry}, {Harry}, {Hart}, {Hartman}, {Haster}, {Haughian}, {Heidmann}, {Heintze}, {Heitmann}, {Hello}, {Hemming}, {Hendry}, {Heng}, {Hennig}, {Heptonstall}, {Heurs}, {Hild}, {Hoak}, {Hodge}, {Hofman}, {Hollitt}, {Holt}, {Holz},
  {Hopkins}, {Hosken}, {Hough}, {Houston}, {Howell}, {Hu}, {Huang}, {Huerta}, {Huet}, {Hughey}, {Husa}, {Huttner}, {Huynh-Dinh}, {Idrisy}, {Indik}, {Ingram}, {Inta}, {Isa}, {Isac}, {Isi}, {Islas}, {Isogai}, {Iyer}, {Izumi}, {Jacqmin}, {Jang}, {Jani}, {Jaranowski}, {Jawahar}, {Jim{\'e}nez-Forteza}, {Johnson}, {Jones}, {Jones}, {Jonker}, {Ju}, {Haris}, {Kalaghatgi}, {Kalmus}, {Kalogera}, {Kamaretsos}, {Kandhasamy}, {Kang}, {Kanner}, {Karki}, {Kasprzack}, {Katsavounidis}, {Katzman}, {Kaufer}, {Kaur}, {Kawabe}, {Kawazoe}, {K{\'e}f{\'e}lian}, {Kehl}, {Keitel}, {Kelley}, {Kells}, {Kennedy}, {Key}, {Khalaidovski}, {Khalili}, {Khan}, {Khan}, {Khan}, {Khazanov}, {Kijbunchoo}, {Kim}, {Kim}, {Kim}, {Kim}, {Kim}, {Kim}, {King}, {King}, {Kinzel}, {Kissel}, {Kleybolte}, {Klimenko}, {Koehlenbeck}, {Kokeyama}, {Koley}, {Kondrashov}, {Kontos}, {Korobko}, {Korth}, {Kowalska}, {Kozak}, {Kringel}, {Krishnan}, {Kr{\'o}lak}, {Krueger}, {Kuehn}, {Kumar}, {Kuo}, {Kutynia}, {Lackey}, {Landry}, {Lange}, {Lantz}, {Lasky}, {Lazzarini},
  {Lazzaro}, {Leaci}, {Leavey}, {Lebigot}, {Lee}, {Lee}, {Lee}, {Lee}, {Lenon}, {Leonardi}, {Leong}, {Leroy}, {Letendre}, {Levin}, {Levine}, {Li}, {Libson}, {Littenberg}, {Lockerbie}, {Loew}, {Logue}, {Lombardi}, {Lord}, {Lorenzini}, {Loriette}, {Lormand}, {Losurdo}, {Lough}, {L{\"u}ck}, {Lundgren}, {Luo}, {Lynch}, {Ma}, {MacDonald}, {Machenschalk}, {MacInnis}, {Macleod}, {Maga{\~n}a-Sandoval}, {Magee}, {Mageswaran}, {Majorana}, {Maksimovic}, {Malvezzi}, {Man}, {Mandel}, {Mandic}, {Mangano}, {Mansell}, {Manske}, {Mantovani}, {Marchesoni}, {Marion}, {M{\'a}rka}, {M{\'a}rka}, {Markosyan}, {Maros}, {Martelli}, {Martellini}, {Martin}, {Martin}, {Martynov}, {Marx}, {Mason}, {Masserot}, {Massinger}, {Masso-Reid}, {Mastrogiovanni}, {Matichard}, {Matone}, {Mavalvala}, {Mazumder}, {Mazzolo}, {McCarthy}, {McClelland}, {McCormick}, {McGuire}, {McIntyre}, {McIver}, {McManus}, {McWilliams}, {Meacher}, {Meadors}, {Meidam}, {Melatos}, {Mendell}, {Mendoza-Gandara}, {Mercer}, {Merilh}, {Merzougui}, {Meshkov}, {Messenger},
  {Messick}, {Metzdorff}, {Meyers}, {Mezzani}, {Miao}, {Michel}, {Middleton}, {Mikhailov}, {Milano}, {Miller}, {Miller}, {Millhouse}, {Minenkov}, {Ming}, {Mirshekari}, {Mishra}, {Mitra}, {Mitrofanov}, {Mitselmakher}, {Mittleman}, {Moggi}, {Mohan}, {Mohapatra}, {Montani}, {Moore}, {Moore}, {Moraru}, {Moreno}, {Morriss}, {Mossavi}, {Mours}, {Mow-Lowry}, {Mueller}, {Mueller}, {Muir}, {Mukherjee}, {Mukherjee}, {Mukherjee}, {Mukund}, {Mullavey}, {Munch}, {Murphy}, {Murray}, {Mytidis}, {Nardecchia}, {Naticchioni}, {Nayak}, {Necula}, {Nedkova}, {Nelemans}, {Neri}, {Neunzert}, {Newton}, {Nguyen}, {Nielsen}, {Nissanke}, {Nitz}, {Nocera}, {Nolting}, {Normandin}, {Nuttall}, {Oberling}, {Ochsner}, {O'Dell}, {Oelker}, {Ogin}, {Oh}, {Oh}, {Ohme}, {Oliver}, {Oppermann}, {Oram}, {O'Reilly}, {O'Shaughnessy}, {Ott}, {Ottaway}, {Ottens}, {Overmier}, {Owen}, {Pai}, {Pai}, {Palamos}, {Palashov}, {Palomba}, {Pal-Singh}, {Pan}, {Pankow}, {Pannarale}, {Pant}, {Paoletti}, {Paoli}, {Papa}, {Paris}, {Parker}, {Pascucci}, {Pasqualetti},
  {Passaquieti}, {Passuello}, {Patricelli}, {Patrick}, {Pearlstone}, {Pedraza}, {Pedurand}, {Pekowsky}, {Pele}, {Penn}, {Pereira}, {Perreca}, {Phelps}, {Piccinni}, {Pichot}, {Piergiovanni}, {Pierro}, {Pillant}, {Pinard}, {Pinto}, {Pitkin}, {Poggiani}, {Popolizio}, {Post}, {Powell}, {Prasad}, {Predoi}, {Premachandra}, {Prestegard}, {Price}, {Prijatelj}, {Principe}, {Privitera}, {Prix}, {Prodi}, {Prokhorov}, {Puncken}, {Punturo}, {Puppo}, {P{\"u}rrer}, {Qi}, {Qin}, {Quetschke}, {Quintero}, {Quitzow-James}, {Raab}, {Rabeling}, {Radkins}, {Raffai}, {Raja}, {Rakhmanov}, {Rapagnani}, {Raymond}, {Razzano}, {Re}, {Read}, {Reed}, {Regimbau}, {Rei}, {Reid}, {Reitze}, {Rew}, {Ricci}, {Riles}, {Robertson}, {Robie}, {Robinet}, {Rocchi}, {Rolland}, {Rollins}, {Roma}, {Romano}, {Romano}, {Romanov}, {Romie}, {Rosi{\'n}ska}, {Rowan}, {R{\"u}diger}, {Ruggi}, {Ryan}, {Sachdev}, {Sadecki}, {Sadeghian}, {Salconi}, {Saleem}, {Salemi}, {Samajdar}, {Sammut}, {Sanchez}, {Sandberg}, {Sandeen}, {Sanders}, {Santamaria}, {Sassolas},
  {Sathyaprakash}, {Saulson}, {Sauter}, {Savage}, {Sawadsky}, {Schale}, {Schilling}, {Schmidt}, {Schmidt}, {Schnabel}, {Schofield}, {Sch{\"o}nbeck}, {Schreiber}, {Schuette}, {Schutz}, {Scott}, {Scott}, {Sellers}, {Sentenac}, {Sequino}, {Sergeev}, {Serna}, {Setyawati}, {Sevigny}, {Shaddock}, {Shahriar}, {Shaltev}, {Shao}, {Shapiro}, {Shawhan}, {Sheperd}, {Shoemaker}, {Shoemaker}, {Siellez}, {Siemens}, {Sieniawska}, {Sigg}, {Silva}, {Simakov}, {Singer}, {Singer}, {Singh}, {Singh}, {Singhal}, {Sintes}, {Slagmolen}, {Smith}, {Smith}, {Smith}, {Son}, {Sorazu}, {Sorrentino}, {Souradeep}, {Srivastava}, {Staley}, {Steinke}, {Steinlechner}, {Steinlechner}, {Steinmeyer}, {Stephens}, {Stone}, {Strain}, {Straniero}, {Stratta}, {Strauss}, {Strigin}, {Sturani}, {Stuver}, {Summerscales}, {Sun}, {Sutton}, {Swinkels}, {Szczepa{\'n}czyk}, {Tacca}, {Talukder}, {Tanner}, {T{\'a}pai}, {Tarabrin}, {Taracchini}, {Taylor}, {Theeg}, {Thirugnanasambandam}, {Thomas}, {Thomas}, {Thomas}, {Thorne}, {Thorne}, {Thrane}, {Tiwari}, {Tiwari},
  {Tokmakov}, {Tomlinson}, {Tonelli}, {Torres}, {Torrie}, {T{\"o}yr{\"a}}, {Travasso}, {Traylor}, {Trifir{\`o}}, {Tringali}, {Trozzo}, {Tse}, {Turconi}, {Tuyenbayev}, {Ugolini}, {Unnikrishnan}, {Urban}, {Usman}, {Vahlbruch}, {Vajente}, {Valdes}, {van Bakel}, {van Beuzekom}, {van den Brand}, {Van Den Broeck}, {Vander-Hyde}, {van der Schaaf}, {van Heijningen}, {van Veggel}, {Vardaro}, {Vass}, {Vas{\'u}th}, {Vaulin}, {Vecchio}, {Vedovato}, {Veitch}, {Veitch}, {Venkateswara}, {Verkindt}, {Vetrano}, {Vicer{\'e}}, {Vinciguerra}, {Vine}, {Vinet}, {Vitale}, {Vo}, {Vocca}, {Vorvick}, {Voss}, {Vousden}, {Vyatchanin}, {Wade}, {Wade}, {Wade}, {Walker}, {Wallace}, {Walsh}, {Wang}, {Wang}, {Wang}, {Wang}, {Wang}, {Ward}, {Warner}, {Was}, {Weaver}, {Wei}, {Weinert}, {Weinstein}, {Weiss}, {Welborn}, {Wen}, {We{\ss}els}, {Westphal}, {Wette}, {Whelan}, {Whitcomb}, {White}, {Whiting}, {Williams}, {Williamson}, {Willis}, {Willke}, {Wimmer}, {Winkler}, {Wipf}, {Wittel}, {Woan}, {Worden}, {Wright}, {Wu}, {Yablon}, {Yam},
  {Yamamoto}, {Yancey}, {Yap}, {Yu}, {Yvert}, {Zadro{\.Z}ny}, {Zangrando}, {Zanolin}, {Zendri}, {Zevin}, {Zhang}, {Zhang}, {Zhang}, {Zhang}, {Zhao}, {Zhou}, {Zhou}, {Zhu}, {Zucker}, {Zuraw}, {Zweizig}, {LIGO Scientific Collaboration}, \& {Virgo Collaboration}}]{abb16}
{Abbott}, B.~P., {Abbott}, R., {Abbott}, T.~D., {et~al.} 2016, \prd, 94, 102001, \dodoi{10.1103/PhysRevD.94.102001}

\bibitem[{Abbott {et~al.}(2017{\natexlab{a}})Abbott, Abbott, Abbott, Acernese, Ackley, Adams, Adams, Addesso, Adhikari, Adya, Affeldt, Afrough, Agarwal, Agathos, Agatsuma, Aggarwal, Aguiar, Aiello, Ain, Ajith, Allen, Allen, Allocca, Altin, Amato, Ananyeva, Anderson, Anderson, Angelova, Antier, Appert, Arai, Araya, Areeda, Arnaud, Arun, Ascenzi, Ashton, Ast, Aston, Astone, Atallah, Aufmuth, Aulbert, AultONeal, Austin, Avila-Alvarez, Babak, Bacon, Bader, Bae, Baker, Baldaccini, Ballardin, Ballmer, Banagiri, Barayoga, Barclay, Barish, Barker, Barkett, Barone, Barr, Barsotti, Barsuglia, Barta, Barthelmy, Bartlett, Bartos, Bassiri, Basti, Batch, Bawaj, Bayley, Bazzan, Bécsy, Beer, Bejger, Belahcene, Bell, Berger, Bergmann, Bero, Berry, Bersanetti, Bertolini, Betzwieser, Bhagwat, Bhandare, Bilenko, Billingsley, Billman, Birch, Birney, Birnholtz, Biscans, Biscoveanu, Bisht, Bitossi, Biwer, Bizouard, Blackburn, Blackman, Blair, Blair, Blair, Bloemen, Bock, Bode, Boer, Bogaert, Bohe, Bondu, Bonilla, Bonnand, Boom,
  Bork, Boschi, Bose, Bossie, Bouffanais, Bozzi, Bradaschia, Brady, Branchesi, Brau, Briant, Brillet, Brinkmann, Brisson, Brockill, Broida, Brooks, Brown, Brown, Brunett, Buchanan, Buikema, Bulik, Bulten, Buonanno, Buskulic, Buy, Byer, Cabero, Cadonati, Cagnoli, Cahillane, Bustillo, Callister, Calloni, Camp, Canepa, Canizares, Cannon, Cao, Cao, Capano, Capocasa, Carbognani, Caride, Carney, Diaz, Casentini, Caudill, Cavaglià, Cavalier, Cavalieri, Cella, Cepeda, Cerdá-Durán, Cerretani, Cesarini, Chamberlin, Chan, Chao, Charlton, Chase, Chassande-Mottin, Chatterjee, Chatziioannou, Cheeseboro, Chen, Chen, Chen, Cheng, Chia, Chincarini, Chiummo, Chmiel, Cho, Cho, Chow, Christensen, Chu, Chua, Chua, Chung, Chung, Ciani, Ciolfi, Cirelli, Cirone, Clara, Clark, Clearwater, Cleva, Cocchieri, Coccia, Cohadon, Cohen, Colla, Collette, Cominsky, Jr., Conti, Cooper, Corban, Corbitt, Cordero-Carrión, Corley, Cornish, Corsi, Cortese, Costa, Coughlin, Coughlin, Coulon, Countryman, Couvares, Covas, Cowan, Coward, Cowart,
  Coyne, Coyne, Creighton, Creighton, Cripe, Crowder, Cullen, Cumming, Cunningham, Cuoco, Canton, Dálya, Danilishin, D’Antonio, Danzmann, Dasgupta, Da~Silva~Costa, Dattilo, Dave, Davier, Davis, Daw, Day, De, DeBra, Degallaix, Laurentis, Deléglise, Pozzo, Demos, Denker, Dent, Pietri, Dergachev, Rosa, DeRosa, Rossi, DeSalvo, Varona, Devenson, Dhurandhar, Díaz, Fiore, Giovanni, Girolamo, Lieto, Pace, Palma, Renzo, Doctor, Dolique, Donovan, Dooley, Doravari, Dorrington, Douglas, Álvarez, Downes, Drago, Dreissigacker, Driggers, Du, Ducrot, Dupej, Dwyer, Edo, Edwards, Effler, Ehrens, Eichholz, Eikenberry, Eisenstein, Essick, Estevez, Etienne, Etzel, Evans, Evans, Factourovich, Fafone, Fair, Fairhurst, Fan, Farinon, Farr, Farr, Fauchon-Jones, Favata, Fays, Fee, Fehrmann, Feicht, Fejer, Fernandez-Galiana, Ferrante, Ferreira, Ferrini, Fidecaro, Finstad, Fiori, Fiorucci, Fishbach, Fisher, Fitz-Axen, Flaminio, Fletcher, Fong, Font, Forsyth, Forsyth, Fournier, Frasca, Frasconi, Frei, Freise, Frey, Frey, Fries,
  Fritschel, Frolov, Fulda, Fyffe, Gabbard, Gadre, Gaebel, Gair, Gammaitoni, Ganija, Gaonkar, Garcia-Quiros, Garufi, Gateley, Gaudio, Gaur, Gayathri, Gehrels, Gemme, Genin, Gennai, George, George, Gergely, Germain, Ghonge, Ghosh, Ghosh, Ghosh, Giaime, Giardina, Giazotto, Gill, Glover, Goetz, Goetz, Gomes, Goncharov, González, Castro, Gopakumar, Gorodetsky, Gossan, Gosselin, Gouaty, Grado, Graef, Granata, Grant, Gras, Gray, Greco, Green, Gretarsson, Griswold, Groot, Grote, Grunewald, Gruning, Guidi, Guo, Gupta, Gupta, Gushwa, Gustafson, Gustafson, Halim, Hall, Hall, Hamilton, Hammond, Haney, Hanke, Hanks, Hanna, Hannam, Hannuksela, Hanson, Hardwick, Harms, Harry, Harry, Hart, Haster, Haughian, Healy, Heidmann, Heintze, Heitmann, Hello, Hemming, Hendry, Heng, Hennig, Heptonstall, Heurs, Hild, Hinderer, Hoak, Hofman, Holt, Holz, Hopkins, Horst, Hough, Houston, Howell, Hreibi, Hu, Huerta, Huet, Hughey, Husa, Huttner, Huynh-Dinh, Indik, Inta, Intini, Isa, Isac, Isi, Iyer, Izumi, Jacqmin, Jani, Jaranowski,
  Jawahar, Jiménez-Forteza, Johnson, Jones, Jones, Jonker, Ju, Junker, Kalaghatgi, Kalogera, Kamai, Kandhasamy, Kang, Kanner, Kapadia, Karki, Karvinen, Kasprzack, Katolik, Katsavounidis, Katzman, Kaufer, Kawabe, Kéfélian, Keitel, Kemball, Kennedy, Kent, Key, Khalili, Khan, Khan, Khan, Khazanov, Kijbunchoo, Kim, Kim, Kim, Kim, Kim, Kim, Kimbrell, King, King, Kinley-Hanlon, Kirchhoff, Kissel, Kleybolte, Klimenko, Knowles, Koch, Koehlenbeck, Koley, Kondrashov, Kontos, Korobko, Korth, Kowalska, Kozak, Krämer, Kringel, Krishnan, Królak, Kuehn, Kumar, Kumar, Kumar, Kuo, Kutynia, Kwang, Lackey, Lai, Landry, Lang, Lange, Lantz, Lanza, Larson, Lartaux-Vollard, Lasky, Laxen, Lazzarini, Lazzaro, Leaci, Leavey, Lee, Lee, Lee, Lee, Lee, Lehmann, Lenon, Leonardi, Leroy, Letendre, Levin, Li, Linker, Littenberg, Liu, Lo, Lockerbie, London, Lord, Lorenzini, Loriette, Lormand, Losurdo, Lough, Lousto, Lovelace, Lück, Lumaca, Lundgren, Lynch, Ma, Macas, Macfoy, Machenschalk, MacInnis, Macleod, Hernandez, Magaña-Sandoval,
  Zertuche, Magee, Majorana, Maksimovic, Man, Mandic, Mangano, Mansell, Manske, Mantovani, Marchesoni, Marion, Márka, Márka, Markakis, Markosyan, Markowitz, Maros, Marquina, Marsh, Martelli, Martellini, Martin, Martin, Martynov, Mason, Massera, Masserot, Massinger, Masso-Reid, Mastrogiovanni, Matas, Matichard, Matone, Mavalvala, Mazumder, McCarthy, McClelland, McCormick, McCuller, McGuire, McIntyre, McIver, McManus, McNeill, McRae, McWilliams, Meacher, Meadors, Mehmet, Meidam, Mejuto-Villa, Melatos, Mendell, Mercer, Merilh, Merzougui, Meshkov, Messenger, Messick, Metzdorff, Meyers, Miao, Michel, Middleton, Mikhailov, Milano, Miller, Miller, Miller, Millhouse, Milovich-Goff, Minazzoli, Minenkov, Ming, Mishra, Mitra, Mitrofanov, Mitselmakher, Mittleman, Moffa, Moggi, Mogushi, Mohan, Mohapatra, Montani, Moore, Moraru, Moreno, Morriss, Mours, Mow-Lowry, Mueller, Muir, Mukherjee, Mukherjee, Mukherjee, Mukund, Mullavey, Munch, Muñiz, Muratore, Murray, Napier, Nardecchia, Naticchioni, Nayak, Neilson, Nelemans,
  Nelson, Nery, Neunzert, Nevin, Newport, Newton, Ng, Nguyen, Nguyen, Nichols, Nielsen, Nissanke, Nitz, Noack, Nocera, Nolting, North, Nuttall, Oberling, O’Dea, Ogin, Oh, Oh, Ohme, Okada, Oliver, Oppermann, Oram, O’Reilly, Ormiston, Ortega, O’Shaughnessy, Ossokine, Ottaway, Overmier, Owen, Pace, Page, Page, Pai, Pai, Palamos, Palashov, Palomba, Pal-Singh, Pan, Pan, Pang, Pang, Pankow, Pannarale, Pant, Paoletti, Paoli, Papa, Parida, Parker, Pascucci, Pasqualetti, Passaquieti, Passuello, Patil, Patricelli, Pearlstone, Pedraza, Pedurand, Pekowsky, Pele, Penn, Perez, Perreca, Perri, Pfeiffer, Phelps, Piccinni, Pichot, Piergiovanni, Pierro, Pillant, Pinard, Pinto, Pirello, Pitkin, Poe, Poggiani, Popolizio, Porter, Post, Powell, Prasad, Pratt, Pratten, Predoi, Prestegard, Price, Prijatelj, Principe, Privitera, Prodi, Prokhorov, Puncken, Punturo, Puppo, Pürrer, Qi, Quetschke, Quintero, Quitzow-James, Raab, Rabeling, Radkins, Raffai, Raja, Rajan, Rajbhandari, Rakhmanov, Ramirez, Ramos-Buades, Rapagnani,
  Raymond, Razzano, Read, Regimbau, Rei, Reid, Reitze, Ren, Reyes, Ricci, Ricker, Rieger, Riles, Rizzo, Robertson, Robie, Robinet, Rocchi, Rolland, Rollins, Roma, Romano, Romel, Romie, Rosińska, Ross, Rowan, Rüdiger, Ruggi, Rutins, Ryan, Sachdev, Sadecki, Sadeghian, Sakellariadou, Salconi, Saleem, Salemi, Samajdar, Sammut, Sampson, Sanchez, Sanchez, Sanchis-Gual, Sandberg, Sanders, Sassolas, Sathyaprakash, Saulson, Sauter, Savage, Sawadsky, Schale, Scheel, Scheuer, Schmidt, Schmidt, Schnabel, Schofield, Schönbeck, Schreiber, Schuette, Schulte, Schutz, Schwalbe, Scott, Scott, Seidel, Sellers, Sengupta, Sentenac, Sequino, Sergeev, Shaddock, Shaffer, Shah, Shahriar, Shaner, Shao, Shapiro, Shawhan, Sheperd, Shoemaker, Shoemaker, Siellez, Siemens, Sieniawska, Sigg, Silva, Singer, Singh, Singhal, Sintes, Slagmolen, Smith, Smith, Smith, Somala, Son, Sonnenberg, Sorazu, Sorrentino, Souradeep, Spencer, Srivastava, Staats, Staley, Steinke, Steinlechner, Steinlechner, Steinmeyer, Stevenson, Stone, Stops, Strain,
  Stratta, Strigin, Strunk, Sturani, Stuver, Summerscales, Sun, Sunil, Suresh, Sutton, Swinkels, Szczepańczyk, Tacca, Tait, Talbot, Talukder, Tanner, Tápai, Taracchini, Tasson, Taylor, Taylor, Tewari, Theeg, Thies, Thomas, Thomas, Thomas, Thorne, Thorne, Thrane, Tiwari, Tiwari, Tokmakov, Toland, Tonelli, Tornasi, Torres-Forné, Torrie, Töyrä, Travasso, Traylor, Trinastic, Tringali, Trozzo, Tsang, Tse, Tso, Tsukada, Tsuna, Tuyenbayev, Ueno, Ugolini, Unnikrishnan, Urban, Usman, Vahlbruch, Vajente, Valdes, Bakel, Beuzekom, van~den Brand, Van Den~Broeck, Vander-Hyde, van~der Schaaf, Heijningen, Veggel, Vardaro, Varma, Vass, Vasúth, Vecchio, Vedovato, Veitch, Veitch, Venkateswara, Venugopalan, Verkindt, Vetrano, Viceré, Viets, Vinciguerra, Vine, Vinet, Vitale, Vo, Vocca, Vorvick, Vyatchanin, Wade, Wade, Wade, Walet, Walker, Wallace, Walsh, Wang, Wang, Wang, Wang, Wang, Ward, Warner, Was, Watchi, Weaver, Wei, Weinert, Weinstein, Weiss, Wen, Wessel, Wessels, Westerweck, Westphal, Wette, Whelan, Whitcomb,
  Whiting, Whittle, Wilken, Williams, Williams, Williamson, Willis, Willke, Wimmer, Winkler, Wipf, Wittel, Woan, Woehler, Wofford, Wong, Worden, Wright, Wu, Wysocki, Xiao, Yamamoto, Yancey, Yang, Yap, Yazback, Yu, Yu, Yvert, Zadrożny, Zanolin, Zelenova, Zendri, Zevin, Zhang, Zhang, Zhang, Zhang, Zhao, Zhou, Zhou, Zhu, Zhu, Zimmerman, Zucker, Zweizig, Wilson-Hodge, Bissaldi, Blackburn, Briggs, Burns, Cleveland, Connaughton, Gibby, Giles, Goldstein, Hamburg, Jenke, Hui, Kippen, Kocevski, McBreen, Meegan, Paciesas, Poolakkil, Preece, Racusin, Roberts, Stanbro, Veres, von Kienlin, Savchenko, Ferrigno, Kuulkers, Bazzano, Bozzo, Brandt, Chenevez, Courvoisier, Diehl, Domingo, Hanlon, Jourdain, Laurent, Lebrun, Lutovinov, Martin-Carrillo, Mereghetti, Natalucci, Rodi, Roques, Sunyaev, Ubertini, Aartsen, Ackermann, Adams, Aguilar, Ahlers, Ahrens, Samarai, Altmann, Andeen, Anderson, Ansseau, Anton, Argüelles, Auffenberg, Axani, Bagherpour, Bai, Barron, Barwick, Baum, Bay, Beatty, Tjus, Bernardini, Besson, Binder,
  Bindig, Blaufuss, Blot, Bohm, Börner, Bos, Bose, Böser, Botner, Bourbeau, Bourbeau, Bradascio, Braun, Brayeur, Brenzke, Bretz, Bron, Brostean-Kaiser, Burgman, Carver, Casey, Casier, Cheung, Chirkin, Christov, Clark, Classen, Coenders, Collin, Conrad, Cowen, Cross, Day, André, Clercq, DeLaunay, Dembinski, Ridder, Desiati, Vries, Wasseige, With, DeYoung, Díaz-Vélez, Lorenzo, Dujmovic, Dumm, Dunkman, Dvorak, Eberhardt, Ehrhardt, Eichmann, Eller, Evenson, Fahey, Fazely, Felde, Filimonov, Finley, Flis, Franckowiak, Friedman, Fuchs, Gaisser, Gallagher, Gerhardt, Ghorbani, Giang, Glauch, Glüsenkamp, Goldschmidt, Gonzalez, Grant, Griffith, Haack, Hallgren, Halzen, Hanson, Hebecker, Heereman, Helbing, Hellauer, Hickford, Hignight, Hill, Hoffman, Hoffmann, Hokanson-Fasig, Hoshina, Huang, Huber, Hultqvist, Hünnefeld, In, Ishihara, Jacobi, Japaridze, Jeong, Jero, Jones, Kalaczynski, Kang, Kappes, Karg, Karle, Kauer, Keivani, Kelley, Kheirandish, Kim, Kim, Kintscher, Kiryluk, Kittler, Klein, Kohnen, Koirala,
  Kolanoski, Köpke, Kopper, Kopper, Koschinsky, Koskinen, Kowalski, Krings, Kroll, Krückl, Kunnen, Kunwar, Kurahashi, Kuwabara, Kyriacou, Labare, Lanfranchi, Larson, Lauber, Lesiak-Bzdak, Leuermann, Liu, Lu, Lünemann, Luszczak, Madsen, Maggi, Mahn, Mancina, Maruyama, Mase, Maunu, McNally, Meagher, Medici, Meier, Menne, Merino, Meures, Miarecki, Micallef, Momenté, Montaruli, Moore, Moulai, Nahnhauer, Nakarmi, Naumann, Neer, Niederhausen, Nowicki, Nygren, Pollmann, Olivas, O’Murchadha, Palczewski, Pandya, Pankova, Peiffer, Pepper, Pérez de~los Heros, Pieloth, Pinat, Price, Przybylski, Raab, Rädel, Rameez, Rawlins, Rea, Reimann, Relethford, Relich, Resconi, Rhode, Richman, Robertson, Rongen, Rott, Ruhe, Ryckbosch, Rysewyk, Sälzer, Herrera, Sandrock, Sandroos, Santander, Sarkar, Sarkar, Satalecka, Schlunder, Schmidt, Schneider, Schoenen, Schöneberg, Schumacher, Seckel, Seunarine, Soedingrekso, Soldin, Song, Spiczak, Spiering, Stachurska, Stamatikos, Stanev, Stasik, Stettner, Steuer, Stezelberger,
  Stokstad, Stössl, Strotjohann, Stuttard, Sullivan, Sutherland, Taboada, Tatar, Tenholt, Ter-Antonyan, Terliuk, Tešić, Tilav, Toale, Tobin, Toscano, Tosi, Tselengidou, Tung, Turcati, Turley, Ty, Unger, Usner, Vandenbroucke, Driessche, Eijndhoven, Vanheule, Santen, Vehring, Vogel, Vraeghe, Walck, Wallace, Wallraff, Wandler, Wandkowsky, Waza, Weaver, Weiss, Wendt, Werthebach, Whelan, Wiebe, Wiebusch, Wille, Williams, Wills, Wolf, Wood, Woolsey, Woschnagg, Xu, Xu, Xu, Yanez, Yodh, Yoshida, Yuan, Zoll, Balasubramanian, Mate, Bhalerao, Bhattacharya, Vibhute, Dewangan, Rao, Vadawale, Svinkin, Hurley, Aptekar, Frederiks, Golenetskii, Kozlova, Lysenko, Oleynik, Tsvetkova, Ulanov, Cline, Li, Xiong, Zhang, Lu, Song, Cao, Chang, Chen, Chen, Chen, Chen, Chen, Chen, Cui, Cui, Deng, Dong, Du, Fu, Gao, Gao, Gao, Ge, Gu, Guan, Guo, Han, Hu, Huang, Huo, Jia, Jiang, Jiang, Jin, Jin, Li, Li, Li, Li, Li, Li, Li, Li, Li, Li, Li, Liang, Liao, Liu, Liu, Liu, Liu, Liu, Liu, Liu, Lu, Lu, Luo, Ma, Meng, Nang, Nie, Ou, Qu, Sai,
  Sun, Tan, Tao, Tao, Tuo, Wang, Wang, Wang, Wang, Wang, Wen, Wu, Wu, Xiao, Xu, Xu, Yan, Yang, Yang, Yang, Zhang, Zhang, Zhang, Zhang, Zhang, Zhang, Zhang, Zhang, Zhang, Zhang, Zhang, Zhang, Zhang, Zhang, Zhang, Zhang, Zhang, Zhang, Zhao, Zhao, Zhao, Zheng, Zhu, Zhu, Zou, Albert, André, Anghinolfi, Ardid, Aubert, Aublin, Avgitas, Baret, Barrios-Martí, Basa, Belhorma, Bertin, Biagi, Bormuth, Bourret, Bouwhuis, Brânzaş, Bruijn, Brunner, Busto, Capone, Caramete, Carr, Celli, Cherkaoui El~Moursli, Chiarusi, Circella, Coelho, Coleiro, Coniglione, Costantini, Coyle, Creusot, Díaz, Deschamps, Bonis, Distefano, Palma, Domi, Donzaud, Dornic, Drouhin, Eberl, El~Bojaddaini, El~Khayati, Elsässer, Enzenhöfer, Ettahiri, Fassi, Felis, Fusco, Gay, Giordano, Glotin, Grégoire, Ruiz, Graf, Hallmann, Haren, Heijboer, Hello, Hernández-Rey, Hössl, Hofestädt, Hugon, Illuminati, James, Jong, Jongen, Kadler, Kalekin, Katz, Kiessling, Kouchner, Kreter, Kreykenbohm, Kulikovskiy, Lachaud, Lahmann, Lefèvre, Leonora, Lotze,
  Loucatos, Marcelin, Margiotta, Marinelli, Martínez-Mora, Mele, Melis, Michael, Migliozzi, Moussa, Navas, Nezri, Organokov, Păvălaş, Pellegrino, Perrina, Piattelli, Popa, Pradier, Quinn, Racca, Riccobene, Sánchez-Losa, Saldaña, Salvadori, Samtleben, Sanguineti, Sapienza, Sieger, Spurio, Stolarczyk, Taiuti, Tayalati, Trovato, Turpin, Tönnis, Vallage, Elewyck, Versari, Vivolo, Vizzoca, Wilms, Zornoza, Zúñiga, Beardmore, Breeveld, Burrows, Cenko, Cusumano, D’Aì, de~Pasquale, Emery, Evans, Giommi, Gronwall, Kennea, Krimm, Kuin, Lien, Marshall, Melandri, Nousek, Oates, Osborne, Pagani, Page, Palmer, Perri, Siegel, Sbarufatti, Tagliaferri, Tohuvavohu, Tavani, Verrecchia, Bulgarelli, Evangelista, Pacciani, Feroci, Pittori, Giuliani, Monte, Donnarumma, Argan, Trois, Ursi, Cardillo, Piano, Longo, Lucarelli, Munar-Adrover, Fuschino, Labanti, Marisaldi, Minervini, Fioretti, Parmiggiani, Gianotti, Trifoglio, Persio, Antonelli, Barbiellini, Caraveo, Cattaneo, Costa, Colafrancesco, D’Amico, Ferrari,
  Morselli, Paoletti, Picozza, Pilia, Rappoldi, Soffitta, Vercellone, Foley, Coulter, Kilpatrick, Drout, Piro, Shappee, Siebert, Simon, Ulloa, Kasen, Madore, Murguia-Berthier, Pan, Prochaska, Ramirez-Ruiz, Rest, Rojas-Bravo, Berger, Soares-Santos, Annis, Alexander, Allam, Balbinot, Blanchard, Brout, Butler, Chornock, Cook, Cowperthwaite, Diehl, Drlica-Wagner, Drout, Durret, Eftekhari, Finley, Fong, Frieman, Fryer, García-Bellido, Gruendl, Hartley, Herner, Kessler, Lin, Lopes, Lourenço, Margutti, Marshall, Matheson, Medina, Metzger, Muñoz, Muir, Nicholl, Nugent, Palmese, Paz-Chinchón, Quataert, Sako, Sauseda, Schlegel, Scolnic, Secco, Smith, Sobreira, Villar, Vivas, Wester, Williams, Yanny, Zenteno, Zhang, Abbott, Banerji, Bechtol, Benoit-Lévy, Bertin, Brooks, Buckley-Geer, Burke, Capozzi, Rosell, Kind, Castander, Crocce, Cunha, D’Andrea, da~Costa, Davis, DePoy, Desai, Dietrich, Eifler, Fernandez, Flaugher, Fosalba, Gaztanaga, Gerdes, Giannantonio, Goldstein, Gruen, Gschwend, Gutierrez, Honscheid,
  James, Jeltema, Johnson, Johnson, Kent, Krause, Kron, Kuehn, Lahav, Lima, Maia, March, Martini, McMahon, Menanteau, Miller, Miquel, Mohr, Nichol, Ogando, Plazas, Romer, Roodman, Rykoff, Sanchez, Scarpine, Schindler, Schubnell, Sevilla-Noarbe, Sheldon, Smith, Smith, Stebbins, Suchyta, Swanson, Tarle, Thomas, Troxel, Tucker, Vikram, Walker, Wechsler, Weller, Carlin, Gill, Li, Marriner, Neilsen, Haislip, Kouprianov, Reichart, Sand, Tartaglia, Valenti, Yang, Benetti, Brocato, Campana, Cappellaro, Covino, D’Avanzo, D’Elia, Getman, Ghirlanda, Ghisellini, Limatola, Nicastro, Palazzi, Pian, Piranomonte, Possenti, Rossi, Salafia, Tomasella, Amati, Antonelli, Bernardini, Bufano, Capaccioli, Casella, Dadina, Cesare, Paola, Giuffrida, Giunta, Israel, Lisi, Maiorano, Mapelli, Masetti, Pescalli, Pulone, Salvaterra, Schipani, Spera, Stamerra, Stella, Testa, Turatto, Vergani, Aresu, Bachetti, Buffa, Burgay, Buttu, Caria, Carretti, Casasola, Castangia, Carboni, Casu, Concu, Corongiu, Deiana, Egron, Fara, Gaudiomonte,
  Gusai, Ladu, Loru, Leurini, Marongiu, Melis, Melis, Migoni, Milia, Navarrini, Orlati, Ortu, Palmas, Pellizzoni, Perrodin, Pisanu, Poppi, Righini, Saba, Serra, Serrau, Stagni, Surcis, Vacca, Vargiu, Hunt, Jin, Klose, Kouveliotou, Mazzali, Møller, Nava, Piran, Selsing, Vergani, Wiersema, Toma, Higgins, Mundell, di~Serego~Alighieri, Gótz, Gao, Gomboc, Kaper, Kobayashi, Kopac, Mao, Starling, Steele, van~der Horst, Acero, Atwood, Baldini, Barbiellini, Bastieri, Berenji, Bellazzini, Bissaldi, Blandford, Bloom, Bonino, Bottacini, Bregeon, Buehler, Buson, Cameron, Caputo, Caraveo, Cavazzuti, Chekhtman, Cheung, Chiang, Ciprini, Cohen-Tanugi, Cominsky, Costantin, Cuoco, D’Ammando, Palma, Digel, Lalla, Mauro, Venere, Dubois, Fegan, Focke, Franckowiak, Fukazawa, Funk, Fusco, Gargano, Gasparrini, Giglietto, Giordano, Giroletti, Glanzman, Green, Grondin, Guillemot, Guiriec, Harding, Horan, Jóhannesson, Kamae, Kensei, Kuss, Mura, Latronico, Lemoine-Goumard, Longo, Loparco, Lovellette, Lubrano, Magill, Maldera,
  Manfreda, Mazziotta, McEnery, Meyer, Michelson, Mirabal, Monzani, Moretti, Morselli, Moskalenko, Negro, Nuss, Ojha, Omodei, Orienti, Orlando, Palatiello, Paliya, Paneque, Pesce-Rollins, Piron, Porter, Principe, Rainò, Rando, Razzano, Razzaque, Reimer, Reimer, Reposeur, Rochester, Parkinson, Sgrò, Siskind, Spada, Spandre, Suson, Takahashi, Tanaka, Thayer, Thayer, Thompson, Tibaldo, Torres, Torresi, Troja, Venters, Vianello, Zaharijas, Allison, Bannister, Dobie, Kaplan, Lenc, Lynch, Murphy, Sadler, Hotan, James, Oslowski, Raja, Shannon, Whiting, Arcavi, Howell, McCully, Hosseinzadeh, Hiramatsu, Poznanski, Barnes, Zaltzman, Vasylyev, Maoz, Cooke, Bailes, Wolf, Deller, Lidman, Wang, Gendre, Andreoni, Ackley, Pritchard, Bessell, Chang, Möller, Onken, Scalzo, Ridden-Harper, Sharp, Tucker, Farrell, Elmer, Johnston, Krishnan, Keane, Green, Jameson, Hu, Ma, Sun, Wu, Wang, Shang, Hu, Ashley, Yuan, Li, Tao, Zhu, Zhang, Suntzeff, Zhou, Yang, Orange, Morris, Cucchiara, Giblin, Klotz, Staff, Thierry, Schmidt, Tanvir,
  Levan, Cano, de~Ugarte-Postigo, González-Fernández, Greiner, Hjorth, Irwin, Krühler, Mandel, Milvang-Jensen, O’Brien, Rol, Rosetti, Rosswog, Rowlinson, Steeghs, Thöne, Ulaczyk, Watson, Bruun, Cutter, Figuera~Jaimes, Fujii, Fruchter, Gompertz, Jakobsson, Hodosan, Jèrgensen, Kangas, Kann, Rabus, Schrøder, Stanway, Wijers, Lipunov, Gorbovskoy, Kornilov, Tyurina, Balanutsa, Kuznetsov, Vlasenko, Podesta, Lopez, Podesta, Levato, Saffe, Mallamaci, Budnev, Gress, Kuvshinov, Gorbunov, Vladimirov, Zimnukhov, Gabovich, Yurkov, Sergienko, Rebolo, Serra-Ricart, Tlatov, Ishmuhametova, Abe, Aoki, Aoki, Asakura, Baar, Barway, Bond, Doi, Finet, Fujiyoshi, Furusawa, Honda, Itoh, Kanda, Kawabata, Kawabata, Kim, Koshida, Kuroda, Lee, Liu, Matsubayashi, Miyazaki, Morihana, Morokuma, Motohara, Murata, Nagai, Nagashima, Nagayama, Nakaoka, Nakata, Ohsawa, Ohshima, Ohta, Okita, Saito, Saito, Sako, Sekiguchi, Sumi, Tajitsu, Takahashi, Takayama, Tamura, Tanaka, Tanaka, Terai, Tominaga, Tristram, Uemura, Utsumi, Yamaguchi,
  Yasuda, Yoshida, Zenko, Adams, Anupama, Bally, Barway, Bellm, Blagorodnova, Cannella, Chandra, Chatterjee, Clarke, Cobb, Cook, Copperwheat, De, Emery, Feindt, Foster, Fox, Frail, Fremling, Frohmaier, Garcia, Ghosh, Giacintucci, Goobar, Gottlieb, Grefenstette, Hallinan, Harrison, Heida, Helou, Ho, Horesh, Hotokezaka, Ip, Itoh, Jacobs, Jencson, Kasen, Kasliwal, Kassim, Kim, Kiran, Kuin, Kulkarni, Kupfer, Lau, Madsen, Mazzali, Miller, Miyasaka, Mooley, Myers, Nakar, Ngeow, Nugent, Ofek, Palliyaguru, Pavana, Perley, Peters, Pike, Piran, Qi, Quimby, Rana, Rosswog, Rusu, Sadler, Sistine, Sollerman, Xu, Yan, Yatsu, Yu, Zhang, Zhao, Chambers, Huber, Schultz, Bulger, Flewelling, Magnier, Lowe, Wainscoat, Waters, Willman, Ebisawa, Hanyu, Harita, Hashimoto, Hidaka, Hori, Ishikawa, Isobe, Iwakiri, Kawai, Kawai, Kawamuro, Kawase, Kitaoka, Makishima, Matsuoka, Mihara, Morita, Morita, Nakahira, Nakajima, Nakamura, Negoro, Oda, Sakamaki, Sasaki, Serino, Shidatsu, Shimomukai, Sugawara, Sugita, Sugizaki, Tachibana, Takao,
  Tanimoto, Tomida, Tsuboi, Tsunemi, Ueda, Ueno, Yamada, Yamaoka, Yamauchi, Yatabe, Yoneyama, Yoshii, Coward, Crisp, Macpherson, Andreoni, Laugier, Noysena, Klotz, Gendre, Thierry, Turpin, Im, Choi, Kim, Yoon, Lim, Lee, Lee, Kim, Ko, Joe, Kwon, Kim, Lim, Choi, Fynbo, Malesani, Xu, Smartt, Jerkstrand, Kankare, Sim, Fraser, Inserra, Maguire, Leloudas, Magee, Shingles, Smith, Young, Kotak, Gal-Yam, Lyman, Homan, Agliozzo, Anderson, Angus, Ashall, Barbarino, Bauer, Berton, Botticella, Bulla, Cannizzaro, Cartier, Cikota, Clark, De~Cia, Della~Valle, Dennefeld, Dessart, Dimitriadis, Elias-Rosa, Firth, Flörs, Frohmaier, Galbany, González-Gaitán, Gromadzki, Gutiérrez, Hamanowicz, Harmanen, Heintz, Hernandez, Hodgkin, Hook, Izzo, James, Jonker, Kerzendorf, Kostrzewa-Rutkowska, Kromer, Kuncarayakti, Lawrence, Manulis, Mattila, McBrien, Müller, Nordin, O’Neill, Onori, Palmerio, Pastorello, Patat, Pignata, Podsiadlowski, Razza, Reynolds, Roy, Ruiter, Rybicki, Salmon, Pumo, Prentice, Seitenzahl, Smith, Sollerman,
  Sullivan, Szegedi, Taddia, Taubenberger, Terreran, Van~Soelen, Vos, Walton, Wright, Wyrzykowski, Yaron, Chen, Krühler, Schady, Wiseman, Greiner, Rau, Schweyer, Klose, Nicuesa~Guelbenzu, Palliyaguru, Shara, Williams, Vaisanen, Potter, Colmenero, Crawford, Buckley, Mao, Díaz, Macri, García~Lambas, Mendes~de Oliveira, Nilo~Castellón, Ribeiro, Sánchez, Schoenell, Abramo, Akras, Alcaniz, Artola, Beroiz, Bonoli, Cabral, Camuccio, Chavushyan, Coelho, Colazo, Costa-Duarte, Cuevas~Larenas, Domínguez~Romero, Dultzin, Fernández, García, Girardini, Gonçalves, Gonçalves, Gurovich, Jiménez-Teja, Kanaan, Lares, Lopes~de Oliveira, López-Cruz, Melia, Molino, Padilla, Peñuela, Placco, Quiñones, Ramírez~Rivera, Renzi, Riguccini, Ríos-López, Rodriguez, Sampedro, Schneiter, Sodré, Starck, Torres-Flores, Tornatore, Zadrożny, Castillo, Castro-Tirado, Tello, Hu, Zhang, Cunniffe, Castellón, Hiriart, Caballero-García, Jelínek, Kubánek, Pérez~del Pulgar, Park, Jeong, Castro~Cerón, Pandey, Yock, Querel, Fan,
  Wang, Beardsley, Brown, Crosse, Emrich, Franzen, Gaensler, Horsley, Johnston-Hollitt, Kenney, Morales, Pallot, Sokolowski, Steele, Tingay, Trott, Walker, Wayth, Williams, Wu, Yoshida, Sakamoto, Kawakubo, Yamaoka, Takahashi, Asaoka, Ozawa, Torii, Shimizu, Tamura, Ishizaki, Cherry, Ricciarini, Penacchioni, Marrocchesi, Pozanenko, Volnova, Mazaeva, Minaev, Krugov, Kusakin, Reva, Moskvitin, Rumyantsev, Inasaridze, Klunko, Tungalag, Schmalz, Burhonov, Abdalla, Abramowski, Aharonian, Benkhali, Angüner, Arakawa, Arrieta, Aubert, Backes, Balzer, Barnard, Becherini, Tjus, Berge, Bernhard, Bernlöhr, Blackwell, Böttcher, Boisson, Bolmont, Bonnefoy, Bordas, Bregeon, Brun, Brun, Bryan, Büchele, Bulik, Capasso, Caroff, Carosi, Casanova, Cerruti, Chakraborty, Chaves, Chen, Chevalier, Colafrancesco, Condon, Conrad, Davids, Decock, Deil, Devin, deWilt, Dirson, Djannati-Ataï, Donath, O’C.~Drury, Dutson, Dyks, Edwards, Egberts, Emery, Ernenwein, Eschbach, Farnier, Fegan, Fernandes, Fiasson, Fontaine, Funk, Füssling,
  Gabici, Gallant, Garrigoux, Gaté, Giavitto, Giebels, Glawion, Glicenstein, Gottschall, Grondin, Hahn, Haupt, Hawkes, Heinzelmann, Henri, Hermann, Hinton, Hofmann, Hoischen, Holch, Holler, Horns, Ivascenko, Iwasaki, Jacholkowska, Jamrozy, Jankowsky, Jankowsky, Jingo, Jouvin, Jung-Richardt, Kastendieck, Katarzyński, Katsuragawa, Kerszberg, Khangulyan, Khélifi, King, Klepser, Klochkov, Kluźniak, Komin, Kosack, Krakau, Kraus, Krüger, Laffon, Lamanna, Lau, Lees, Lefaucheur, Lemière, Lemoine-Goumard, Lenain, Leser, Lohse, Lorentz, Liu, Lypova, Malyshev, Marandon, Marcowith, Mariaud, Marx, Maurin, Maxted, Mayer, Meintjes, Meyer, Mitchell, Moderski, Mohamed, Mohrmann, Morå, Moulin, Murach, Nakashima, Naurois, Ndiyavala, Niederwanger, Niemiec, Oakes, O’Brien, Odaka, Ohm, Ostrowski, Oya, Padovani, Panter, Parsons, Pekeur, Pelletier, Perennes, Petrucci, Peyaud, Piel, Pita, Poireau, Poon, Prokhorov, Prokoph, Pühlhofer, Punch, Quirrenbach, Raab, Rauth, Reimer, Reimer, Renaud, de~los Reyes, Rieger, Rinchiuso,
  Romoli, Rowell, Rudak, Rulten, Sahakian, Saito, Sanchez, Santangelo, Sasaki, Schlickeiser, Schüssler, Schulz, Schwanke, Schwemmer, Seglar-Arroyo, Settimo, Seyffert, Shafi, Shilon, Shiningayamwe, Simoni, Sol, Spanier, Spir-Jacob, Stawarz, Steenkamp, Stegmann, Steppa, Sushch, Takahashi, Tavernet, Tavernier, Taylor, Terrier, Tibaldo, Tiziani, Tluczykont, Trichard, Tsirou, Tsuji, Tuffs, Uchiyama, van~der Walt, Eldik, Rensburg, Soelen, Vasileiadis, Veh, Venter, Viana, Vincent, Vink, Voisin, Völk, Vuillaume, Wadiasingh, Wagner, Wagner, Wagner, White, Wierzcholska, Willmann, Wörnlein, Wouters, Yang, Zaborov, Zacharias, Zanin, Zdziarski, Zech, Zefi, Ziegler, Zorn, Żywucka, Fender, Broderick, Rowlinson, Wijers, Stewart, ter Veen, Shulevski, Kavic, Simonetti, League, Tsai, Obenberger, Nathaniel, Taylor, Dowell, Liebling, Estes, Lippert, Sharma, Vincent, Farella, Abeysekara, Albert, Alfaro, Alvarez, Arceo, Arteaga-Velázquez, Avila~Rojas, Ayala~Solares, Barber, Becerra~Gonzalez, Becerril, Belmont-Moreno, BenZvi,
  Berley, Bernal, Braun, Brisbois, Caballero-Mora, Capistrán, Carramiñana, Casanova, Castillo, Cotti, Cotzomi, Coutiño~de León, De~León, De~la Fuente, Diaz~Hernandez, Dichiara, Dingus, DuVernois, Díaz-Vélez, Ellsworth, Engel, Enríquez-Rivera, Fiorino, Fleischhack, Fraija, García-González, Garfias, Gerhardt, Gonzõlez~Muñoz, González, Goodman, Hampel-Arias, Harding, Hernandez, Hernandez-Almada, Hona, Hüntemeyer, Iriarte, Jardin-Blicq, Joshi, Kaufmann, Kieda, Lara, Lauer, Lennarz, León~Vargas, Linnemann, Longinotti, Luis~Raya, Luna-García, López-Coto, Malone, Marinelli, Martinez, Martinez-Castellanos, Martínez-Castro, Martínez-Huerta, Matthews, Miranda-Romagnoli, Moreno, Mostafá, Nellen, Newbold, Nisa, Noriega-Papaqui, Pelayo, Pretz, Pérez-Pérez, Ren, Rho, Rivière, Rosa-González, Rosenberg, Ruiz-Velasco, Salazar, Salesa~Greus, Sandoval, Schneider, Schoorlemmer, Sinnis, Smith, Springer, Surajbali, Tibolla, Tollefson, Torres, Ukwatta, Weisgarber, Westerhoff, Wisher, Wood, Yapici, Yodh,
  Younk, Zhou, Álvarez, Aab, Abreu, Aglietta, Albuquerque, Albury, Allekotte, Almela, Alvarez~Castillo, Alvarez-Muñiz, Anastasi, Anchordoqui, Andrada, Andringa, Aramo, Arsene, Asorey, Assis, Avila, Badescu, Balaceanu, Barbato, Barreira~Luz, Becker, Bellido, Berat, Bertaina, Bertou, Biermann, Biteau, Blaess, Blanco, Blazek, Bleve, Boháčová, Bonifazi, Borodai, Botti, Brack, Brancus, Bretz, Bridgeman, Briechle, Buchholz, Bueno, Buitink, Buscemi, Caballero-Mora, Caccianiga, Cancio, Canfora, Caruso, Castellina, Catalani, Cataldi, Cazon, Chavez, Chinellato, Chudoba, Clay, Cobos~Cerutti, Colalillo, Coleman, Collica, Coluccia, Conceição, Consolati, Contreras, Cooper, Coutu, Covault, Cronin, D’Amico, Daniel, Dasso, Daumiller, Dawson, Day, Almeida, Jong, Mauro, de~Mello~Neto, Mitri, Oliveira, Souza, Debatin, Deligny, Díaz~Castro, Diogo, Dobrigkeit, D’Olivo, Dorosti, Dos~Anjos, Dova, Dundovic, Ebr, Engel, Erdmann, Erfani, Escobar, Espadanal, Etchegoyen, Falcke, Farmer, Farrar, Fauth, Fazzini, Feldbusch,
  Fenu, Fick, Figueira, Filipčič, Freire, Fujii, Fuster, Gaïor, García, Gaté, Gemmeke, Gherghel-Lascu, Ghia, Giaccari, Giammarchi, Giller, Głas, Glaser, Golup, Gómez~Berisso, Gómez~Vitale, González, Gorgi, Gottowik, Grillo, Grubb, Guarino, Guedes, Halliday, Hampel, Hansen, Harari, Harrison, Harvey, Haungs, Hebbeker, Heck, Heimann, Herve, Hill, Hojvat, Holt, Homola, Hörandel, Horvath, Hrabovský, Huege, Hulsman, Insolia, Isar, Jandt, Johnsen, Josebachuili, Jurysek, Kääpä, Kampert, Keilhauer, Kemmerich, Kemp, Kieckhafer, Klages, Kleifges, Kleinfeller, Krause, Krohm, Kuempel, Kukec~Mezek, Kunka, Kuotb~Awad, Lago, LaHurd, Lang, Lauscher, Legumina, Leigui~de Oliveira, Letessier-Selvon, Lhenry-Yvon, Link, Lo~Presti, Lopes, López, López~Casado, Lorek, Luce, Lucero, Malacari, Mallamaci, Mandat, Mantsch, Mariazzi, Maris, Marsella, Martello, Martinez, Martínez~Bravo, Masías~Meza, Mathes, Mathys, Matthews, Matthiae, Mayotte, Mazur, Medina, Medina-Tanco, Melo, Menshikov, Merenda, Michal, Micheletti,
  Middendorf, Miramonti, Mitrica, Mockler, Mollerach, Montanet, Morello, Morlino, Müller, Müller, Muller, Müller, Mussa, Naranjo, Nguyen, Niculescu-Oglinzanu, Niechciol, Niemietz, Niggemann, Nitz, Nosek, Novotny, Nožka, Núñez, Oikonomou, Olinto, Palatka, Pallotta, Papenbreer, Parente, Parra, Paul, Pech, Pedreira, Pȩkala, Peña-Rodriguez, Pereira, Perlin, Perrone, Peters, Petrera, Phuntsok, Pierog, Pimenta, Pirronello, Platino, Plum, Poh, Porowski, Prado, Privitera, Prouza, Quel, Querchfeld, Quinn, Ramos-Pollan, Rautenberg, Ravignani, Ridky, Riehn, Risse, Ristori, Rizi, Rodrigues~de Carvalho, Rodriguez~Fernandez, Rodriguez~Rojo, Roncoroni, Roth, Roulet, Rovero, Ruehl, Saffi, Saftoiu, Salamida, Salazar, Saleh, Salina, Sánchez, Sanchez-Lucas, Santos, Santos, Sarazin, Sarmento, Sarmiento-Cano, Sato, Schauer, Scherini, Schieler, Schimp, Schmidt, Scholten, Schovánek, Schröder, Schröder, Schulz, Schumacher, Sciutto, Segreto, Shadkam, Shellard, Sigl, Silli, Šmída, Snow, Sommers, Sonntag, Soriano,
  Squartini, Stanca, Stanič, Stasielak, Stassi, Stolpovskiy, Strafella, Streich, Suarez, Suarez-Durán, Sudholz, Suomijärvi, Supanitsky, Šupík, Swain, Szadkowski, Taboada, Taborda, Timmermans, Todero~Peixoto, Tomankova, Tomé, Torralba~Elipe, Travnicek, Trini, Tueros, Ulrich, Unger, Urban, Valdés~Galicia, Valiño, Valore, Aar, Bodegom, van~den Berg, Vliet, Varela, Cárdenas, Vázquez, Veberič, Ventura, Vergara~Quispe, Verzi, Vicha, Villaseñor, Vorobiov, Wahlberg, Wainberg, Walz, Watson, Weber, Weindl, Wiedeński, Wiencke, Wilczyński, Wirtz, Wittkowski, Wundheiler, Yang, Yushkov, Zas, Zavrtanik, Zavrtanik, Zepeda, Zimmermann, Ziolkowski, Zong, Zuccarello, Kim, Schulze, Bauer, Corral-Santana, de~Gregorio-Monsalvo, González-López, Hartmann, Ishwara-Chandra, Martín, Mehner, Misra, Michałowski, Resmi, Paragi, Agudo, An, Beswick, Casadio, Frey, Jonker, Kettenis, Marcote, Moldon, Szomoru, van Langevelde, Yang, Cwiek, Cwiok, Czyrkowski, Dabrowski, Kasprowicz, Mankiewicz, Nawrocki, Opiela, Piotrowski,
  Wrochna, Zaremba, Żarnecki, Haggard, Nynka, Ruan, Bland, Booler, Devillepoix, Gois, Hancock, Howie, Paxman, Sansom, Towner, Tonry, Coughlin, Stubbs, Denneau, Heinze, Stalder, Weiland, Eatough, Kramer, Kraus, Troja, Piro, González, Butler, Fox, Khandrika, Kutyrev, Lee, Ricci, Ryan~Jr., Sánchez-Ramírez, Veilleux, Watson, Wieringa, Burgess, Eerten, Fontes, Fryer, Korobkin, Wollaeger, Camilo, Foley, Goedhart, Makhathini, Oozeer, Smirnov, Fender, \& Woudt}]{abb17}
Abbott, B.~P., Abbott, R., Abbott, T.~D., {et~al.} 2017{\natexlab{a}}, The Astrophysical Journal Letters, 848, L12, \dodoi{10.3847/2041-8213/aa91c9}

\bibitem[{Abbott {et~al.}(2017{\natexlab{b}})Abbott, Abbott, Abbott, Acernese, Ackley, Adams, Adams, Addesso, Adhikari, Adya, Affeldt, Afrough, Agarwal, Agathos, Agatsuma, Aggarwal, Aguiar, Aiello, Ain, Ajith, Allen, Allen, Allocca, Altin, Amato, Ananyeva, Anderson, Anderson, Angelova, Antier, Appert, Arai, Araya, Areeda, Arnaud, Arun, Ascenzi, Ashton, Ast, Aston, Astone, Atallah, Aufmuth, Aulbert, AultONeal, Austin, Avila-Alvarez, Babak, Bacon, Bader, Bae, Bailes, Baker, Baldaccini, Ballardin, Ballmer, Banagiri, Barayoga, Barclay, Barish, Barker, Barkett, Barone, Barr, Barsotti, Barsuglia, Barta, Barthelmy, Bartlett, Bartos, Bassiri, Basti, Batch, Bawaj, Bayley, Bazzan, B\'ecsy, Beer, Bejger, Belahcene, Bell, Berger, Bergmann, Bernuzzi, Bero, Berry, Bersanetti, Bertolini, Betzwieser, Bhagwat, Bhandare, Bilenko, Billingsley, Billman, Birch, Birney, Birnholtz, Biscans, Biscoveanu, Bisht, Bitossi, Biwer, Bizouard, Blackburn, Blackman, Blair, Blair, Blair, Bloemen, Bock, Bode, Boer, Bogaert, Bohe, Bondu,
  Bonilla, Bonnand, Boom, Bork, Boschi, Bose, Bossie, Bouffanais, Bozzi, Bradaschia, Brady, Branchesi, Brau, Briant, Brillet, Brinkmann, Brisson, Brockill, Broida, Brooks, Brown, Brown, Brunett, Buchanan, Buikema, Bulik, Bulten, Buonanno, Buskulic, Buy, Byer, Cabero, Cadonati, Cagnoli, Cahillane, Calder\'on~Bustillo, Callister, Calloni, Camp, Canepa, Canizares, Cannon, Cao, Cao, Capano, Capocasa, Carbognani, Caride, Carney, Carullo, Casanueva~Diaz, Casentini, Caudill, Cavagli\`a, Cavalier, Cavalieri, Cella, Cepeda, Cerd\'a-Dur\'an, Cerretani, Cesarini, Chamberlin, Chan, Chao, Charlton, Chase, Chassande-Mottin, Chatterjee, Chatziioannou, Cheeseboro, Chen, Chen, Chen, Cheng, Chia, Chincarini, Chiummo, Chmiel, Cho, Cho, Chow, Christensen, Chu, Chua, Chua, Chung, Chung, Ciani, Ciolfi, Cirelli, Cirone, Clara, Clark, Clearwater, Cleva, Cocchieri, Coccia, Cohadon, Cohen, Colla, Collette, Cominsky, Constancio, Conti, Cooper, Corban, Corbitt, Cordero-Carri\'on, Corley, Cornish, Corsi, Cortese, Costa, Coughlin,
  Coughlin, Coulon, Countryman, Couvares, Covas, Cowan, Coward, Cowart, Coyne, Coyne, Creighton, Creighton, Cripe, Crowder, Cullen, Cumming, Cunningham, Cuoco, Dal~Canton, D\'alya, Danilishin, D'Antonio, Danzmann, Dasgupta, Da~Silva~Costa, Dattilo, Dave, Davier, Davis, Daw, Day, De, DeBra, Degallaix, De~Laurentis, Del\'eglise, Del~Pozzo, Demos, Denker, Dent, De~Pietri, Dergachev, De~Rosa, DeRosa, De~Rossi, DeSalvo, de~Varona, Devenson, Dhurandhar, D\'{\i}az, Dietrich, Di~Fiore, Di~Giovanni, Di~Girolamo, Di~Lieto, Di~Pace, Di~Palma, Di~Renzo, Doctor, Dolique, Donovan, Dooley, Doravari, Dorrington, Douglas, Dovale~\'Alvarez, Downes, Drago, Dreissigacker, Driggers, Du, Ducrot, Dudi, Dupej, Dwyer, Edo, Edwards, Effler, Eggenstein, Ehrens, Eichholz, Eikenberry, Eisenstein, Essick, Estevez, Etienne, Etzel, Evans, Evans, Factourovich, Fafone, Fair, Fairhurst, Fan, Farinon, Farr, Farr, Fauchon-Jones, Favata, Fays, Fee, Fehrmann, Feicht, Fejer, Fernandez-Galiana, Ferrante, Ferreira, Ferrini, Fidecaro, Finstad, Fiori,
  Fiorucci, Fishbach, Fisher, Fitz-Axen, Flaminio, Fletcher, Fong, Font, Forsyth, Forsyth, Fournier, Frasca, Frasconi, Frei, Freise, Frey, Frey, Fries, Fritschel, Frolov, Fulda, Fyffe, Gabbard, Gadre, Gaebel, Gair, Gammaitoni, Ganija, Gaonkar, Garcia-Quiros, Garufi, Gateley, Gaudio, Gaur, Gayathri, Gehrels, Gemme, Genin, Gennai, George, George, Gergely, Germain, Ghonge, Ghosh, Ghosh, Ghosh, Giaime, Giardina, Giazotto, Gill, Glover, Goetz, Goetz, Gomes, Goncharov, Gonz\'alez, Gonzalez~Castro, Gopakumar, Gorodetsky, Gossan, Gosselin, Gouaty, Grado, Graef, Granata, Grant, Gras, Gray, Greco, Green, Gretarsson, Groot, Grote, Grunewald, Gruning, Guidi, Guo, Gupta, Gupta, Gushwa, Gustafson, Gustafson, Halim, Hall, Hall, Hamilton, Hammond, Haney, Hanke, Hanks, Hanna, Hannam, Hannuksela, Hanson, Hardwick, Harms, Harry, Harry, Hart, Haster, Haughian, Healy, Heidmann, Heintze, Heitmann, Hello, Hemming, Hendry, Heng, Hennig, Heptonstall, Heurs, Hild, Hinderer, Ho, Hoak, Hofman, Holt, Holz, Hopkins, Horst, Hough, Houston,
  Howell, Hreibi, Hu, Huerta, Huet, Hughey, Husa, Huttner, Huynh-Dinh, Indik, Inta, Intini, Isa, Isac, Isi, Iyer, Izumi, Jacqmin, Jani, Jaranowski, Jawahar, Jim\'enez-Forteza, Johnson, Johnson-McDaniel, Jones, Jones, Jonker, Ju, Junker, Kalaghatgi, Kalogera, Kamai, Kandhasamy, Kang, Kanner, Kapadia, Karki, Karvinen, Kasprzack, Kastaun, Katolik, Katsavounidis, Katzman, Kaufer, Kawabe, K\'ef\'elian, Keitel, Kemball, Kennedy, Kent, Key, Khalili, Khan, Khan, Khan, Khazanov, Kijbunchoo, Kim, Kim, Kim, Kim, Kim, Kim, Kimbrell, King, King, Kinley-Hanlon, Kirchhoff, Kissel, Kleybolte, Klimenko, Knowles, Koch, Koehlenbeck, Koley, Kondrashov, Kontos, Korobko, Korth, Kowalska, Kozak, Kr\"amer, Kringel, Krishnan, Kr\'olak, Kuehn, Kumar, Kumar, Kumar, Kuo, Kutynia, Kwang, Lackey, Lai, Landry, Lang, Lange, Lantz, Lanza, Larson, Lartaux-Vollard, Lasky, Laxen, Lazzarini, Lazzaro, Leaci, Leavey, Lee, Lee, Lee, Lee, Lee, Lehmann, Lenon, Leon, Leonardi, Leroy, Letendre, Levin, Li, Linker, Littenberg, Liu, Liu, Lo, Lockerbie,
  London, Lord, Lorenzini, Loriette, Lormand, Losurdo, Lough, Lousto, Lovelace, L\"uck, Lumaca, Lundgren, Lynch, Ma, Macas, Macfoy, Machenschalk, MacInnis, Macleod, Maga\~na Hernandez, Maga\~na Sandoval, Maga\~na Zertuche, Magee, Majorana, Maksimovic, Man, Mandic, Mangano, Mansell, Manske, Mantovani, Marchesoni, Marion, M\'arka, M\'arka, Markakis, Markosyan, Markowitz, Maros, Marquina, Marsh, Martelli, Martellini, Martin, Martin, Martynov, Marx, Mason, Massera, Masserot, Massinger, Masso-Reid, Mastrogiovanni, Matas, Matichard, Matone, Mavalvala, Mazumder, McCarthy, McClelland, McCormick, McCuller, McGuire, McIntyre, McIver, McManus, McNeill, McRae, McWilliams, Meacher, Meadors, Mehmet, Meidam, Mejuto-Villa, Melatos, Mendell, Mercer, Merilh, Merzougui, Meshkov, Messenger, Messick, Metzdorff, Meyers, Miao, Michel, Middleton, Mikhailov, Milano, Miller, Miller, Miller, Millhouse, Milovich-Goff, Minazzoli, Minenkov, Ming, Mishra, Mitra, Mitrofanov, Mitselmakher, Mittleman, Moffa, Moggi, Mogushi, Mohan, Mohapatra,
  Molina, Montani, Moore, Moraru, Moreno, Morisaki, Morriss, Mours, Mow-Lowry, Mueller, Muir, Mukherjee, Mukherjee, Mukherjee, Mukund, Mullavey, Munch, Mu\~niz, Muratore, Murray, Nagar, Napier, Nardecchia, Naticchioni, Nayak, Neilson, Nelemans, Nelson, Nery, Neunzert, Nevin, Newport, Newton, Ng, Nguyen, Nguyen, Nichols, Nielsen, Nissanke, Nitz, Noack, Nocera, Nolting, North, Nuttall, Oberling, O'Dea, Ogin, Oh, Oh, Ohme, Okada, Oliver, Oppermann, Oram, O'Reilly, Ormiston, Ortega, O'Shaughnessy, Ossokine, Ottaway, Overmier, Owen, Pace, Page, Page, Pai, Pai, Palamos, Palashov, Palomba, Pal-Singh, Pan, Pan, Pang, Pang, Pankow, Pannarale, Pant, Paoletti, Paoli, Papa, Parida, Parker, Pascucci, Pasqualetti, Passaquieti, Passuello, Patil, Patricelli, Pearlstone, Pedraza, Pedurand, Pekowsky, Pele, Penn, Perez, Perreca, Perri, Pfeiffer, Phelps, Piccinni, Pichot, Piergiovanni, Pierro, Pillant, Pinard, Pinto, Pirello, Pitkin, Poe, Poggiani, Popolizio, Porter, Post, Powell, Prasad, Pratt, Pratten, Predoi, Prestegard,
  Prijatelj, Principe, Privitera, Prix, Prodi, Prokhorov, Puncken, Punturo, Puppo, P\"urrer, Qi, Quetschke, Quintero, Quitzow-James, Raab, Rabeling, Radkins, Raffai, Raja, Rajan, Rajbhandari, Rakhmanov, Ramirez, Ramos-Buades, Rapagnani, Raymond, Razzano, Read, Regimbau, Rei, Reid, Reitze, Ren, Reyes, Ricci, Ricker, Rieger, Riles, Rizzo, Robertson, Robie, Robinet, Rocchi, Rolland, Rollins, Roma, Romano, Romano, Romel, Romie, Rosi\ifmmode~\acute{n}\else \'{n}\fi{}ska, Ross, Rowan, R\"udiger, Ruggi, Rutins, Ryan, Sachdev, Sadecki, Sadeghian, Sakellariadou, Salconi, Saleem, Salemi, Samajdar, Sammut, Sampson, Sanchez, Sanchez, Sanchis-Gual, Sandberg, Sanders, Sassolas, Sathyaprakash, Saulson, Sauter, Savage, Sawadsky, Schale, Scheel, Scheuer, Schmidt, Schmidt, Schnabel, Schofield, Sch\"onbeck, Schreiber, Schuette, Schulte, Schutz, Schwalbe, Scott, Scott, Seidel, Sellers, Sengupta, Sentenac, Sequino, Sergeev, Shaddock, Shaffer, Shah, Shahriar, Shaner, Shao, Shapiro, Shawhan, Sheperd, Shoemaker, Shoemaker, Siellez,
  Siemens, Sieniawska, Sigg, Silva, Singer, Singh, Singhal, Sintes, Slagmolen, Smith, Smith, Smith, Somala, Son, Sonnenberg, Sorazu, Sorrentino, Souradeep, Spencer, Srivastava, Staats, Staley, Steinke, Steinlechner, Steinlechner, Steinmeyer, Stevenson, Stone, Stops, Strain, Stratta, Strigin, Strunk, Sturani, Stuver, Summerscales, Sun, Sunil, Suresh, Sutton, Swinkels, Szczepa\ifmmode~\acute{n}\else \'{n}\fi{}czyk, Tacca, Tait, Talbot, Talukder, Tanner, T\'apai, Taracchini, Tasson, Taylor, Taylor, Tewari, Theeg, Thies, Thomas, Thomas, Thomas, Thorne, Thorne, Thrane, Tiwari, Tiwari, Tokmakov, Toland, Tonelli, Tornasi, Torres-Forn\'e, Torrie, T\"oyr\"a, Travasso, Traylor, Trinastic, Tringali, Trozzo, Tsang, Tse, Tso, Tsukada, Tsuna, Tuyenbayev, Ueno, Ugolini, Unnikrishnan, Urban, Usman, Vahlbruch, Vajente, Valdes, Vallisneri, van Bakel, van Beuzekom, van~den Brand, Van Den~Broeck, Vander-Hyde, van~der Schaaf, van Heijningen, van Veggel, Vardaro, Varma, Vass, Vas\'uth, Vecchio, Vedovato, Veitch, Veitch,
  Venkateswara, Venugopalan, Verkindt, Vetrano, Vicer\'e, Viets, Vinciguerra, Vine, Vinet, Vitale, Vo, Vocca, Vorvick, Vyatchanin, Wade, Wade, Wade, Walet, Walker, Wallace, Walsh, Wang, Wang, Wang, Wang, Wang, Ward, Warner, Was, Watchi, Weaver, Wei, Weinert, Weinstein, Weiss, Wen, Wessel, We\ss{}els, Westerweck, Westphal, Wette, Whelan, Whitcomb, Whiting, Whittle, Wilken, Williams, Williams, Williamson, Willis, Willke, Wimmer, Winkler, Wipf, Wittel, Woan, Woehler, Wofford, Wong, Worden, Wright, Wu, Wysocki, Xiao, Yamamoto, Yancey, Yang, Yap, Yazback, Yu, Yu, Yvert, Zadro\ifmmode~\dot{z}\else \.{z}\fi{}ny, Zanolin, Zelenova, Zendri, Zevin, Zhang, Zhang, Zhang, Zhang, Zhao, Zhou, Zhou, Zhu, Zhu, Zimmerman, Zucker, \& Zweizig}]{abb17a}
---. 2017{\natexlab{b}}, Phys. Rev. Lett., 119, 161101, \dodoi{10.1103/PhysRevLett.119.161101}

\bibitem[{Abbott {et~al.}(2020)Abbott, Abbott, Abbott, Abraham, Acernese, Ackley, Adams, Adhikari, Adya, Affeldt, Agathos, Agatsuma, Aggarwal, Aguiar, Aiello, Ain, Ajith, Allen, Allocca, Aloy, Altin, Amato, Anand, Ananyeva, Anderson, Anderson, Angelova, Antier, Appert, Arai, Araya, Areeda, Arène, Arnaud, Aronson, Arun, Ascenzi, Ashton, Aston, Astone, Aubin, Aufmuth, AultONeal, Austin, Avendano, Avila-Alvarez, Babak, Bacon, Badaracco, Bader, Bae, Baird, Baker, Baldaccini, Ballardin, Ballmer, Bals, Banagiri, Barayoga, Barbieri, Barclay, Barish, Barker, Barkett, Barnum, Barone, Barr, Barsotti, Barsuglia, Barta, Bartlett, Bartos, Bassiri, Basti, Bawaj, Bayley, Baylor, Bazzan, Bécsy, Bejger, Belahcene, Bell, Beniwal, Benjamin, Berger, Bergmann, Bernuzzi, Berry, Bersanetti, Bertolini, Betzwieser, Bhandare, Bidler, Biggs, Bilenko, Bilgili, Billingsley, Birney, Birnholtz, Biscans, Bischi, Biscoveanu, Bisht, Bitossi, Bizouard, Blackburn, Blackman, Blair, Blair, Blair, Bloemen, Bobba, Bode, Boer, Boetzel, Bogaert,
  Bondu, Bonnand, Booker, Boom, Bork, Boschi, Bose, Bossilkov, Bosveld, Bouffanais, Bozzi, Bradaschia, Brady, Bramley, Branchesi, Brau, Breschi, Briant, Briggs, Brighenti, Brillet, Brinkmann, Brockill, Brooks, Brooks, Brown, Brunett, Buikema, Bulik, Bulten, Buonanno, Buskulic, Buy, Byer, Cabero, Cadonati, Cagnoli, Cahillane, Bustillo, Callister, Calloni, Camp, Campbell, Canepa, Cannon, Cao, Cao, Carapella, Carbognani, Caride, Carney, Carullo, Diaz, Casentini, Caudill, Cavaglià, Cavalier, Cavalieri, Cella, Cerdá-Durán, Cesarini, Chaibi, Chakravarti, Chamberlin, Chan, Chao, Charlton, Chase, Chassande-Mottin, Chatterjee, Chaturvedi, Chatziioannou, Cheeseboro, Chen, Chen, Chen, Cheng, Cheong, Chia, Chiadini, Chincarini, Chiummo, Cho, Cho, Cho, Christensen, Chu, Chua, Chung, Chung, Ciani, Cieślar, Ciobanu, Ciolfi, Cipriano, Cirone, Clara, Clark, Clearwater, Cleva, Coccia, Cohadon, Cohen, Colleoni, Collette, Collins, Colpi, Cominsky, Constancio, Conti, Cooper, Corban, Corbitt, Cordero-Carrión, Corezzi, Corley,
  Cornish, Corre, Corsi, Cortese, Costa, Cotesta, Coughlin, Coughlin, Coulon, Countryman, Couvares, Covas, Cowan, Coward, Cowart, Coyne, Coyne, Creighton, Creighton, Cripe, Croquette, Crowder, Cullen, Cumming, Cunningham, Cuoco, Canton, Dálya, D’Angelo, Danilishin, D’Antonio, Danzmann, Dasgupta, Costa, Datrier, Dattilo, Dave, Davier, Davis, Daw, DeBra, Deenadayalan, Degallaix, Laurentis, Deléglise, Lillo, Pozzo, DeMarchi, Demos, Dent, Pietri, Rosa, Rossi, DeSalvo, Varona, Dhurandhar, Díaz, Dietrich, Fiore, DiFronzo, Giorgio, Giovanni, Giovanni, Girolamo, Lieto, Ding, Pace, Palma, Renzo, Divakarla, Dmitriev, Doctor, Donovan, Dooley, Doravari, Dorrington, Downes, Drago, Driggers, Du, Ducoin, Dudi, Dupej, Durante, Dwyer, Easter, Eddolls, Edo, Effler, Ehrens, Eichholz, Eikenberry, Eisenmann, Eisenstein, Errico, Essick, Estelles, Estevez, Etienne, Etzel, Evans, Evans, Fafone, Fairhurst, Fan, Farinon, Farr, Farr, Fauchon-Jones, Favata, Fays, Fazio, Fee, Feicht, Fejer, Feng, Fernandez-Galiana, Ferrante,
  Ferreira, Ferreira, Fidecaro, Fiori, Fiorucci, Fishbach, Fisher, Fishner, Fittipaldi, Fitz-Axen, Fiumara, Flaminio, Fletcher, Floden, Flynn, Fong, Font, Forsyth, Fournier, Vivanco, Frasca, Frasconi, Frei, Freise, Frey, Frey, Fritschel, Frolov, Fronzè, Fulda, Fyffe, Gabbard, Gadre, Gaebel, Gair, Gamba, Gammaitoni, Gaonkar, García-Quirós, Garufi, Gateley, Gaudio, Gaur, Gayathri, Gemme, Genin, Gennai, George, George, George, Gergely, Ghonge, Ghosh, Ghosh, Ghosh, Giacomazzo, Giaime, Giardina, Gibson, Gill, Glover, Gniesmer, Godwin, Goetz, Goetz, Goncharov, González, Castro, Gopakumar, Gossan, Gosselin, Gouaty, Grace, Grado, Granata, Grant, Gras, Grassia, Gray, Gray, Greco, Green, Green, Gretarsson, Grimaldi, Grimm, Groot, Grote, Grunewald, Gruning, Guidi, Gulati, Guo, Gupta, Gupta, Gupta, Gustafson, Gustafson, Haegel, Halim, Hall, Hall, Hamilton, Hammond, Haney, Hanke, Hanks, Hanna, Hannam, Hannuksela, Hansen, Hanson, Harder, Hardwick, Haris, Harms, Harry, Harry, Hasskew, Haster, Haughian, Hayes, Healy,
  Heidmann, Heintze, Heitmann, Hellman, Hello, Hemming, Hendry, Heng, Hennig, Heurs, Hild, Hinderer, Ho, Hochheim, Hofman, Holgado, Holland, Holt, Holz, Hopkins, Horst, Hough, Howell, Hoy, Huang, Hübner, Huerta, Huet, Hughey, Hui, Husa, Huttner, Huynh-Dinh, Idzkowski, Iess, Inchauspe, Ingram, Inta, Intini, Irwin, Isa, Isac, Isi, Iyer, Jacqmin, Jadhav, Jani, Janthalur, Jaranowski, Jariwala, Jenkins, Jiang, Johnson, Johnson-McDaniel, Jones, Jones, Jones, Jones, Jonker, Ju, Junker, Kalaghatgi, Kalogera, Kamai, Kandhasamy, Kang, Kanner, Kapadia, Karki, Kashyap, Kasprzack, Kastaun, Katsanevas, Katsavounidis, Katzman, Kaufer, Kawabe, Keerthana, Kéfélian, Keitel, Kennedy, Key, Khalili, Khan, Khan, Khazanov, Khetan, Khursheed, Kijbunchoo, Kim, Kim, Kim, Kim, Kim, Kim, Kimball, King, Kinley-Hanlon, Kirchhoff, Kissel, Kleybolte, Klika, Klimenko, Knowles, Koch, Koehlenbeck, Koekoek, Koley, Kondrashov, Kontos, Koper, Korobko, Korth, Kovalam, Kozak, Krämer, Kringel, Krishnendu, Królak, Krupinski, Kuehn, Kumar, Kumar,
  Kumar, Kumar, Kuo, Kutynia, Kwang, Lackey, Laghi, Lai, Lam, Landry, Landry, Lane, Lang, Lange, Lantz, Lanza, Lartaux-Vollard, Lasky, Laxen, Lazzarini, Lazzaro, Leaci, Leavey, Lecoeuche, Lee, Lee, Lee, Lee, Lee, Lee, Lehmann, Lenon, Leroy, Letendre, Levin, Li, Li, Li, Li, Li, Lin, Linde, Linker, Littenberg, Liu, Liu, Llorens-Monteagudo, Lo, London, Longo, Lorenzini, Loriette, Lormand, Losurdo, Lough, Lousto, Lovelace, Lower, Lucaccioni, Lück, Lumaca, Lundgren, Lynch, Ma, Macas, Macfoy, MacInnis, Macleod, Macquet, Hernandez, Magaña-Sandoval, Magee, Majorana, Maksimovic, Malik, Man, Mandic, Mangano, Mansell, Manske, Mantovani, Mapelli, Marchesoni, Marion, Márka, Márka, Markakis, Markosyan, Markowitz, Maros, Marquina, Marsat, Martelli, Martin, Martin, Martinez, Martynov, Masalehdan, Mason, Massera, Masserot, Massinger, Masso-Reid, Mastrogiovanni, Matas, Matichard, Matone, Mavalvala, McCann, McCarthy, McClelland, McCormick, McCuller, McGuire, McIsaac, McIver, McManus, McRae, McWilliams, Meacher, Meadors,
  Mehmet, Mehta, Meidam, Villa, Melatos, Mendell, Mercer, Mereni, Merfeld, Merilh, Merzougui, Meshkov, Messenger, Messick, Messina, Metzdorff, Meyers, Meylahn, Miani, Miao, Michel, Middleton, Milano, Miller, Millhouse, Mills, Milovich-Goff, Minazzoli, Minenkov, Mishkin, Mishra, Mistry, Mitra, Mitrofanov, Mitselmakher, Mittleman, Mo, Moffa, Mogushi, Mohapatra, Molina-Ruiz, Mondin, Montani, Moore, Moraru, Morawski, Moreno, Morisaki, Mours, Mow-Lowry, Muciaccia, Mukherjee, Mukherjee, Mukherjee, Mukherjee, Mukund, Mullavey, Munch, Muñiz, Muratore, Murray, Nagar, Nardecchia, Naticchioni, Nayak, Neil, Neilson, Nelemans, Nelson, Nery, Neunzert, Nevin, Ng, Ng, Nguyen, Nguyen, Nichols, Nichols, Nissanke, Nocera, North, Nuttall, Obergaulinger, Oberling, O’Brien, Oganesyan, Ogin, Oh, Oh, Ohme, Ohta, Okada, Oliver, Oppermann, Oram, O’Reilly, Ormiston, Ortega, O’Shaughnessy, Ossokine, Ottaway, Overmier, Owen, Pace, Pagano, Page, Pagliaroli, Pai, Pai, Palamos, Palashov, Palomba, Pan, Panda, Pang, Pankow, Pannarale,
  Pant, Paoletti, Paoli, Parida, Parker, Pascucci, Pasqualetti, Passaquieti, Passuello, Patil, Patricelli, Payne, Pearlstone, Pechsiri, Pedersen, Pedraza, Pedurand, Pele, Penn, Perego, Perez, Périgois, Perreca, Petermann, Pfeiffer, Phelps, Phukon, Piccinni, Pichot, Piergiovanni, Pierro, Pillant, Pinard, Pinto, Pirello, Pitkin, Plastino, Poggiani, Pong, Ponrathnam, Popolizio, Porter, Powell, Prajapati, Prasad, Prasai, Prasanna, Pratten, Prestegard, Principe, Prodi, Prokhorov, Punturo, Puppo, Pürrer, Qi, Quetschke, Quinonez, Raab, Raaijmakers, Radkins, Radulesco, Raffai, Raja, Rajan, Rajbhandari, Rakhmanov, Ramirez, Ramos-Buades, Rana, Rao, Rapagnani, Raymond, Razzano, Read, Regimbau, Rei, Reid, Reitze, Rettegno, Ricci, Richardson, Richardson, Ricker, Riemenschneider, Riles, Rizzo, Robertson, Robinet, Rocchi, Rolland, Rollins, Roma, Romanelli, Romano, Romel, Romie, Rose, Rose, Rose, Rosell, Rosińska, Rosofsky, Ross, Rowan, Roy, Rüdiger, Ruggi, Rutins, Ryan, Sachdev, Sadecki, Sakellariadou, Salafia, Salconi,
  Saleem, Samajdar, Sammut, Sanchez, Sanchez, Sanchis-Gual, Sanders, Santiago, Santos, Sarin, Sassolas, Sathyaprakash, Sauter, Savage, Schale, Scheel, Scheuer, Schmidt, Schnabel, Schofield, Schönbeck, Schreiber, Schulte, Schutz, Scott, Scott, Seidel, Sellers, Sengupta, Sennett, Sentenac, Sequino, Sergeev, Setyawati, Shaddock, Shaffer, Shahriar, Shaner, Sharma, Sharma, Shawhan, Shen, Shink, Shoemaker, Shoemaker, Shukla, ShyamSundar, Siellez, Sieniawska, Sigg, Singer, Singh, Singh, Singhal, Sintes, Sitmukhambetov, Skliris, Slagmolen, Slaven-Blair, Smith, Smith, Somala, Son, Soni, Sorazu, Sorrentino, Souradeep, Sowell, Spencer, Spera, Srivastava, Srivastava, Staats, Stachie, Standke, Steer, Steinke, Steinlechner, Steinlechner, Steinmeyer, Stevenson, Stocks, Stone, Stops, Strain, Stratta, Strigin, Strunk, Sturani, Stuver, Sudhir, Summerscales, Sun, Sunil, Sur, Suresh, Sutton, Swinkels, Szczepańczyk, Tacca, Tait, Talbot, Tanner, Tao, Tápai, Tapia, Tasson, Taylor, Tenorio, Terkowski, Thomas, Thomas, Thondapu,
  Thorne, Thrane, Tiwari, Tiwari, Tiwari, Toland, Tonelli, Tornasi, Torres-Forné, Torrie, Töyrä, Travasso, Traylor, Tringali, Tripathee, Trovato, Trozzo, Tsang, Tse, Tso, Tsukada, Tsuna, Tsutsui, Tuyenbayev, Ueno, Ugolini, Unnikrishnan, Urban, Usman, Vahlbruch, Vajente, Valdes, Valentini, Bakel, Beuzekom, Brand, Broeck, Vander-Hyde, Schaaf, VanHeijningen, Veggel, Vardaro, Varma, Vass, Vasúth, Vecchio, Vedovato, Veitch, Veitch, Venkateswara, Venugopalan, Verkindt, Vetrano, Viceré, Viets, Vinciguerra, Vine, Vinet, Vitale, Vo, Vocca, Vorvick, Vyatchanin, Wade, Wade, Wade, Walet, Walker, Wallace, Walsh, Wang, Wang, Wang, Wang, Ward, Warden, Warner, Was, Watchi, Weaver, Wei, Weinert, Weinstein, Weiss, Wellmann, Wen, Wessel, Weßels, Westhouse, Wette, Whelan, White, Whiting, Whittle, Wilken, Williams, Williamson, Willis, Willke, Winkler, Wipf, Wittel, Woan, Woehler, Wofford, Wright, Wu, Wysocki, Xiao, Xu, Yamamoto, Yancey, Yang, Yang, Yang, Yap, Yazback, Yeeles, Yu, Yu, Yuen, Zadrożny, Zadrożny, Zanolin,
  Zelenova, Zendri, Zevin, Zhang, Zhang, Zhang, Zhao, Zhao, Zhou, Zhou, Zhu, Zimmerman, Zucker, \& Zweizig}]{abb20a}
---. 2020, The Astrophysical Journal Letters, 892, L3, \dodoi{10.3847/2041-8213/ab75f5}

\bibitem[{{Abbott} {et~al.}(2020){Abbott}, {Abbott}, {Abbott}, {Abraham}, {Acernese}, {Ackley}, {Adams}, {Adya}, {Affeldt}, {Agathos}, {Agatsuma}, {Aggarwal}, {Aguiar}, {Aiello}, {Ain}, {Ajith}, {Allen}, {Allocca}, {Aloy}, {Altin}, {Amato}, {Anand}, {Ananyeva}, {Anderson}, {Anderson}, {Angelova}, {Antier}, {Appert}, {Arai}, {Araya}, {Areeda}, {Ar{\`e}ne}, {Arnaud}, {Aronson}, {Ascenzi}, {Ashton}, {Aston}, {Astone}, {Aubin}, {Aufmuth}, {AultONeal}, {Austin}, {Avendano}, {Avila-Alvarez}, {Babak}, {Bacon}, {Badaracco}, {Bader}, {Bae}, {Baird}, {Baker}, {Baldaccini}, {Ballardin}, {Ballmer}, {Bals}, {Banagiri}, {Barayoga}, {Barbieri}, {Barclay}, {Barish}, {Barker}, {Barkett}, {Barnum}, {Barone}, {Barr}, {Barsotti}, {Barsuglia}, {Barta}, {Bartlett}, {Bartos}, {Bassiri}, {Basti}, {Bawaj}, {Bayley}, {Bazzan}, {B{\'e}csy}, {Bejger}, {Belahcene}, {Bell}, {Beniwal}, {Benjamin}, {Bergmann}, {Bernuzzi}, {Berry}, {Bersanetti}, {Bertolini}, {Betzwieser}, {Bhandare}, {Bidler}, {Biggs}, {Bilenko}, {Bilgili}, {Billingsley},
  {Birney}, {Birnholtz}, {Biscans}, {Bischi}, {Biscoveanu}, {Bisht}, {Bitossi}, {Bizouard}, {Blackburn}, {Blackman}, {Blair}, {Blair}, {Blair}, {Bloemen}, {Bobba}, {Bode}, {Boer}, {Boetzel}, {Bogaert}, {Bondu}, {Bonnand}, {Booker}, {Boom}, {Bork}, {Boschi}, {Bose}, {Bossilkov}, {Bosveld}, {Bouffanais}, {Bozzi}, {Bradaschia}, {Brady}, {Bramley}, {Branchesi}, {Brau}, {Breschi}, {Briant}, {Briggs}, {Brighenti}, {Brillet}, {Brinkmann}, {Brockill}, {Brooks}, {Brooks}, {Brown}, {Brunett}, {Buikema}, {Bulik}, {Bulten}, {Buonanno}, {Buskulic}, {Buy}, {Byer}, {Cabero}, {Cadonati}, {Cagnoli}, {Cahillane}, {Bustillo}, {Callister}, {Calloni}, {Camp}, {Campbell}, {Canepa}, {Cannon}, {Cao}, {Cao}, {Carapella}, {Carbognani}, {Caride}, {Carney}, {Carullo}, {Diaz}, {Casentini}, {Caudill}, {Cavagli{\`a}}, {Cavalier}, {Cavalieri}, {Cella}, {Cerd{\'a}-Dur{\'a}n}, {Cesarini}, {Chaibi}, {Chakravarti}, {Chamberlin}, {Chan}, {Chao}, {Charlton}, {Chase}, {Chassande-Mottin}, {Chatterjee}, {Chaturvedi}, {Cheeseboro}, {Chen}, {Chen},
  {Chen}, {Cheng}, {Cheong}, {Chia}, {Chiadini}, {Chincarini}, {Chiummo}, {Cho}, {Cho}, {Cho}, {Christensen}, {Chu}, {Chua}, {Chung}, {Chung}, {Ciani}, {Cie{\'s}lar}, {Ciobanu}, {Ciolfi}, {Cipriano}, {Cirone}, {Clara}, {Clark}, {Clearwater}, {Cleva}, {Coccia}, {Cohadon}, {Cohen}, {Colleoni}, {Collette}, {Collins}, {Colpi}, {Cominsky}, {Constancio}, {Conti}, {Cooper}, {Corban}, {Corbitt}, {Cordero-Carri{\'o}n}, {Corezzi}, {Corley}, {Cornish}, {Corre}, {Corsi}, {Cortese}, {Costa}, {Cotesta}, {Coughlin}, {Coughlin}, {Coulon}, {Countryman}, {Couvares}, {Covas}, {Cowan}, {Coward}, {Cowart}, {Coyne}, {Coyne}, {Creighton}, {Creighton}, {Cripe}, {Croquette}, {Crowder}, {Cullen}, {Cumming}, {Cunningham}, {Cuoco}, {Canton}, {D{\'a}lya}, {D'Angelo}, {Danilishin}, {D'Antonio}, {Danzmann}, {Dasgupta}, {Costa}, {Datrier}, {Dattilo}, {Dave}, {Davier}, {Davis}, {Daw}, {DeBra}, {Deenadayalan}, {Degallaix}, {De Laurentis}, {Del{\'e}glise}, {Del Pozzo}, {DeMarchi}, {Demos}, {Dent}, {De Pietri}, {De Rosa}, {De Rossi}, {DeSalvo},
  {de Varona}, {Dhurandhar}, {D{\'\i}az}, {Dietrich}, {Di Fiore}, {DiFronzo}, {Di Giorgio}, {Di Giovanni}, {Di Giovanni}, {Di Girolamo}, {Di Lieto}, {Ding}, {Di Pace}, {Di Palma}, {Di Renzo}, {Divakarla}, {Dmitriev}, {Doctor}, {Donovan}, {Dooley}, {Doravari}, {Dorrington}, {Downes}, {Drago}, {Driggers}, {Du}, {Ducoin}, {Dupej}, {Durante}, {Dwyer}, {Easter}, {Eddolls}, {Edo}, {Effler}, {Ehrens}, {Eichholz}, {Eikenberry}, {Eisenmann}, {Eisenstein}, {Errico}, {Essick}, {Estelles}, {Estevez}, {Etienne}, {Etzel}, {Evans}, {Evans}, {Fafone}, {Fairhurst}, {Fan}, {Farinon}, {Farr}, {Farr}, {Fauchon-Jones}, {Favata}, {Fays}, {Fazio}, {Fee}, {Feicht}, {Fejer}, {Feng}, {Fernandez-Galiana}, {Ferrante}, {Ferreira}, {Ferreira}, {Fidecaro}, {Fiori}, {Fiorucci}, {Fishbach}, {Fisher}, {Fishner}, {Fittipaldi}, {Fitz-Axen}, {Fiumara}, {Flaminio}, {Fletcher}, {Floden}, {Flynn}, {Fong}, {Font}, {Forsyth}, {Fournier}, {Vivanco}, {Frasca}, {Frasconi}, {Frei}, {Freise}, {Frey}, {Frey}, {Fritschel}, {Frolov}, {Fronz{\`e}}, {Fulda},
  {Fyffe}, {Gabbard}, {Gadre}, {Gaebel}, {Gair}, {Gammaitoni}, {Gaonkar}, {Garc{\'\i}a-Quir{\'o}s}, {Garufi}, {Gateley}, {Gaudio}, {Gaur}, {Gayathri}, {Gemme}, {Genin}, {Gennai}, {George}, {George}, {Gergely}, {Ghonge}, {Ghosh}, {Ghosh}, {Ghosh}, {Giacomazzo}, {Giaime}, {Giardina}, {Gibson}, {Gill}, {Glover}, {Gniesmer}, {Godwin}, {Goetz}, {Goetz}, {Goncharov}, {Gonz{\'a}lez}, {Castro}, {Gopakumar}, {Gossan}, {Gosselin}, {Gouaty}, {Grace}, {Grado}, {Granata}, {Grant}, {Gras}, {Grassia}, {Gray}, {Gray}, {Greco}, {Green}, {Green}, {Gretarsson}, {Grimaldi}, {Grimm}, {Groot}, {Grote}, {Grunewald}, {Gruning}, {Guidi}, {Gulati}, {Guo}, {Gupta}, {Gupta}, {Gupta}, {Gustafson}, {Gustafson}, {Haegel}, {Halim}, {Hall}, {Hall}, {Hamilton}, {Hammond}, {Haney}, {Hanke}, {Hanks}, {Hanna}, {Hannam}, {Hannuksela}, {Hansen}, {Hanson}, {Harder}, {Hardwick}, {Haris}, {Harms}, {Harry}, {Harry}, {Hasskew}, {Haster}, {Haughian}, {Hayes}, {Healy}, {Heidmann}, {Heintze}, {Heitmann}, {Hellman}, {Hello}, {Hemming}, {Hendry}, {Heng},
  {Hennig}, {Heurs}, {Hild}, {Hinderer}, {Hochheim}, {Hofman}, {Holgado}, {Holland}, {Holt}, {Holz}, {Hopkins}, {Horst}, {Hough}, {Howell}, {Hoy}, {Huang}, {H{\"u}bner}, {Huerta}, {Huet}, {Hughey}, {Hui}, {Husa}, {Huttner}, {Huynh-Dinh}, {Idzkowski}, {Iess}, {Inchauspe}, {Ingram}, {Inta}, {Intini}, {Irwin}, {Isa}, {Isac}, {Isi}, {Iyer}, {Jacqmin}, {Jadhav}, {Jani}, {Janthalur}, {Jaranowski}, {Jariwala}, {Jenkins}, {Jiang}, {Johnson}, {Jones}, {Jones}, {Jones}, {Jones}, {Jonker}, {Ju}, {Junker}, {Kalaghatgi}, {Kalogera}, {Kamai}, {Kandhasamy}, {Kang}, {Kanner}, {Kapadia}, {Karki}, {Kashyap}, {Kasprzack}, {Katsanevas}, {Katsavounidis}, {Katzman}, {Kaufer}, {Kawabe}, {Keerthana}, {K{\'e}f{\'e}lian}, {Keitel}, {Kennedy}, {Key}, {Khalili}, {Khan}, {Khan}, {Khazanov}, {Khetan}, {Khursheed}, {Kijbunchoo}, {Kim}, {Kim}, {Kim}, {Kim}, {Kim}, {Kim}, {Kimball}, {King}, {Kinley-Hanlon}, {Kirchhoff}, {Kissel}, {Kleybolte}, {Klika}, {Klimenko}, {Knowles}, {Koch}, {Koehlenbeck}, {Koekoek}, {Koley}, {Kondrashov}, {Kontos},
  {Koper}, {Korobko}, {Korth}, {Kovalam}, {Kozak}, {Kr{\"a}mer}, {Kringel}, {Krishnendu}, {Kr{\'o}lak}, {Krupinski}, {Kuehn}, {Kumar}, {Kumar}, {Kumar}, {Kumar}, {Kuo}, {Kutynia}, {Kwang}, {Lackey}, {Laghi}, {Lai}, {Lam}, {Landry}, {Lane}, {Lang}, {Lange}, {Lantz}, {Lanza}, {Lartaux-Vollard}, {Lasky}, {Laxen}, {Lazzarini}, {Lazzaro}, {Leaci}, {Leavey}, {Lecoeuche}, {Lee}, {Lee}, {Lee}, {Lee}, {Lee}, {Lee}, {Lehmann}, {Lenon}, {Leroy}, {Letendre}, {Levin}, {Li}, {Li}, {Li}, {Li}, {Li}, {Lin}, {Linde}, {Linker}, {Littenberg}, {Liu}, {Liu}, {Llorens-Monteagudo}, {Lo}, {London}, {Longo}, {Lorenzini}, {Loriette}, {Lormand}, {Losurdo}, {Lough}, {Lousto}, {Lovelace}, {Lower}, {L{\"u}ck}, {Lumaca}, {Lundgren}, {Lynch}, {Ma}, {Macas}, {Macfoy}, {MacInnis}, {Macleod}, {Macquet}, {Hernandez}, {Maga{\~n}a-Sandoval}, {Magee}, {Majorana}, {Maksimovic}, {Malik}, {Man}, {Mandic}, {Mangano}, {Mansell}, {Manske}, {Mantovani}, {Mapelli}, {Marchesoni}, {Marion}, {M{\'a}rka}, {M{\'a}rka}, {Markakis}, {Markosyan}, {Markowitz},
  {Maros}, {Marquina}, {Marsat}, {Martelli}, {Martin}, {Martin}, {Martinez}, {Martynov}, {Masalehdan}, {Mason}, {Massera}, {Masserot}, {Massinger}, {Masso-Reid}, {Mastrogiovanni}, {Matas}, {Matichard}, {Matone}, {Mavalvala}, {McCann}, {McCarthy}, {McClelland}, {McCormick}, {McCuller}, {McGuire}, {McIsaac}, {McIver}, {McManus}, {McRae}, {McWilliams}, {Meacher}, {Meadors}, {Mehmet}, {Mehta}, {Meidam}, {Villa}, {Melatos}, {Mendell}, {Mercer}, {Mereni}, {Merfeld}, {Merilh}, {Merzougui}, {Meshkov}, {Messenger}, {Messick}, {Messina}, {Metzdorff}, {Meyers}, {Meylahn}, {Miani}, {Miao}, {Michel}, {Middleton}, {Milano}, {Miller}, {Millhouse}, {Mills}, {Milovich-Goff}, {Minazzoli}, {Minenkov}, {Mishkin}, {Mishra}, {Mistry}, {Mitra}, {Mitrofanov}, {Mitselmakher}, {Mittleman}, {Mo}, {Moffa}, {Mogushi}, {Mohapatra}, {Molina-Ruiz}, {Mondin}, {Montani}, {Moore}, {Moraru}, {Morawski}, {Moreno}, {Morisaki}, {Mours}, {Mow-Lowry}, {Muciaccia}, {Mukherjee}, {Mukherjee}, {Mukherjee}, {Mukherjee}, {Mukund}, {Mullavey}, {Munch},
  {Mu{\~n}iz}, {Muratore}, {Murray}, {Nardecchia}, {Naticchioni}, {Nayak}, {Neil}, {Neilson}, {Nelemans}, {Nelson}, {Nery}, {Neunzert}, {Nevin}, {Ng}, {Ng}, {Nguyen}, {Nguyen}, {Nichols}, {Nichols}, {Nissanke}, {Nocera}, {North}, {Nuttall}, {Obergaulinger}, {Oberling}, {O'Brien}, {Oganesyan}, {Ogin}, {Oh}, {Oh}, {Ohme}, {Ohta}, {Okada}, {Oliver}, {Oppermann}, {Oram}, {O'Reilly}, {Ormiston}, {Ortega}, {O'Shaughnessy}, {Ossokine}, {Ottaway}, {Overmier}, {Owen}, {Pace}, {Pagano}, {Page}, {Pagliaroli}, {Pai}, {Pai}, {Palamos}, {Palashov}, {Palomba}, {Pan}, {Panda}, {Pang}, {Pankow}, {Pannarale}, {Pant}, {Paoletti}, {Paoli}, {Parida}, {Parker}, {Pascucci}, {Pasqualetti}, {Passaquieti}, {Passuello}, {Patil}, {Patricelli}, {Payne}, {Pearlstone}, {Pechsiri}, {Pedersen}, {Pedraza}, {Pedurand}, {Pele}, {Penn}, {Perego}, {Perez}, {P{\'e}rigois}, {Perreca}, {Petermann}, {Pfeiffer}, {Phelps}, {Phukon}, {Piccinni}, {Pichot}, {Piergiovanni}, {Pierro}, {Pillant}, {Pinard}, {Pinto}, {Pirello}, {Pitkin}, {Plastino},
  {Poggiani}, {Pong}, {Ponrathnam}, {Popolizio}, {Porter}, {Powell}, {Prajapati}, {Prasad}, {Prasai}, {Prasanna}, {Pratten}, {Prestegard}, {Principe}, {Prodi}, {Prokhorov}, {Punturo}, {Puppo}, {P{\"u}rrer}, {Qi}, {Quetschke}, {Quinonez}, {Raab}, {Raaijmakers}, {Radkins}, {Radulesco}, {Raffai}, {Raja}, {Rajan}, {Rajbhandari}, {Rakhmanov}, {Ramirez}, {Ramos-Buades}, {Rana}, {Rao}, {Rapagnani}, {Raymond}, {Razzano}, {Read}, {Regimbau}, {Rei}, {Reid}, {Reitze}, {Rettegno}, {Ricci}, {Richardson}, {Richardson}, {Ricker}, {Riemenschneider}, {Riles}, {Rizzo}, {Robertson}, {Robinet}, {Rocchi}, {Rolland}, {Rollins}, {Roma}, {Romanelli}, {Romano}, {Romel}, {Romie}, {Rose}, {Rose}, {Rose}, {Rosi{\'n}ska}, {Rosofsky}, {Ross}, {Rowan}, {R{\"u}diger}, {Ruggi}, {Rutins}, {Ryan}, {Sachdev}, {Sadecki}, {Sakellariadou}, {Salafia}, {Salconi}, {Saleem}, {Samajdar}, {Sammut}, {Sanchez}, {Sanchez}, {Sanchis-Gual}, {Sanders}, {Santiago}, {Santos}, {Sarin}, {Sassolas}, {Sauter}, {Savage}, {Schale}, {Scheel}, {Scheuer}, {Schmidt},
  {Schnabel}, {Schofield}, {Sch{\"o}nbeck}, {Schreiber}, {Schulte}, {Schutz}, {Scott}, {Scott}, {Seidel}, {Sellers}, {Sengupta}, {Sennett}, {Sentenac}, {Sequino}, {Sergeev}, {Setyawati}, {Shaddock}, {Shaffer}, {Shahriar}, {Shaner}, {Sharma}, {Sharma}, {Shawhan}, {Shen}, {Shink}, {Shoemaker}, {Shoemaker}, {Shukla}, {ShyamSundar}, {Siellez}, {Sieniawska}, {Sigg}, {Singer}, {Singh}, {Singh}, {Singhal}, {Sintes}, {Sitmukhambetov}, {Skliris}, {Slagmolen}, {Slaven-Blair}, {Smith}, {Smith}, {Somala}, {Son}, {Soni}, {Sorazu}, {Sorrentino}, {Souradeep}, {Sowell}, {Spencer}, {Spera}, {Srivastava}, {Srivastava}, {Staats}, {Stachie}, {Standke}, {Steer}, {Steinke}, {Steinlechner}, {Steinlechner}, {Steinmeyer}, {Stevenson}, {Stocks}, {Stone}, {Stops}, {Strain}, {Stratta}, {Strigin}, {Strunk}, {Sturani}, {Stuver}, {Sudhir}, {Summerscales}, {Sun}, {Sunil}, {Sur}, {Suresh}, {Sutton}, {Swinkels}, {Szczepa{\'n}czyk}, {Tacca}, {Tait}, {Talbot}, {Tanner}, {Tao}, {T{\'a}pai}, {Tapia}, {Tasson}, {Taylor}, {Tenorio}, {Terkowski},
  {Thomas}, {Thomas}, {Thondapu}, {Thorne}, {Thrane}, {Tiwari}, {Tiwari}, {Tiwari}, {Toland}, {Tonelli}, {Tornasi}, {Torres-Forn{\'e}}, {Torrie}, {T{\"o}yr{\"a}}, {Travasso}, {Traylor}, {Tringali}, {Tripathee}, {Trovato}, {Trozzo}, {Tsang}, {Tse}, {Tso}, {Tsukada}, {Tsuna}, {Tsutsui}, {Tuyenbayev}, {Ueno}, {Ugolini}, {Unnikrishnan}, {Urban}, {Usman}, {Vahlbruch}, {Vajente}, {Valdes}, {Valentini}, {van Bakel}, {van Beuzekom}, {van den Brand}, {Van Den Broeck}, {Vander-Hyde}, {van der Schaaf}, {VanHeijningen}, {van Veggel}, {Vardaro}, {Varma}, {Vass}, {Vas{\'u}th}, {Vecchio}, {Vedovato}, {Veitch}, {Veitch}, {Venkateswara}, {Venugopalan}, {Verkindt}, {Vetrano}, {Vicer{\'e}}, {Viets}, {Vinciguerra}, {Vine}, {Vinet}, {Vitale}, {Vo}, {Vocca}, {Vorvick}, {Vyatchanin}, {Wade}, {Wade}, {Wade}, {Walet}, {Walker}, {Wallace}, {Walsh}, {Wang}, {Wang}, {Wang}, {Wang}, {Wang}, {Ward}, {Warden}, {Warner}, {Was}, {Watchi}, {Weaver}, {Wei}, {Weinert}, {Weinstein}, {Weiss}, {Wellmann}, {Wen}, {Wessel}, {We{\ss}els},
  {Westhouse}, {Wette}, {Whelan}, {Whiting}, {Whittle}, {Wilken}, {Williams}, {Williamson}, {Willis}, {Willke}, {Winkler}, {Wipf}, {Wittel}, {Woan}, {Woehler}, {Wofford}, {Wright}, {Wu}, {Wysocki}, {Xiao}, {Xu}, {Yamamoto}, {Yancey}, {Yang}, {Yang}, {Yang}, {Yap}, {Yazback}, {Yeeles}, {Yu}, {Yu}, {Yuen}, {Zadro{\.Z}ny}, {Zadro{\.Z}ny}, {Zanolin}, {Zelenova}, {Zendri}, {Zevin}, {Zhang}, {Zhang}, {Zhang}, {Zhao}, {Zhao}, {Zhou}, {Zhou}, {Zhu}, {Zucker}, {Zweizig}, {Holoien}, {Kochanek}, {Prieto}, {Shappee}, {Stanek}, {Haislip}, {Kouprianov}, {Reichart}, {Sand}, {Tartaglia}, {Valenti}, {Wyatt}, {Yang}, {Salemi}, {LIGO Scientific Collaboration}, \& {Virgo Collaboration}}]{abb20c}
{Abbott}, B.~P., {Abbott}, R., {Abbott}, T.~D., {et~al.} 2020, \prd, 101, 084002, \dodoi{10.1103/PhysRevD.101.084002}

\bibitem[{Abbott {et~al.}(2020)Abbott, Abbott, Abraham, Acernese, Ackley, Adams, Adhikari, Adya, Affeldt, Agathos, Agatsuma, Aggarwal, Aguiar, Aich, Aiello, Ain, Ajith, Akcay, Allen, Allocca, Altin, Amato, Anand, Ananyeva, Anderson, Anderson, Angelova, Ansoldi, Antier, Appert, Arai, Araya, Areeda, Arène, Arnaud, Aronson, Arun, Asali, Ascenzi, Ashton, Aston, Astone, Aubin, Aufmuth, AultONeal, Austin, Avendano, Babak, Bacon, Badaracco, Bader, Bae, Baer, Baird, Baldaccini, Ballardin, Ballmer, Bals, Balsamo, Baltus, Banagiri, Bankar, Bankar, Barayoga, Barbieri, Barish, Barker, Barkett, Barneo, Barone, Barr, Barsotti, Barsuglia, Barta, Bartlett, Bartos, Bassiri, Basti, Bawaj, Bayley, Bazzan, Bécsy, Bejger, Belahcene, Bell, Beniwal, Benjamin, Benkel, Bentley, Bergamin, Berger, Bergmann, Bernuzzi, Berry, Bersanetti, Bertolini, Betzwieser, Bhandare, Bhandari, Bidler, Biggs, Bilenko, Billingsley, Birney, Birnholtz, Biscans, Bischi, Biscoveanu, Bisht, Bissenbayeva, Bitossi, Bizouard, Blackburn, Blackman, Blair,
  Blair, Blair, Bobba, Bode, Boer, Boetzel, Bogaert, Bondu, Bonilla, Bonnand, Booker, Boom, Bork, Boschi, Bose, Bossilkov, Bosveld, Bouffanais, Bozzi, Bradaschia, Brady, Bramley, Branchesi, Brau, Breschi, Briant, Briggs, Brighenti, Brillet, Brinkmann, Brito, Brockill, Brooks, Brooks, Brown, Brunett, Bruno, Bruntz, Buikema, Bulik, Bulten, Buonanno, Buskulic, Byer, Cabero, Cadonati, Cagnoli, Cahillane, Bustillo, Callaghan, Callister, Calloni, Camp, Canepa, Cannon, Cao, Cao, Carapella, Carbognani, Caride, Carney, Carullo, Diaz, Casentini, Castañeda, Caudill, Cavaglià, Cavalier, Cavalieri, Cella, Cerdá-Durán, Cesarini, Chaibi, Chakravarti, Chan, Chan, Chao, Charlton, Chase, Chassande-Mottin, Chatterjee, Chaturvedi, Chatziioannou, Chen, Chen, Chen, Cheng, Cheong, Chia, Chiadini, Chierici, Chincarini, Chiummo, Cho, Cho, Cho, Christensen, Chu, Chua, Chung, Chung, Ciani, Ciecielag, Cieślar, Ciobanu, Ciolfi, Cipriano, Cirone, Clara, Clark, Clearwater, Clesse, Cleva, Coccia, Cohadon, Cohen, Colleoni, Collette,
  Collins, Colpi, Constancio, Conti, Cooper, Corban, Corbitt, Cordero-Carrión, Corezzi, Corley, Cornish, Corre, Corsi, Cortese, Costa, Cotesta, Coughlin, Coughlin, Coulon, Countryman, Couvares, Covas, Coward, Cowart, Coyne, Coyne, Creighton, Creighton, Cripe, Croquette, Crowder, Cudell, Cullen, Cumming, Cummings, Cunningham, Cuoco, Curylo, Canton, Dálya, Dana, Daneshgaran-Bajastani, D’Angelo, Danilishin, D’Antonio, Danzmann, Darsow-Fromm, Dasgupta, Datrier, Dattilo, Dave, Davier, Davies, Davis, Daw, DeBra, Deenadayalan, Degallaix, Laurentis, Deléglise, Delfavero, Lillo, Pozzo, DeMarchi, D’Emilio, Demos, Dent, Pietri, Rosa, Rossi, DeSalvo, Varona, Dhurandhar, Díaz, Diaz-Ortiz, Dietrich, Fiore, Fronzo, Giorgio, Giovanni, Giovanni, Girolamo, Lieto, Ding, Pace, Palma, Renzo, Divakarla, Dmitriev, Doctor, Donovan, Dooley, Doravari, Dorrington, Downes, Drago, Driggers, Du, Ducoin, Dupej, Durante, D’Urso, Dwyer, Easter, Eddolls, Edelman, Edo, Edy, Effler, Ehrens, Eichholz, Eikenberry, Eisenmann,
  Eisenstein, Ejlli, Errico, Essick, Estelles, Estevez, Etienne, Etzel, Evans, Evans, Ewing, Fafone, Fairhurst, Fan, Farinon, Farr, Farr, Fauchon-Jones, Favata, Fays, Fazio, Feicht, Fejer, Feng, Fenyvesi, Ferguson, Fernandez-Galiana, Ferrante, Ferreira, Ferreira, Fidecaro, Fiori, Fiorucci, Fishbach, Fisher, Fittipaldi, Fitz-Axen, Fiumara, Flaminio, Floden, Flynn, Fong, Font, Forsyth, Fournier, Frasca, Frasconi, Frei, Freise, Frey, Frey, Fritschel, Frolov, Fronzè, Fulda, Fyffe, Gabbard, Gadre, Gaebel, Gair, Galaudage, Ganapathy, Ganguly, Gaonkar, García-Quirós, Garufi, Gateley, Gaudio, Gayathri, Gemme, Genin, Gennai, George, George, Gergely, Ghonge, Ghosh, Ghosh, Ghosh, Giacomazzo, Giaime, Giardina, Gibson, Gier, Gill, Glanzer, Gniesmer, Godwin, Goetz, Goetz, Gohlke, Goncharov, González, Gopakumar, Gossan, Gosselin, Gouaty, Grace, Grado, Granata, Grant, Gras, Grassia, Gray, Gray, Greco, Green, Green, Gretarsson, Griggs, Grignani, Grimaldi, Grimm, Grote, Grunewald, Gruning, Guidi, Guimaraes, Guixé, Gulati,
  Guo, Gupta, Gupta, Gupta, Gustafson, Gustafson, Haegel, Halim, Hall, Hamilton, Hammond, Haney, Hanke, Hanks, Hanna, Hannam, Hannuksela, Hansen, Hanson, Harder, Hardwick, Haris, Harms, Harry, Harry, Hasskew, Haster, Haughian, Hayes, Healy, Heidmann, Heintze, Heinze, Heitmann, Hellman, Hello, Hemming, Hendry, Heng, Hennes, Hennig, Heurs, Hild, Hinderer, Hoback, Hochheim, Hofgard, Hofman, Holgado, Holland, Holt, Holz, Hopkins, Horst, Hough, Howell, Hoy, Huang, Hübner, Huerta, Huet, Hughey, Hui, Husa, Huttner, Huxford, Huynh-Dinh, Idzkowski, Iess, Inchauspe, Ingram, Intini, Isac, Isi, Iyer, Jacqmin, Jadhav, Jadhav, James, Jani, Janthalur, Jaranowski, Jariwala, Jaume, Jenkins, Jiang, Johns, Johnson-McDaniel, Jones, Jones, Jones, Jones, Jones, Jonker, Ju, Junker, Kalaghatgi, Kalogera, Kamai, Kandhasamy, Kang, Kanner, Kapadia, Karki, Kashyap, Kasprzack, Kastaun, Katsanevas, Katsavounidis, Katzman, Kaufer, Kawabe, Kéfélian, Keitel, Keivani, Kennedy, Key, Khadka, Khalili, Khan, Khan, Khan, Khazanov, Khetan,
  Khursheed, Kijbunchoo, Kim, Kim, Kim, Kim, Kim, Kim, Kim, Kimball, King, Kinley-Hanlon, Kirchhoff, Kissel, Kleybolte, Klimenko, Knowles, Knyazev, Koch, Koehlenbeck, Koekoek, Koley, Kondrashov, Kontos, Koper, Korobko, Korth, Kovalam, Kozak, Kringel, Krishnendu, Królak, Krupinski, Kuehn, Kumar, Kumar, Kumar, Kumar, Kumar, Kuo, Kutynia, Lackey, Laghi, Lalande, Lam, Lamberts, Landry, Landry, Lane, Lang, Lange, Lantz, Lanza, Rosa, Lartaux-Vollard, Lasky, Laxen, Lazzarini, Lazzaro, Leaci, Leavey, Lecoeuche, Lee, Lee, Lee, Lee, Lee, Lehmann, Leroy, Letendre, Levin, Li, Li, li, Li, Li, Linde, Linker, Linley, Littenberg, Liu, Liu, Llorens-Monteagudo, Lo, Lockwood, London, Longo, Lorenzini, Loriette, Lormand, Losurdo, Lough, Lousto, Lovelace, Lück, Lumaca, Lundgren, Ma, Macas, Macfoy, MacInnis, Macleod, MacMillan, Macquet, Hernandez, Magaña-Sandoval, Magee, Majorana, Maksimovic, Malik, Man, Mandic, Mangano, Mansell, Manske, Mantovani, Mapelli, Marchesoni, Marion, Márka, Márka, Markakis, Markosyan, Markowitz,
  Maros, Marquina, Marsat, Martelli, Martin, Martin, Martinez, Martynov, Masalehdan, Mason, Massera, Masserot, Massinger, Masso-Reid, Mastrogiovanni, Matas, Matichard, Mavalvala, Maynard, McCann, McCarthy, McClelland, McCormick, McCuller, McGuire, McIsaac, McIver, McManus, McRae, McWilliams, Meacher, Meadors, Mehmet, Mehta, Villa, Melatos, Mendell, Mercer, Mereni, Merfeld, Merilh, Merritt, Merzougui, Meshkov, Messenger, Messick, Metzdorff, Meyers, Meylahn, Mhaske, Miani, Miao, Michaloliakos, Michel, Middleton, Milano, Miller, Millhouse, Mills, Milotti, Milovich-Goff, Minazzoli, Minenkov, Mishkin, Mishra, Mistry, Mitra, Mitrofanov, Mitselmakher, Mittleman, Mo, Mogushi, Mohapatra, Mohite, Molina-Ruiz, Mondin, Montani, Moore, Moraru, Morawski, Moreno, Morisaki, Mours, Mow-Lowry, Mozzon, Muciaccia, Mukherjee, Mukherjee, Mukherjee, Mukherjee, Mukund, Mullavey, Munch, Muñiz, Murray, Nagar, Nardecchia, Naticchioni, Nayak, Neil, Neilson, Nelemans, Nelson, Nery, Neunzert, Ng, Ng, Nguyen, Nguyen, Nichols, Nichols,
  Nissanke, Nocera, Noh, North, Nothard, Nuttall, Oberling, O’Brien, Oganesyan, Ogin, Oh, Oh, Ohme, Ohta, Okada, Oliver, Olivetto, Oppermann, Oram, O’Reilly, Ormiston, Ortega, O’Shaughnessy, Ossokine, Osthelder, Ottaway, Overmier, Owen, Pace, Pagano, Page, Pagliaroli, Pai, Pai, Palamos, Palashov, Palomba, Pan, Panda, Pang, Pankow, Pannarale, Pant, Paoletti, Paoli, Parida, Parker, Pascucci, Pasqualetti, Passaquieti, Passuello, Patricelli, Payne, Pearlstone, Pechsiri, Pedersen, Pedraza, Pele, Penn, Perego, Perez, Périgois, Perreca, Perriès, Petermann, Pfeiffer, Phelps, Phukon, Piccinni, Pichot, Piendibene, Piergiovanni, Pierro, Pillant, Pinard, Pinto, Piotrzkowski, Pirello, Pitkin, Plastino, Poggiani, Pong, Ponrathnam, Popolizio, Porter, Powell, Prajapati, Prasai, Prasanna, Pratten, Prestegard, Principe, Prodi, Prokhorov, Punturo, Puppo, Pürrer, Qi, Quetschke, Quinonez, Raab, Raaijmakers, Radkins, Radulesco, Raffai, Rafferty, Raja, Rajan, Rajbhandari, Rakhmanov, Ramirez, Ramos-Buades, Rana, Rao,
  Rapagnani, Raymond, Razzano, Read, Regimbau, Rei, Reid, Reitze, Rettegno, Ricci, Richardson, Richardson, Ricker, Riemenschneider, Riles, Rizzo, Robertson, Robinet, Rocchi, Rodriguez-Soto, Rolland, Rollins, Roma, Romanelli, Romano, Romel, Romero-Shaw, Romie, Rose, Rose, Rose, Rosińska, Rosofsky, Ross, Rowan, Rowlinson, Roy, Roy, Roy, Ruggi, Rutins, Ryan, Sachdev, Sadecki, Sakellariadou, Salafia, Salconi, Saleem, Salemi, Samajdar, Sanchez, Sanchez, Sanchis-Gual, Sanders, Santiago, Santos, Sarin, Sassolas, Sathyaprakash, Sauter, Savage, Savant, Sawant, Sayah, Schaetzl, Schale, Scheel, Scheuer, Schmidt, Schnabel, Schofield, Schönbeck, Schreiber, Schulte, Schutz, Schwarm, Schwartz, Scott, Scott, Seidel, Sellers, Sengupta, Sennett, Sentenac, Sequino, Sergeev, Setyawati, Shaddock, Shaffer, Shahriar, Sharma, Sharma, Shawhan, Shen, Shikauchi, Shink, Shoemaker, Shoemaker, Shukla, ShyamSundar, Siellez, Sieniawska, Sigg, Singer, Singh, Singh, Singha, Singhal, Sintes, Sipala, Skliris, Slagmolen, Slaven-Blair, Smetana,
  Smith, Smith, Somala, Son, Soni, Sorazu, Sordini, Sorrentino, Souradeep, Sowell, Spencer, Spera, Srivastava, Srivastava, Staats, Stachie, Standke, Steer, Steinhoff, Steinke, Steinlechner, Steinlechner, Steinmeyer, Stevenson, Stocks, Stops, Stover, Strain, Stratta, Strunk, Sturani, Stuver, Sudhagar, Sudhir, Summerscales, Sun, Sunil, Sur, Suresh, Sutton, Swinkels, Szczepańczyk, Tacca, Tait, Talbot, Tanasijczuk, Tanner, Tao, Tápai, Tapia, San~Martin, Tasson, Taylor, Tenorio, Terkowski, Thirugnanasambandam, Thomas, Thomas, Thompson, Thondapu, Thorne, Thrane, Tinsman, Saravanan, Tiwari, Tiwari, Tiwari, Toland, Tonelli, Tornasi, Torres-Forné, Torrie, e~Melo, Töyrä, Trail, Travasso, Traylor, Tringali, Tripathee, Trovato, Trudeau, Tsang, Tse, Tso, Tsukada, Tsuna, Tsutsui, Turconi, Ubhi, Ueno, Ugolini, Unnikrishnan, Urban, Usman, Utina, Vahlbruch, Vajente, Valdes, Valentini, Bakel, Beuzekom, Brand, Broeck, Vander-Hyde, Schaaf, Heijningen, Veggel, Vardaro, Varma, Vass, Vasúth, Vecchio, Vedovato, Veitch, Veitch,
  Venkateswara, Venugopalan, Verkindt, Veske, Vetrano, Viceré, Viets, Vinciguerra, Vine, Vinet, Vitale, Vivanco, Vo, Vocca, Vorvick, Vyatchanin, Wade, Wade, Wade, Walet, Walker, Wallace, Wallace, Walsh, Wang, Wang, Wang, Ward, Warden, Warner, Was, Watchi, Weaver, Wei, Weinert, Weinstein, Weiss, Wellmann, Wen, Weßels, Westhouse, Wette, Whelan, Whiting, Whittle, Wilken, Williams, Willis, Willke, Winkler, Wipf, Wittel, Woan, Woehler, Wofford, Wong, Wright, Wu, Wysocki, Xiao, Yamamoto, Yang, Yang, Yang, Yap, Yazback, Yeeles, Yu, Yu, Yuen, Zadrożny, Zadrożny, Zanolin, Zelenova, Zendri, Zevin, Zhang, Zhang, Zhang, Zhao, Zhao, Zhou, Zhou, Zhu, Zimmerman, Zucker, \& Zweizig}]{abb20b}
Abbott, R., Abbott, T.~D., Abraham, S., {et~al.} 2020, The Astrophysical Journal Letters, 896, L44, \dodoi{10.3847/2041-8213/ab960f}

\bibitem[{Abbott {et~al.}(2023)Abbott, Abbott, Acernese, Ackley, Adams, Adhikari, Adhikari, Adya, Affeldt, Agarwal, Agathos, Agatsuma, Aggarwal, Aguiar, Aiello, Ain, Ajith, Akcay, Akutsu, Albanesi, Allocca, Altin, Amato, Anand, Anand, Ananyeva, Anderson, Anderson, Ando, Andrade, Andres, Andri\ifmmode~\acute{c}\else \'{c}\fi{}, Angelova, Ansoldi, Antelis, Antier, Appert, Arai, Arai, Arai, Araki, Araya, Araya, Areeda, Ar\`ene, Aritomi, Arnaud, Arogeti, Aronson, Arun, Asada, Asali, Ashton, Aso, Assiduo, Aston, Astone, Aubin, Austin, Babak, Badaracco, Bader, Badger, Bae, Bae, Baer, Bagnasco, Bai, Baiotti, Baird, Bajpai, Ball, Ballardin, Ballmer, Balsamo, Baltus, Banagiri, Bankar, Barayoga, Barbieri, Barish, Barker, Barneo, Barone, Barr, Barsotti, Barsuglia, Barta, Bartlett, Barton, Bartos, Bassiri, Basti, Bawaj, Bayley, Baylor, Bazzan, B\'ecsy, Bedakihale, Bejger, Belahcene, Benedetto, Beniwal, Bennett, Bentley, BenYaala, Bergamin, Berger, Bernuzzi, Berry, Bersanetti, Bertolini, Betzwieser, Beveridge, Bhandare,
  Bhardwaj, Bhattacharjee, Bhaumik, Bilenko, Billingsley, Bini, Birney, Birnholtz, Biscans, Bischi, Biscoveanu, Bisht, Biswas, Bitossi, Bizouard, Blackburn, Blair, Blair, Blair, Bobba, Bode, Boer, Bogaert, Boldrini, Bonavena, Bondu, Bonilla, Bonnand, Booker, Boom, Bork, Boschi, Bose, Bose, Bossilkov, Boudart, Bouffanais, Bozzi, Bradaschia, Brady, Bramley, Branch, Branchesi, Brandt, Brau, Breschi, Briant, Briggs, Brillet, Brinkmann, Brockill, Brooks, Brooks, Brown, Brunett, Bruno, Bruntz, Bryant, Bulik, Bulten, Buonanno, Buscicchio, Buskulic, Buy, Byer, Davies, Cadonati, Cagnoli, Cahillane, Bustillo, Callaghan, Callister, Calloni, Cameron, Camp, Canepa, Canevarolo, Cannavacciuolo, Cannon, Cao, Cao, Capocasa, Capote, Carapella, Carbognani, Carlin, Carney, Carpinelli, Carrillo, Carullo, Carver, Diaz, Casentini, Castaldi, Caudill, Cavagli\`a, Cavalier, Cavalieri, Ceasar, Cella, Cerd\'a-Dur\'an, Cesarini, Chaibi, Chakravarti, Subrahmanya, Champion, Chan, Chan, Chan, Chan, Chan, Chandra, Chanial, Chao,
  Chapman-Bird, Charlton, Chase, Chassande-Mottin, Chatterjee, Chatterjee, Chatterjee, Chaturvedi, Chaty, Chatziioannou, Chen, Chen, Chen, Chen, Chen, Chen, Chen, Chen, Cheng, Cheong, Cheung, Chia, Chiadini, Chiang, Chiarini, Chierici, Chincarini, Chiofalo, Chiummo, Cho, Cho, Choudhary, Choudhary, Christensen, Chu, Chu, Chu, Chua, Chung, Ciani, Ciecielag, Cie\ifmmode~\acute{s}\else \'{s}\fi{}lar, Cifaldi, Ciobanu, Ciolfi, Cipriano, Cirone, Clara, Clark, Clark, Clarke, Clearwater, Clesse, Cleva, Coccia, Codazzo, Cohadon, Cohen, Cohen, Colleoni, Collette, Colombo, Colpi, Compton, Constancio, Conti, Cooper, Corban, Corbitt, Cordero-Carri\'on, Corezzi, Corley, Cornish, Corre, Corsi, Cortese, Costa, Cotesta, Coughlin, Coulon, Countryman, Cousins, Couvares, Coward, Cowart, Coyne, Coyne, Creighton, Creighton, Criswell, Croquette, Crowder, Cudell, Cullen, Cumming, Cummings, Cunningham, Cuoco, Cury\l{}o, Dabadie, Canton, Dall'Osso, D\'alya, Dana, DaneshgaranBajastani, D'Angelo, Danila, Danilishin, D'Antonio, Danzmann,
  Darsow-Fromm, Dasgupta, Datrier, Dattilo, Dave, Davier, Davis, Davis, Daw, de~Alarc\'on, Dean, DeBra, Deenadayalan, Degallaix, De~Laurentis, Del\'eglise, Del~Favero, De~Lillo, De~Lillo, Del~Pozzo, DeMarchi, De~Matteis, D'Emilio, Demos, Dent, Depasse, De~Pietri, De~Rosa, De~Rossi, DeSalvo, De~Simone, Dhurandhar, D\'{\i}az, Diaz-Ortiz, Didio, Dietrich, Di~Fiore, Di~Fronzo, Di~Giorgio, Di~Giovanni, Di~Giovanni, Di~Girolamo, Di~Lieto, Ding, Di~Pace, Di~Palma, Di~Renzo, Divakarla, Dmitriev, Doctor, D'Onofrio, Donovan, Dooley, Doravari, Dorrington, Drago, Driggers, Drori, Ducoin, Dupej, Durante, D'Urso, Duverne, Dwyer, Eassa, Easter, Ebersold, Eckhardt, Eddolls, Edelman, Edo, Edy, Effler, Eguchi, Eichholz, Eikenberry, Eisenmann, Eisenstein, Ejlli, Engelby, Enomoto, Errico, Essick, Estell\'es, Estevez, Etienne, Etzel, Evans, Evans, Ewing, Fafone, Fair, Fairhurst, Farah, Farinon, Farr, Farr, Farrow, Fauchon-Jones, Favaro, Favata, Fays, Fazio, Feicht, Fejer, Fenyvesi, Ferguson, Fernandez-Galiana, Ferrante, Ferreira,
  Fidecaro, Figura, Fiori, Fishbach, Fisher, Fittipaldi, Fiumara, Flaminio, Floden, Fong, Font, Fornal, Forsyth, Franke, Frasca, Frasconi, Frederick, Freed, Frei, Freise, Frey, Fritschel, Frolov, Fronz\'e, Fujii, Fujikawa, Fukunaga, Fukushima, Fulda, Fyffe, Gabbard, Gabella, Gadre, Gair, Gais, Galaudage, Gamba, Ganapathy, Ganguly, Gao, Gaonkar, Garaventa, Garc\'{\i}a, Garc\'{\i}a-N\'u\~nez, Garc\'{\i}a-Quir\'os, Garufi, Gateley, Gaudio, Gayathri, Ge, Gemme, Gennai, George, George, Gerberding, Gergely, Gewecke, Ghonge, Ghosh, Ghosh, Ghosh, Ghosh, Giacomazzo, Giacoppo, Giaime, Giardina, Gibson, Gier, Giesler, Giri, Gissi, Glanzer, Gleckl, Godwin, Goetz, Goetz, Gohlke, Golomb, Goncharov, Gonz\'alez, Gopakumar, Gosselin, Gouaty, Gould, Grace, Grado, Granata, Granata, Grant, Gras, Grassia, Gray, Gray, Greco, Green, Green, Gretarsson, Gretarsson, Griffith, Griffiths, Griggs, Grignani, Grimaldi, Grimm, Grote, Grunewald, Gruning, Guerra, Guidi, Guimaraes, Guix\'e, Gulati, Guo, Guo, Gupta, Gupta, Gupta, Gustafson,
  Gustafson, Guzman, Ha, Haegel, Hagiwara, Haino, Halim, Hall, Hamilton, Hammond, Han, Haney, Hanks, Hanna, Hannam, Hannuksela, Hansen, Hansen, Hanson, Harder, Hardwick, Haris, Harms, Harry, Harry, Hartwig, Hasegawa, Haskell, Hasskew, Haster, Hattori, Haughian, Hayakawa, Hayama, Hayes, Healy, Heidmann, Heidt, Heintze, Heinze, Heinzel, Heitmann, Hellman, Hello, Helmling-Cornell, Hemming, Hendry, Heng, Hennes, Hennig, Hennig, Hernandez, Hernandez~Vivanco, Heurs, Hild, Hill, Himemoto, Hines, Hiranuma, Hirata, Hirose, Hochheim, Hofman, Hohmann, Holcomb, Holland, Holley-Bockelmann, Hollows, Holmes, Holt, Holz, Hong, Hopkins, Hough, Hourihane, Howell, Hoy, Hoyland, Hreibi, Hsieh, Hsu, Huang, Huang, Huang, Huang, Huang, Huang, H\"ubner, Huddart, Hughey, Hui, Hui, Husa, Huttner, Huxford, Huynh-Dinh, Ide, Idzkowski, Iess, Ikenoue, Imam, Inayoshi, Ingram, Inoue, Ioka, Isi, Isleif, Ito, Itoh, Iyer, Izumi, JaberianHamedan, Jacqmin, Jadhav, Jadhav, James, Jan, Jani, Janquart, Janssens, Janthalur, Jaranowski, Jariwala,
  Jaume, Jenkins, Jenner, Jeon, Jeunon, Jia, Jin, Johns, Johnson-McDaniel, Jones, Jones, Jones, Jones, Jones, Jonker, Ju, Jung, Jung, Junker, Juste, Kaihotsu, Kajita, Kakizaki, Kalaghatgi, Kalogera, Kamai, Kamiizumi, Kanda, Kandhasamy, Kang, Kanner, Kao, Kapadia, Kapasi, Karat, Karathanasis, Karki, Kashyap, Kasprzack, Kastaun, Katsanevas, Katsavounidis, Katzman, Kaur, Kawabe, Kawaguchi, Kawai, Kawasaki, K\'ef\'elian, Keitel, Key, Khadka, Khalili, Khan, Khazanov, Khetan, Khursheed, Kijbunchoo, Kim, Kim, Kim, Kim, Kim, Kim, Kimball, Kimura, Kinley-Hanlon, Kirchhoff, Kissel, Kita, Kitazawa, Kleybolte, Klimenko, Knee, Knowles, Knyazev, Koch, Koekoek, Kojima, Kokeyama, Koley, Kolitsidou, Kolstein, Komori, Kondrashov, Kong, Kontos, Koper, Korobko, Kotake, Kovalam, Kozak, Kozakai, Kozu, Kringel, Krishnendu, Kr\'olak, Kuehn, Kuei, Kuijer, Kulkarni, Kumar, Kumar, Kumar, Kumar, Kume, Kuns, Kuo, Kuo, Kuromiya, Kuroyanagi, Kusayanagi, Kuwahara, Kwak, Lagabbe, Laghi, Lalande, Lam, Lamberts, Landry, Lane, Lang, Lange,
  Lantz, La~Rosa, Lartaux-Vollard, Lasky, Laxen, Lazzarini, Lazzaro, Leaci, Leavey, Lecoeuche, Lee, Lee, Lee, Lee, Lee, Lee, Lehmann, Lema\^{\i}tre, Leonardi, Leroy, Letendre, Levesque, Levin, Leviton, Leyde, Li, Li, Li, Li, Li, Li, Lin, Lin, Lin, Lin, Lin, Linde, Linker, Linley, Littenberg, Liu, Liu, Liu, Liu, Llamas, Llorens-Monteagudo, Lo, Lockwood, Loh, London, Longo, Lopez, Portilla, Lorenzini, Loriette, Lormand, Losurdo, Lott, Lough, Lousto, Lovelace, Lucaccioni, L\"uck, Lumaca, Lundgren, Luo, Lynam, Macas, MacInnis, Macleod, MacMillan, Macquet, Hernandez, Magazz\`u, Magee, Maggiore, Magnozzi, Mahesh, Majorana, Makarem, Maksimovic, Maliakal, Malik, Man, Mandic, Mangano, Mango, Mansell, Manske, Mantovani, Mapelli, Marchesoni, Marchio, Marion, Mark, M\'arka, M\'arka, Markakis, Markosyan, Markowitz, Maros, Marquina, Marsat, Martelli, Martin, Martin, Martinez, Martinez, Martinez, Martinovic, Martynov, Marx, Masalehdan, Mason, Massera, Masserot, Massinger, Masso-Reid, Mastrogiovanni, Matas, Mateu-Lucena,
  Matichard, Matiushechkina, Mavalvala, McCann, McCarthy, McClelland, McClincy, McCormick, McCuller, McGhee, McGuire, McIsaac, McIver, McRae, McWilliams, Meacher, Mehmet, Mehta, Meijer, Melatos, Melchor, Mendell, Menendez-Vazquez, Menoni, Mercer, Mereni, Merfeld, Merilh, Merritt, Merzougui, Meshkov, Messenger, Messick, Meyers, Meylahn, Mhaske, Miani, Miao, Michaloliakos, Michel, Michimura, Middleton, Milano, Miller, Miller, Miller, Millhouse, Mills, Milotti, Minazzoli, Minenkov, Mio, Mir, Miravet-Ten\'es, Mishra, Mishra, Mistry, Mitra, Mitrofanov, Mitselmakher, Mittleman, Miyakawa, Miyamoto, Miyazaki, Miyo, Miyoki, Mo, Modafferi, Moguel, Mogushi, Mohapatra, Mohite, Molina, Molina-Ruiz, Mondin, Montani, Moore, Moraru, Morawski, More, Moreno, Moreno, Mori, Morisaki, Moriwaki, Morr\'as, Mours, Mow-Lowry, Mozzon, Muciaccia, Mukherjee, Mukherjee, Mukherjee, Mukherjee, Mukherjee, Mukund, Mullavey, Munch, Mu\~niz, Murray, Musenich, Muusse, Nadji, Nagano, Nagano, Nagar, Nakamura, Nakano, Nakano, Nakashima, Nakayama,
  Napolano, Nardecchia, Narikawa, Naticchioni, Nayak, Nayak, Negishi, Neil, Neilson, Nelemans, Nelson, Nery, Neubauer, Neunzert, Ng, Ng, Nguyen, Nguyen, Nguyen, Quynh, Ni, Nichols, Nishizawa, Nissanke, Nitoglia, Nocera, Norman, North, Nozaki, Siles, Nuttall, Oberling, O'Brien, Obuchi, O'Dell, Oelker, Ogaki, Oganesyan, Oh, Oh, Oh, Ohashi, Ohishi, Ohkawa, Ohme, Ohta, Okada, Okutani, Okutomi, Olivetto, Oohara, Ooi, Oram, O'Reilly, Ormiston, Ormsby, Ortega, O'Shaughnessy, O'Shea, Oshino, Ossokine, Osthelder, Otabe, Ottaway, Overmier, Pace, Pagano, Page, Pagliaroli, Pai, Pai, Palamos, Palashov, Palomba, Pan, Pan, Panda, Pang, Pang, Pankow, Pannarale, Pant, Panther, Paoletti, Paoli, Paolone, Parisi, Park, Park, Parker, Pascucci, Pasqualetti, Passaquieti, Passuello, Patel, Pathak, Patricelli, Patron, Paul, Payne, Pedraza, Pegoraro, Pele, Arellano, Penn, Perego, Pereira, Pereira, Perez, P\'erigois, Perkins, Perreca, Perri\`es, Petermann, Petterson, Pfeiffer, Pham, Phukon, Piccinni, Pichot, Piendibene, Piergiovanni,
  Pierini, Pierro, Pillant, Pillas, Pilo, Pinard, Pinto, Pinto, Piotrzkowski, Piotrzkowski, Pirello, Pitkin, Placidi, Planas, Plastino, Pluchar, Poggiani, Polini, Pong, Ponrathnam, Popolizio, Porter, Poulton, Powell, Pracchia, Pradier, Prajapati, Prasai, Prasanna, Pratten, Principe, Prodi, Prokhorov, Prosposito, Prudenzi, Puecher, Punturo, Puosi, Puppo, P\"urrer, Qi, Quetschke, Quitzow-James, Qutob, Raab, Raaijmakers, Radkins, Radulesco, Raffai, Rail, Raja, Rajan, Ramirez, Ramirez, Ramos-Buades, Rana, Rapagnani, Rapol, Ray, Raymond, Raza, Razzano, Read, Rees, Regimbau, Rei, Reid, Reid, Reitze, Relton, Renzini, Rettegno, Reza, Rezac, Ricci, Richards, Richardson, Richardson, Riemenschneider, Riles, Rinaldi, Rink, Rizzo, Robertson, Robie, Robinet, Rocchi, Rodriguez, Rolland, Rollins, Romanelli, Romano, Romel, Romero-Rodr\'{\i}guez, Romero-Shaw, Romie, Ronchini, Rosa, Rose, Rosi\ifmmode~\acute{n}\else \'{n}\fi{}ska, Ross, Rowan, Rowlinson, Roy, Roy, Roy, Rozza, Ruggi, Ruiz-Rocha, Ryan, Sachdev, Sadecki, Sadiq,
  Sago, Saito, Saito, Sakai, Sakai, Sakellariadou, Sakuno, Salafia, Salconi, Saleem, Salemi, Samajdar, Sanchez, Sanchez, Sanchez, Sanchis-Gual, Sanders, Sanuy, Saravanan, Sarin, Sassolas, Satari, Sathyaprakash, Sato, Sato, Sauter, Savage, Sawada, Sawant, Sawant, Sayah, Schaetzl, Scheel, Scheuer, Schiworski, Schmidt, Schmidt, Schnabel, Schneewind, Schofield, Sch\"onbeck, Schulte, Schutz, Schwartz, Scott, Scott, Seglar-Arroyo, Sekiguchi, Sekiguchi, Sellers, Sengupta, Sentenac, Seo, Sequino, Sergeev, Setyawati, Shaffer, Shahriar, Shams, Shao, Sharma, Sharma, Shawhan, Shcheblanov, Shibagaki, Shikauchi, Shimizu, Shimoda, Shimode, Shinkai, Shishido, Shoda, Shoemaker, Shoemaker, ShyamSundar, Sieniawska, Sigg, Singer, Singh, Singh, Singha, Sintes, Sipala, Skliris, Slagmolen, Slaven-Blair, Smetana, Smith, Smith, Soldateschi, Somala, Somiya, Son, Soni, Soni, Sordini, Sorrentino, Sorrentino, Sotani, Soulard, Souradeep, Sowell, Spagnuolo, Spencer, Spera, Srinivasan, Srivastava, Srivastava, Staats, Stachie, Steer,
  Steinhoff, Steinlechner, Steinlechner, Stevenson, Stops, Stover, Strain, Strang, Stratta, Strunk, Sturani, Stuver, Sudhagar, Sudhir, Sugimoto, Suh, Sullivan, Sullivan, Summerscales, Sun, Sun, Sunil, Sur, Suresh, Sutton, Suzuki, Suzuki, Swinkels, Szczepa\ifmmode~\acute{n}\else \'{n}\fi{}czyk, Szewczyk, Tacca, Tagoshi, Tait, Takahashi, Takahashi, Takamori, Takano, Takeda, Takeda, Talbot, Talbot, Tanaka, Tanaka, Tanaka, Tanaka, Tanaka, Tanasijczuk, Tanioka, Tanner, Tao, Tao, Mart\'{\i}n, Taranto, Tasson, Telada, Tenorio, Terhune, Terkowski, Thirugnanasambandam, Thomas, Thomas, Thomas, Thompson, Thondapu, Thorne, Thrane, Tiwari, Tiwari, Tiwari, Toivonen, Toland, Tolley, Tomaru, Tomigami, Tomura, Tonelli, Torres-Forn\'e, Torrie, e~Melo, T\"oyr\"a, Trapananti, Travasso, Traylor, Trevor, Tringali, Tripathee, Troiano, Trovato, Trozzo, Trudeau, Tsai, Tsai, Tsang, Tsang, Tsao, Tse, Tso, Tsubono, Tsuchida, Tsukada, Tsuna, Tsutsui, Tsuzuki, Turbang, Turconi, Tuyenbayev, Ubhi, Uchikata, Uchiyama, Udall, Ueda, Uehara,
  Ueno, Ueshima, Unnikrishnan, Uraguchi, Urban, Ushiba, Utina, Vahlbruch, Vajente, Vajpeyi, Valdes, Valentini, Valsan, van Bakel, van Beuzekom, van~den Brand, Van Den~Broeck, Vander-Hyde, van~der Schaaf, van Heijningen, Vanosky, van Putten, van Remortel, Vardaro, Vargas, Varma, Vas\'uth, Vecchio, Vedovato, Veitch, Veitch, Venneberg, Venugopalan, Verkindt, Verma, Verma, Veske, Vetrano, Vicer\'e, Vidyant, Viets, Vijaykumar, Villa-Ortega, Vinet, Virtuoso, Vitale, Vo, Vocca, von Reis, von Wrangel, Vorvick, Vyatchanin, Wade, Wade, Wagner, Walet, Walker, Wallace, Wallace, Walsh, Wang, Wang, Wang, Ward, Warner, Was, Washimi, Washington, Watchi, Weaver, Webster, Weinert, Weinstein, Weiss, Weller, Weller, Wellmann, Wen, We\ss{}els, Wette, Whelan, White, Whiting, Whittle, Wilken, Williams, Williams, Williams, Williamson, Willis, Willke, Wilson, Winkler, Wipf, Wlodarczyk, Woan, Woehler, Wofford, Wong, Wu, Wu, Wu, Wu, Wysocki, Xiao, Xu, Yamada, Yamamoto, Yamamoto, Yamamoto, Yamamoto, Yamashita, Yamazaki, Yang, Yang,
  Yang, Yang, Yang, Yap, Yeeles, Yelikar, Ying, Yokogawa, Yokoyama, Yokozawa, Yoo, Yoshioka, Yu, Yu, Yuzurihara, Zadro\ifmmode~\dot{z}\else \.{z}\fi{}ny, Zanolin, Zeidler, Zelenova, Zendri, Zevin, Zhan, Zhang, Zhang, Zhang, Zhang, Zhang, Zhao, Zhao, Zhao, Zhao, Zheng, Zhou, Zhou, Zhu, Zhu, Zimmerman, Zlochower, Zucker, \& Zweizig}]{abb23}
Abbott, R., Abbott, T.~D., Acernese, F., {et~al.} 2023, Phys. Rev. X, 13, 041039, \dodoi{10.1103/PhysRevX.13.041039}

\bibitem[{Aghaei~Abchouyeh {et~al.}(2023)Aghaei~Abchouyeh, van Putten, \& Amati}]{abc23}
Aghaei~Abchouyeh, M., van Putten, M. H. P.~M., \& Amati, L. 2023, The Astrophysical Journal, 952, 157, \dodoi{10.3847/1538-4357/acd114}

\bibitem[{Alp {et~al.}(2018)Alp, Larsson, Fransson, Indebetouw, Jerkstrand, Ahola, Burrows, Challis, Cigan, Cikota, Kirshner, van Loon, Mattila, Ng, Park, Spyromilio, Woosley, Baes, Bouchet, Chevalier, Frank, Gaensler, Gomez, Janka, Leibundgut, Lundqvist, Marcaide, Matsuura, Sollerman, Sonneborn, Staveley-Smith, Zanardo, Gabler, Taddia, \& Wheeler}]{alp18}
Alp, D., Larsson, J., Fransson, C., {et~al.} 2018, The Astrophysical Journal, 864, 174, \dodoi{10.3847/1538-4357/aad739}

\bibitem[{{Amaro-Seoane} {et~al.}(2023){Amaro-Seoane}, {Andrews}, {Arca Sedda}, {Askar}, {Baghi}, {Balasov}, {Bartos}, {Bavera}, {Bellovary}, {Berry}, {Berti}, {Bianchi}, {Blecha}, {Blondin}, {Bogdanovi{\'c}}, {Boissier}, {Bonetti}, {Bonoli}, {Bortolas}, {Breivik}, {Capelo}, {Caramete}, {Cattorini}, {Charisi}, {Chaty}, {Chen}, {Chru{\'s}li{\'n}ska}, {Chua}, {Church}, {Colpi}, {D'Orazio}, {Danielski}, {Davies}, {Dayal}, {De Rosa}, {Derdzinski}, {Destounis}, {Dotti}, {Dutan}, {Dvorkin}, {Fabj}, {Foglizzo}, {Ford}, {Fouvry}, {Franchini}, {Fragos}, {Fryer}, {Gaspari}, {Gerosa}, {Graziani}, {Groot}, {Habouzit}, {Haggard}, {Haiman}, {Han}, {Istrate}, {Johansson}, {Khan}, {Kimpson}, {Kokkotas}, {Kong}, {Korol}, {Kremer}, {Kupfer}, {Lamberts}, {Larson}, {Lau}, {Liu}, {Lloyd-Ronning}, {Lodato}, {Lupi}, {Ma}, {Maccarone}, {Mandel}, {Mangiagli}, {Mapelli}, {Mathis}, {Mayer}, {McGee}, {McKernan}, {Miller}, {Mota}, {Mumpower}, {Nasim}, {Nelemans}, {Noble}, {Pacucci}, {Panessa}, {Paschalidis}, {Pfister}, {Porquet},
  {Quenby}, {Ricarte}, {R{\"o}pke}, {Regan}, {Rosswog}, {Ruiter}, {Ruiz}, {Runnoe}, {Schneider}, {Schnittman}, {Secunda}, {Sesana}, {Seto}, {Shao}, {Shapiro}, {Sopuerta}, {Stone}, {Suvorov}, {Tamanini}, {Tamfal}, {Tauris}, {Temmink}, {Tomsick}, {Toonen}, {Torres-Orjuela}, {Toscani}, {Tsokaros}, {Unal}, {V{\'a}zquez-Aceves}, {Valiante}, {van Putten}, {van Roestel}, {Vignali}, {Volonteri}, {Wu}, {Younsi}, {Yu}, {Zane}, {Zwick}, {Antonini}, {Baibhav}, {Barausse}, {Bonilla Rivera}, {Branchesi}, {Branduardi-Raymont}, {Burdge}, {Chakraborty}, {Cuadra}, {Dage}, {Davis}, {de Mink}, {Decarli}, {Doneva}, {Escoffier}, {Gandhi}, {Haardt}, {Lousto}, {Nissanke}, {Nordhaus}, {O'Shaughnessy}, {Portegies Zwart}, {Pound}, {Schussler}, {Sergijenko}, {Spallicci}, {Vernieri}, \& {Vigna-G{\'o}mez}}]{ama23}
{Amaro-Seoane}, P., {Andrews}, J., {Arca Sedda}, M., {et~al.} 2023, Living Reviews in Relativity, 26, 2, \dodoi{10.1007/s41114-022-00041-y}

\bibitem[{Amati(2006)}]{ama06}
Amati, L. 2006, Monthly Notices of the Royal Astronomical Society, 372, 233, \dodoi{10.1111/j.1365-2966.2006.10840.x}

\bibitem[{Amati(2021)}]{ama21}
---. 2021, Nature Astronomy, 5, 877, \dodoi{10.1038/s41550-021-01401-4}

\bibitem[{Amati {et~al.}(2009)Amati, Frontera, \& Guidorzi}]{ama09}
Amati, L., Frontera, F., \& Guidorzi, C. 2009, Astronomy \& Astrophysics, 508, 173, \dodoi{10.1051/0004-6361/200912788}

\bibitem[{Amati {et~al.}(2002)Amati, Frontera, Tavani, in~’t Zand, Antonelli, Costa, Feroci, Guidorzi, Heise, Masetti, Montanari, Nicastro, Palazzi, Pian, Piro, \& Soffitta}]{ama02}
Amati, L., Frontera, F., Tavani, M., {et~al.} 2002, Astronomy \& Astrophysics, 390, 81, \dodoi{10.1051/0004-6361:20020722}

\bibitem[{Amati {et~al.}(2018)Amati, O’Brien, Götz, Bozzo, Tenzer, Frontera, Ghirlanda, Labanti, Osborne, Stratta, Tanvir, Willingale, Attina, Campana, Castro-Tirado, Contini, Fuschino, Gomboc, Hudec, Orleanski, Renotte, Rodic, Bagoly, Blain, Callanan, Covino, Ferrara, Le~Floch, Marisaldi, Mereghetti, Rosati, Vacchi, D’Avanzo, Giommi, Piranomonte, Piro, Reglero, Rossi, Santangelo, Salvaterra, Tagliaferri, Vergani, Vinciguerra, Briggs, Campolongo, Ciolfi, Connaughton, Cordier, Morelli, Orlandini, Adami, Argan, Atteia, Auricchio, Balazs, Baldazzi, Basa, Basak, Bellutti, Bernardini, Bertuccio, Braga, Branchesi, Brandt, Brocato, Budtz-Jorgensen, Bulgarelli, Burderi, Camp, Capozziello, Caruana, Casella, Cenko, Chardonnet, Ciardi, Colafrancesco, Dainotti, D’Elia, De~Martino, De~Pasquale, Del~Monte, Della~Valle, Drago, Evangelista, Feroci, Finelli, Fiorini, Fynbo, Gal-Yam, Gendre, Ghisellini, Grado, Guidorzi, Hafizi, Hanlon, Hjorth, Izzo, Kiss, Kumar, Kuvvetli, Lavagna, Li, Longo, Lyutikov, Maio, Maiorano,
  Malcovati, Malesani, Margutti, Martin-Carrillo, Masetti, McBreen, Mignani, Morgante, Mundell, Nargaard-Nielsen, Nicastro, Palazzi, Paltani, Panessa, Pareschi, Pe’er, Penacchioni, Pian, Piedipalumbo, Piran, Rauw, Razzano, Read, Rezzolla, Romano, Ruffini, Savaglio, Sguera, Schady, Skidmore, Song, Stanway, Starling, Topinka, Troja, van Putten, Vanzella, Vercellone, Wilson-Hodge, Yonetoku, Zampa, Zampa, Zhang, Zhang, Zhang, Zhang, Antonelli, Bianco, Boci, Boer, Botticella, Boulade, Butler, Campana, Capitanio, Celotti, Chen, Colpi, Comastri, Cuby, Dadina, De~Luca, Dong, Ettori, Gandhi, Geza, Greiner, Guiriec, Harms, Hernanz, Hornstrup, Hutchinson, Israel, Jonker, Kaneko, Kawai, Wiersema, Korpela, Lebrun, Lu, MacFadyen, Malaguti, Maraschi, Melandri, Modjaz, Morris, Omodei, Paizis, Páta, Petrosian, Rachevski, Rhoads, Ryde, Sabau-Graziati, Shigehiro, Sims, Soomin, Szécsi, Urata, Uslenghi, Valenziano, Vianello, Vojtech, Watson, \& Zicha}]{ama18}
Amati, L., O’Brien, P., Götz, D., {et~al.} 2018, Advances in Space Research, 62, 191, \dodoi{10.1016/j.asr.2018.03.010}

\bibitem[{Arcavi {et~al.}(2011)Arcavi, Gal-Yam, Yaron, Sternberg, Rabinak, Waxman, Kasliwal, Quimby, Ofek, Horesh, Kulkarni, Filippenko, Silverman, Cenko, Li, Bloom, Sullivan, Nugent, Poznanski, Gorbikov, Fulton, Howell, Bersier, Riou, Lamotte-Bailey, Griga, Cohen, Hachinger, Polishook, Xu, Ben-Ami, Manulis, Walker, Maguire, Pan, Matheson, Mazzali, Pian, Fox, Gehrels, Law, James, Marchant, Smith, Mottram, Barnsley, Kandrashoff, \& Clubb}]{arc11}
Arcavi, I., Gal-Yam, A., Yaron, O., {et~al.} 2011, The Astrophysical Journal, 742, L18, \dodoi{10.1088/2041-8205/742/2/l18}

\bibitem[{Armitage \& Natarajan(1999)}]{arm99}
Armitage, P.~J., \& Natarajan, P. 1999, The Astrophysical Journal, 525, 909, \dodoi{10.1086/307955}

\bibitem[{Arnett {et~al.}(1989)Arnett, Bahcall, Kirshner, \& Woosley}]{arn89}
Arnett, W.~D., Bahcall, J.~N., Kirshner, R.~P., \& Woosley, S.~E. 1989, Annual Review of Astronomy and Astrophysics, 27, 629, \dodoi{10.1146/annurev.aa.27.090189.003213}

\bibitem[{Ashoka {et~al.}(1987)Ashoka, Anupama, Prabhu, Giridhar, Ghosh, Jain, Pati, \& Rao}]{ash87}
Ashoka, B.~N., Anupama, G.~C., Prabhu, T.~P., {et~al.} 1987, Journal of Astrophysics and Astronomy, 8, 195, \dodoi{10.1007/bf02714317}

\bibitem[{{Baiotti} {et~al.}(2008){Baiotti}, {Giacomazzo}, \& {Rezzolla}}]{bai08}
{Baiotti}, L., {Giacomazzo}, B., \& {Rezzolla}, L. 2008, \prd, 78, 084033, \dodoi{10.1103/PhysRevD.78.084033}

\bibitem[{Banerjee {et~al.}(2019)Banerjee, Chakraborty, \& Bhattacharyya}]{ban19}
Banerjee, S., Chakraborty, C., \& Bhattacharyya, S. 2019, Monthly Notices of the Royal Astronomical Society, 487, 3488, \dodoi{10.1093/mnras/stz1518}

\bibitem[{Bianco {et~al.}(2024)Bianco, Mirtorabi, Moradi, Rastegarnia, Rueda, Ruffini, Wang, Della~Valle, Li, \& Zhang}]{bia24}
Bianco, C.~L., Mirtorabi, M.~T., Moradi, R., {et~al.} 2024, The Astrophysical Journal, 966, 219, \dodoi{10.3847/1538-4357/ad2fa9}

\bibitem[{{Biretta}(1996)}]{bir95}
{Biretta}, J.~A. 1996, in Astronomical Society of the Pacific Conference Series, Vol. 100, Energy Transport in Radio Galaxies and Quasars, ed. P.~E. {Hardee}, A.~H. {Bridle}, \& J.~A. {Zensus}, 187

\bibitem[{Bisnovatyi-Kogan(1970)}]{bis70}
Bisnovatyi-Kogan, G.~S. 1970, \azh, 47, 813.
\newblock \url{https://ui.adsabs.harvard.edu/abs/1970AZh....47..813B}

\bibitem[{Bisnovatyi-Kogan(1971)}]{bis71}
---. 1971, \sovast, 14, 652.
\newblock \url{https://ui.adsabs.harvard.edu/abs/1971SvA....14..652B}

\bibitem[{{Bromberg} {et~al.}(2006){Bromberg}, {Levinson}, \& {van Putten}}]{bro06}
{Bromberg}, O., {Levinson}, A., \& {van Putten}, M. 2006, \na, 11, 619, \dodoi{10.1016/j.newast.2006.03.005}

\bibitem[{{Cappellaro} {et~al.}(1999){Cappellaro}, {Evans}, \& {Turatto}}]{cap99}
{Cappellaro}, E., {Evans}, R., \& {Turatto}, M. 1999, \aap, 351, 459, \dodoi{10.48550/arXiv.astro-ph/9904225}

\bibitem[{{Cappellaro} {et~al.}(2015){Cappellaro}, {Botticella}, {Pignata}, {Grado}, {Greggio}, {Limatola}, {Vaccari}, {Baruffolo}, {Benetti}, {Bufano}, {Capaccioli}, {Cascone}, {Covone}, {De Cicco}, {Falocco}, {Della Valle}, {Jarvis}, {Marchetti}, {Napolitano}, {Paolillo}, {Pastorello}, {Radovich}, {Schipani}, {Spiro}, {Tomasella}, \& {Turatto}}]{cap15}
{Cappellaro}, E., {Botticella}, M.~T., {Pignata}, G., {et~al.} 2015, \aap, 584, A62, \dodoi{10.1051/0004-6361/201526712}

\bibitem[{{Corsi} {et~al.}(2016){Corsi}, {Gal-Yam}, {Kulkarni}, {Frail}, {Mazzali}, {Cenko}, {Kasliwal}, {Cao}, {Horesh}, {Palliyaguru}, {Perley}, {Laher}, {Taddia}, {Leloudas}, {Maguire}, {Nugent}, {Sollerman}, \& {Sullivan}}]{cor16}
{Corsi}, A., {Gal-Yam}, A., {Kulkarni}, S.~R., {et~al.} 2016, \apj, 830, 42, \dodoi{10.3847/0004-637X/830/1/42}

\bibitem[{{Corsi} {et~al.}(2023){Corsi}, {Ho}, {Cenko}, {Kulkarni}, {Anand}, {Yang}, {Sollerman}, {Srinivasaragavan}, {Omand}, {Balasubramanian}, {Frail}, {Fremling}, {Perley}, {Yao}, {Dahiwale}, {De}, {Dugas}, {Hankins}, {Jencson}, {Kasliwal}, {Tzanidakis}, {Bellm}, {Laher}, {Masci}, {Purdum}, \& {Regnault}}]{cor23}
{Corsi}, A., {Ho}, A. Y.~Q., {Cenko}, S.~B., {et~al.} 2023, \apj, 953, 179, \dodoi{10.3847/1538-4357/acd3f2}

\bibitem[{{Cui} {et~al.}(2023){Cui}, {Hada}, {Kawashima}, {Kino}, {Lin}, {Mizuno}, {Ro}, {Honma}, {Yi}, {Yu}, {Park}, {Jiang}, {Shen}, {Kravchenko}, {Algaba}, {Cheng}, {Cho}, {Giovannini}, {Giroletti}, {Jung}, {Lu}, {Niinuma}, {Oh}, {Ohsuga}, {Sawada-Satoh}, {Sohn}, {Takahashi}, {Takamura}, {Tazaki}, {Trippe}, {Wajima}, {Akiyama}, {An}, {Asada}, {Buttaccio}, {Byun}, {Cui}, {Hagiwara}, {Hirota}, {Hodgson}, {Kawaguchi}, {Kim}, {Lee}, {Lee}, {Lee}, {Maccaferri}, {Melis}, {Melnikov}, {Migoni}, {Oh}, {Sugiyama}, {Wang}, {Zhang}, {Chen}, {Hwang}, {Jung}, {Kim}, {Kim}, {Kobayashi}, {Li}, {Li}, {Li}, {Liu}, {Liu}, {Liu}, {Oh}, {Oyama}, {Roh}, {Wang}, {Wang}, {Wang}, {Xia}, {Yan}, {Yeom}, {Yonekura}, {Yuan}, {Zhang}, {Zhao}, \& {Zhong}}]{cui23}
{Cui}, Y., {Hada}, K., {Kawashima}, T., {et~al.} 2023, \nat, 621, 711, \dodoi{10.1038/s41586-023-06479-6}

\bibitem[{Cutler \& Thorne(2002)}]{cut02}
Cutler, C., \& Thorne, K.~S. 2002, An Overview of Gravitational-Wave Sources,  arXiv, \dodoi{10.48550/ARXIV.GR-QC/0204090}

\bibitem[{{De Colle} {et~al.}(2022){De Colle}, {Kumar}, \& {Hoeflich}}]{dec22}
{De Colle}, F., {Kumar}, P., \& {Hoeflich}, P. 2022, \mnras, 512, 3627, \dodoi{10.1093/mnras/stac742}

\bibitem[{{Della Valle}(2006)}]{del06}
{Della Valle}, M. 2006, in American Institute of Physics Conference Series, Vol. 836, Gamma-Ray Bursts in the Swift Era, ed. S.~S. {Holt}, N.~{Gehrels}, \& J.~A. {Nousek} (AIP), 367--379, \dodoi{10.1063/1.2207923}

\bibitem[{{Della Valle} {et~al.}(2006){Della Valle}, {Chincarini}, {Panagia}, {Tagliaferri}, {Malesani}, {Testa}, {Fugazza}, {Campana}, {Covino}, {Mangano}, {Antonelli}, {D'Avanzo}, {Hurley}, {Mirabel}, {Pellizza}, {Piranomonte}, \& {Stella}}]{del06a}
{Della Valle}, M., {Chincarini}, G., {Panagia}, N., {et~al.} 2006, \nat, 444, 1050, \dodoi{10.1038/nature05374}

\bibitem[{D’Avanzo(2015)}]{dav15}
D’Avanzo, P. 2015, Journal of High Energy Astrophysics, 7, 73, \dodoi{10.1016/j.jheap.2015.07.002}

\bibitem[{{Eichler} {et~al.}(1989){Eichler}, {Livio}, {Piran}, \& {Schramm}}]{eic89}
{Eichler}, D., {Livio}, M., {Piran}, T., \& {Schramm}, D.~N. 1989, \nat, 340, 126, \dodoi{10.1038/340126a0}

\bibitem[{{Fanaroff} \& {Riley}(1974)}]{fan74}
{Fanaroff}, B.~L., \& {Riley}, J.~M. 1974, \mnras, 167, 31P, \dodoi{10.1093/mnras/167.1.31P}

\bibitem[{Flanagan \& Hughes(1998)}]{fla98}
Flanagan, E.~E., \& Hughes, S.~A. 1998, Physical Review D, 57, 4535, \dodoi{10.1103/physrevd.57.4535}

\bibitem[{{Foley} {et~al.}(2006){Foley}, {Watson}, {Gorosabel}, {Fynbo}, {Sollerman}, {McGlynn}, {McBreen}, \& {Hjorth}}]{fol06}
{Foley}, S., {Watson}, D., {Gorosabel}, J., {et~al.} 2006, \aap, 447, 891, \dodoi{10.1051/0004-6361:20054382}

\bibitem[{{Frail} {et~al.}(2001){Frail}, {Kulkarni}, {Sari}, {Djorgovski}, {Bloom}, {Galama}, {Reichart}, {Berger}, {Harrison}, {Price}, {Yost}, {Diercks}, {Goodrich}, \& {Chaffee}}]{fra01}
{Frail}, D.~A., {Kulkarni}, S.~R., {Sari}, R., {et~al.} 2001, \apjl, 562, L55, \dodoi{10.1086/338119}

\bibitem[{Fransson {et~al.}(2024)Fransson, Barlow, Kavanagh, Larsson, Jones, Sargent, Meixner, Bouchet, Temim, Wright, Blommaert, Habel, Hirschauer, Hjorth, Lenkić, Tikkanen, Wesson, Coulais, Fox, Gastaud, Glasse, Jaspers, Krause, Lau, Nayak, Rest, Colina, van Dishoeck, Güdel, Henning, Lagage, Östlin, Ray, \& Vandenbussche}]{fra24}
Fransson, C., Barlow, M.~J., Kavanagh, P.~J., {et~al.} 2024, Science, 383, 898, \dodoi{10.1126/science.adj5796}

\bibitem[{Frontera(2019)}]{fro19}
Frontera, F. 2019, Rendiconti Lincei. Scienze Fisiche e Naturali, 30, 171, \dodoi{10.1007/s12210-019-00766-z}

\bibitem[{Fryer {et~al.}(2014)Fryer, Rueda, \& Ruffini}]{fry14}
Fryer, C.~L., Rueda, J.~A., \& Ruffini, R. 2014, The Astrophysical Journal, 793, L36, \dodoi{10.1088/2041-8205/793/2/l36}

\bibitem[{{Fynbo} {et~al.}(2006){Fynbo}, {Watson}, {Th{\"o}ne}, {Sollerman}, {Bloom}, {Davis}, {Hjorth}, {Jakobsson}, {J{\o}rgensen}, {Graham}, {Fruchter}, {Bersier}, {Kewley}, {Cassan}, {Castro Cer{\'o}n}, {Foley}, {Gorosabel}, {Hinse}, {Horne}, {Jensen}, {Klose}, {Kocevski}, {Marquette}, {Perley}, {Ramirez-Ruiz}, {Stritzinger}, {Vreeswijk}, {Wijers}, {Woller}, {Xu}, \& {Zub}}]{fyn06}
{Fynbo}, J. P.~U., {Watson}, D., {Th{\"o}ne}, C.~C., {et~al.} 2006, \nat, 444, 1047, \dodoi{10.1038/nature05375}

\bibitem[{{Gal-Yam} {et~al.}(2006){Gal-Yam}, {Ofek}, {Poznanski}, {Levinson}, {Waxman}, {Frail}, {Soderberg}, {Nakar}, {Li}, \& {Filippenko}}]{gal06}
{Gal-Yam}, A., {Ofek}, E.~O., {Poznanski}, D., {et~al.} 2006, \apj, 639, 331, \dodoi{10.1086/499157}

\bibitem[{{Galama} {et~al.}(1998){Galama}, {Vreeswijk}, {van Paradijs}, {Kouveliotou}, {Augusteijn}, {B{\"o}hnhardt}, {Brewer}, {Doublier}, {Gonzalez}, {Leibundgut}, {Lidman}, {Hainaut}, {Patat}, {Heise}, {in't Zand}, {Hurley}, {Groot}, {Strom}, {Mazzali}, {Iwamoto}, {Nomoto}, {Umeda}, {Nakamura}, {Young}, {Suzuki}, {Shigeyama}, {Koshut}, {Kippen}, {Robinson}, {de Wildt}, {Wijers}, {Tanvir}, {Greiner}, {Pian}, {Palazzi}, {Frontera}, {Masetti}, {Nicastro}, {Feroci}, {Costa}, {Piro}, {Peterson}, {Tinney}, {Boyle}, {Cannon}, {Stathakis}, {Sadler}, {Begam}, \& {Ianna}}]{gal98}
{Galama}, T.~J., {Vreeswijk}, P.~M., {van Paradijs}, J., {et~al.} 1998, \nat, 395, 670, \dodoi{10.1038/27150}

\bibitem[{Galeotti \& Pizzella(2018)}]{gal18}
Galeotti, P., \& Pizzella, G. 2018, Physics of Atomic Nuclei, 81, 105, \dodoi{10.1134/s1063778818010106}

\bibitem[{{Gehrels} {et~al.}(2006){Gehrels}, {Norris}, {Barthelmy}, {Granot}, {Kaneko}, {Kouveliotou}, {Markwardt}, {M{\'e}sz{\'a}ros}, {Nakar}, {Nousek}, {O'Brien}, {Page}, {Palmer}, {Parsons}, {Roming}, {Sakamoto}, {Sarazin}, {Schady}, {Stamatikos}, \& {Woosley}}]{geh06}
{Gehrels}, N., {Norris}, J.~P., {Barthelmy}, S.~D., {et~al.} 2006, \nat, 444, 1044, \dodoi{10.1038/nature05376}

\bibitem[{Ghirlanda {et~al.}(2016)Ghirlanda, Salafia, Pescalli, Ghisellini, Salvaterra, Chassande–Mottin, Colpi, Nappo, D’Avanzo, Melandri, Bernardini, Branchesi, Campana, Ciolfi, Covino, Götz, Vergani, Zennaro, \& Tagliaferri}]{ghi16}
Ghirlanda, G., Salafia, O.~S., Pescalli, A., {et~al.} 2016, Astronomy \& Astrophysics, 594, A84, \dodoi{10.1051/0004-6361/201628993}

\bibitem[{Gill {et~al.}(2019)Gill, Nathanail, \& Rezzolla}]{gil19}
Gill, R., Nathanail, A., \& Rezzolla, L. 2019, The Astrophysical Journal, 876, 139, \dodoi{10.3847/1538-4357/ab16da}

\bibitem[{{Gilmozzi} {et~al.}(1987){Gilmozzi}, {Cassatella}, {Clavel}, {Fransson}, {Gonzalez}, {Gry}, {Panagia}, {Talavera}, \& {Wamsteker}}]{gil87}
{Gilmozzi}, R., {Cassatella}, A., {Clavel}, J., {et~al.} 1987, \nat, 328, 318, \dodoi{10.1038/328318a0}

\bibitem[{González-Díaz {et~al.}(2024)González-Díaz, Galbany, Kangas, García-Benito, Anderson, Lyman, Varela, Oltra, García, Rojo, López-Sanjuan, Pérez-Torres, Rosales-Ortega, Mattila, Kuncarayakti, James, Habergham, Vílchez, Alcaniz, Angulo, Cenarro, Cristóbal-Hornillos, Dupke, Ederoclite, Hernández-Monteagudo, Marín-Franch, Moles, Sodré, \& Ramió}]{gon24}
González-Díaz, R., Galbany, L., Kangas, T., {et~al.} 2024, Astronomy \& Astrophysics, 684, A104, \dodoi{10.1051/0004-6361/202349029}

\bibitem[{Gossan {et~al.}(2016)Gossan, Sutton, Stuver, Zanolin, Gill, \& Ott}]{gos16}
Gossan, S., Sutton, P., Stuver, A., {et~al.} 2016, Physical Review D, 93, 042002, \dodoi{10.1103/physrevd.93.042002}

\bibitem[{Graham {et~al.}(2019)Graham, Kulkarni, Bellm, Adams, Barbarino, Blagorodnova, Bodewits, Bolin, Brady, Cenko, Chang, Coughlin, De, Eadie, Farnham, Feindt, Franckowiak, Fremling, Gezari, Ghosh, Goldstein, Golkhou, Goobar, Ho, Huppenkothen, Ivezić, Jones, Juric, Kaplan, Kasliwal, Kelley, Kupfer, Lee, Lin, Lunnan, Mahabal, Miller, Ngeow, Nugent, Ofek, Prince, Rauch, Roestel, Schulze, Singer, Sollerman, Taddia, Yan, Ye, Yu, Barlow, Bauer, Beck, Belicki, Biswas, Brinnel, Brooke, Bue, Bulla, Burruss, Connolly, Cromer, Cunningham, Dekany, Delacroix, Desai, Duev, Feeney, Flynn, Frederick, Gal-Yam, Giomi, Groom, Hacopians, Hale, Helou, Henning, Hover, Hillenbrand, Howell, Hung, Imel, Ip, Jackson, Kaspi, Kaye, Kowalski, Kramer, Kuhn, Landry, Laher, Mao, Masci, Monkewitz, Murphy, Nordin, Patterson, Penprase, Porter, Rebbapragada, Reiley, Riddle, Rigault, Rodriguez, Rusholme, Santen, Shupe, Smith, Soumagnac, Stein, Surace, Szkody, Terek, Sistine, Velzen, Vestrand, Walters, Ward, Zhang, \& Zolkower}]{gra19}
Graham, M.~J., Kulkarni, S.~R., Bellm, E.~C., {et~al.} 2019, Publications of the Astronomical Society of the Pacific, 131, 078001, \dodoi{10.1088/1538-3873/ab006c}

\bibitem[{Guetta \& Della~Valle(2007)}]{gue07}
Guetta, D., \& Della~Valle, M. 2007, The Astrophysical Journal, 657, L73, \dodoi{10.1086/511417}

\bibitem[{Gutiérrez {et~al.}(2017)Gutiérrez, Anderson, Hamuy, Morrell, González-Gaitan, Stritzinger, Phillips, Galbany, Folatelli, Dessart, Contreras, Valle, Freedman, Hsiao, Krisciunas, Madore, Maza, Suntzeff, Prieto, González, Cappellaro, Navarrete, Pizzella, Ruiz, Smith, \& Turatto}]{gut17}
Gutiérrez, C.~P., Anderson, J.~P., Hamuy, M., {et~al.} 2017, The Astrophysical Journal, 850, 89, \dodoi{10.3847/1538-4357/aa8f52}

\bibitem[{Hamann {et~al.}(2006)Hamann, Gräfener, \& Liermann}]{ham06}
Hamann, W.-R., Gräfener, G., \& Liermann, A. 2006, Astronomy \& Astrophysics, 457, 1015, \dodoi{10.1051/0004-6361:20065052}

\bibitem[{Hanuschik(1991)}]{son87}
Hanuschik, R.~W. 1991, in Reviews in Modern Astronomy, ed. G.~Klare (Berlin, Heidelberg: Springer Berlin Heidelberg), 233--259

\bibitem[{Hirata {et~al.}(1987)Hirata, Kajita, Koshiba, Nakahata, Oyama, Sato, Suzuki, Takita, Totsuka, Kifune, Suda, Takahashi, Tanimori, Miyano, Yamada, Beier, Feldscher, Kim, Mann, Newcomer, Van, Zhang, \& Cortez}]{hir87}
Hirata, K., Kajita, T., Koshiba, M., {et~al.} 1987, Physical Review Letters, 58, 1490, \dodoi{10.1103/physrevlett.58.1490}

\bibitem[{{Hjorth} \& {Bloom}(2012)}]{hjo12}
{Hjorth}, J., \& {Bloom}, J.~S. 2012, in Chapter 9 in ''Gamma-Ray Bursts, 169--190, \dodoi{10.48550/arXiv.1104.2274}

\bibitem[{{Ho} {et~al.}(2020){Ho}, {Kulkarni}, {Perley}, {Cenko}, {Corsi}, {Schulze}, {Lunnan}, {Sollerman}, {Gal-Yam}, {Anand}, {Barbarino}, {Bellm}, {Bruch}, {Burns}, {De}, {Dekany}, {Delacroix}, {Duev}, {Frederiks}, {Fremling}, {Goldstein}, {Golkhou}, {Graham}, {Hale}, {Kasliwal}, {Kupfer}, {Laher}, {Martikainen}, {Masci}, {Neill}, {Ridnaia}, {Rusholme}, {Savchenko}, {Shupe}, {Soumagnac}, {Strotjohann}, {Svinkin}, {Taggart}, {Tartaglia}, {Yan}, \& {Zolkower}}]{ann20}
{Ho}, A. Y.~Q., {Kulkarni}, S.~R., {Perley}, D.~A., {et~al.} 2020, \apj, 902, 86, \dodoi{10.3847/1538-4357/aba630}

\bibitem[{{Hosseinzadeh} {et~al.}(2023){Hosseinzadeh}, {Farah}, {Shrestha}, {Sand}, {Dong}, {Brown}, {Bostroem}, {Valenti}, {Jha}, {Andrews}, {Arcavi}, {Haislip}, {Hiramatsu}, {Hoang}, {Howell}, {Janzen}, {Jencson}, {Kouprianov}, {Lundquist}, {McCully}, {Meza Retamal}, {Modjaz}, {Newsome}, {Padilla Gonzalez}, {Pearson}, {Pellegrino}, {Ravi}, {Reichart}, {Smith}, {Terreran}, \& {Vink{\'o}}}]{hos23}
{Hosseinzadeh}, G., {Farah}, J., {Shrestha}, M., {et~al.} 2023, \apjl, 953, L16, \dodoi{10.3847/2041-8213/ace4c4}

\bibitem[{{Kelly} {et~al.}(2008){Kelly}, {Kirshner}, \& {Pahre}}]{kel08}
{Kelly}, P.~L., {Kirshner}, R.~P., \& {Pahre}, M. 2008, The Astrophysical Journal, 687, 1201, \dodoi{10.1086/591925}

\bibitem[{Kilpatrick {et~al.}(2023)Kilpatrick, Foley, Jacobson-Galán, Piro, Smartt, Drout, Gagliano, Gall, Hjorth, Jones, Mandel, Margutti, Ramirez-Ruiz, Ransome, Villar, Coulter, Gao, Matthews, Taggart, \& Zenati}]{kil23}
Kilpatrick, C.~D., Foley, R.~J., Jacobson-Galán, W.~V., {et~al.} 2023, The Astrophysical Journal Letters, 952, L23, \dodoi{10.3847/2041-8213/ace4ca}

\bibitem[{{Kiuchi} {et~al.}(2011){Kiuchi}, {Shibata}, {Montero}, \& {Font}}]{kiu11}
{Kiuchi}, K., {Shibata}, M., {Montero}, P.~J., \& {Font}, J.~A. 2011, \prl, 106, 251102, \dodoi{10.1103/PhysRevLett.106.251102}

\bibitem[{{Kobayashi} \& {M{\'e}sz{\'a}ros}(2003)}]{kob03}
{Kobayashi}, S., \& {M{\'e}sz{\'a}ros}, P. 2003, \apj, 589, 861, \dodoi{10.1086/374733}

\bibitem[{Kochanek(2014)}]{koc14}
Kochanek, C.~S. 2014, Monthly Notices of the Royal Astronomical Society, 446, 1213, \dodoi{10.1093/mnras/stu2056}

\bibitem[{{Lense} \& {Thirring}(1918)}]{len18}
{Lense}, J., \& {Thirring}, H. 1918, Physikalische Zeitschrift, 19, 156

\bibitem[{Leung {et~al.}(2020)Leung, Blinnikov, Nomoto, Baklanov, Sorokina, \& Tolstov}]{leu20}
Leung, S.-C., Blinnikov, S., Nomoto, K., {et~al.} 2020, The Astrophysical Journal, 903, 66, \dodoi{10.3847/1538-4357/abba33}

\bibitem[{{Levinson} \& {Nakar}(2020)}]{lev20}
{Levinson}, A., \& {Nakar}, E. 2020, \physrep, 866, 1, \dodoi{10.1016/j.physrep.2020.04.003}

\bibitem[{{Levinson} {et~al.}(2015){Levinson}, {van Putten}, \& {Pick}}]{lev15}
{Levinson}, A., {van Putten}, M. H.~P.~M., \& {Pick}, G. 2015, \apj, 812, 124, \dodoi{10.1088/0004-637X/812/2/124}

\bibitem[{{Li} {et~al.}(2011){Li}, {Chornock}, {Leaman}, {Filippenko}, {Poznanski}, {Wang}, {Ganeshalingam}, \& {Mannucci}}]{wei11}
{Li}, W., {Chornock}, R., {Leaman}, J., {et~al.} 2011, \mnras, 412, 1473, \dodoi{10.1111/j.1365-2966.2011.18162.x}

\bibitem[{LIGO Scientific~collaboration(2021)}]{LVK21}
LIGO Scientific~collaboration, Virgo~collaboration, K.~c. 2021, LIGO Document T2100289-v6

\bibitem[{Lin {et~al.}(2023)Lin, Rijal, Lunardini, Morales, \& Zanolin}]{lin23}
Lin, Z., Rijal, A., Lunardini, C., Morales, M.~D., \& Zanolin, M. 2023, Physical Review D, 107, 083017, \dodoi{10.1103/physrevd.107.083017}

\bibitem[{Lucchini {et~al.}(2019)Lucchini, Krauß, \& Markoff}]{luc19}
Lucchini, M., Krauß, F., \& Markoff, S. 2019, Monthly Notices of the Royal Astronomical Society, \dodoi{10.1093/mnras/stz2125}

\bibitem[{Lyman {et~al.}(2016)Lyman, Bersier, James, Mazzali, Eldridge, Fraser, \& Pian}]{lym16}
Lyman, J.~D., Bersier, D., James, P.~A., {et~al.} 2016, Monthly Notices of the Royal Astronomical Society, 457, 328, \dodoi{10.1093/mnras/stv2983}

\bibitem[{MacCray(1996)}]{bou93}
MacCray, R., ed. 1996, Supernovae and supernova remnants, Proceedings of the ... Colloquium of the International Astronomical Union No. 145 (Cambridge: Cambridge University Press)

\bibitem[{Madau \& Dickinson(2014)}]{mad14}
Madau, P., \& Dickinson, M. 2014, Annual Review of Astronomy and Astrophysics, 52, 415, \dodoi{10.1146/annurev-astro-081811-125615}

\bibitem[{Maurer {et~al.}(2009)Maurer, Mazzali, Deng, Filippenko, Hamuy, Kirshner, Matheson, Modjaz, Pian, Stritzinger, Taubenberger, \& Valenti}]{mau09}
Maurer, J.~I., Mazzali, P.~A., Deng, J., {et~al.} 2009, Monthly Notices of the Royal Astronomical Society, 402, 161, \dodoi{10.1111/j.1365-2966.2009.15905.x}

\bibitem[{{Maurer} {et~al.}(2010){Maurer}, {Mazzali}, {Deng}, {Filippenko}, {Hamuy}, {Kirshner}, {Matheson}, {Modjaz}, {Pian}, {Stritzinger}, {Taubenberger}, \& {Valenti}}]{mau10}
{Maurer}, J.~I., {Mazzali}, P.~A., {Deng}, J., {et~al.} 2010, \mnras, 402, 161, \dodoi{10.1111/j.1365-2966.2009.15905.x}

\bibitem[{Melinder {et~al.}(2012)Melinder, Dahlen, Mencía~Trinchant, Östlin, Mattila, Sollerman, Fransson, Hayes, Kankare, \& Nasoudi-Shoar}]{mel12}
Melinder, J., Dahlen, T., Mencía~Trinchant, L., {et~al.} 2012, Astronomy \& Astrophysics, 545, A96, \dodoi{10.1051/0004-6361/201219364}

\bibitem[{Mooley {et~al.}(2017)Mooley, Nakar, Hotokezaka, Hallinan, Corsi, Frail, Horesh, Murphy, Lenc, Kaplan, De, Dobie, Chandra, Deller, Gottlieb, Kasliwal, Kulkarni, Myers, Nissanke, Piran, Lynch, Bhalerao, Bourke, Bannister, \& Singer}]{moo17}
Mooley, K.~P., Nakar, E., Hotokezaka, K., {et~al.} 2017, Nature, 554, 207, \dodoi{10.1038/nature25452}

\bibitem[{Mooley {et~al.}(2018)Mooley, Deller, Gottlieb, Nakar, Hallinan, Bourke, Frail, Horesh, Corsi, \& Hotokezaka}]{moo18}
Mooley, K.~P., Deller, A.~T., Gottlieb, O., {et~al.} 2018, Nature, 561, 355, \dodoi{10.1038/s41586-018-0486-3}

\bibitem[{{Nakar}(2015)}]{nak15}
{Nakar}, E. 2015, \apj, 807, 172, \dodoi{10.1088/0004-637X/807/2/172}

\bibitem[{Nakar \& Sari(2012)}]{nak12}
Nakar, E., \& Sari, R. 2012, The Astrophysical Journal, 747, 88, \dodoi{10.1088/0004-637x/747/2/88}

\bibitem[{{Narayan} {et~al.}(2003){Narayan}, {Igumenshchev}, \& {Abramowicz}}]{nar03}
{Narayan}, R., {Igumenshchev}, I.~V., \& {Abramowicz}, M.~A. 2003, \pasj, 55, L69, \dodoi{10.1093/pasj/55.6.L69}

\bibitem[{Park {et~al.}(2022)Park, Kim, Kim, \& van Putten}]{par2022}
Park, H.-J., Kim, S.-J., Kim, S., \& van Putten, M. H. P.~M. 2022, The Astrophysical Journal, 938, 69, \dodoi{10.3847/1538-4357/ac9300}

\bibitem[{Pasham {et~al.}(2024)Pasham, Zajaček, Nixon, Coughlin, Śniegowska, Janiuk, Czerny, Wevers, Guolo, Ajay, \& Loewenstein}]{pas24}
Pasham, D.~R., Zajaček, M., Nixon, C.~J., {et~al.} 2024, Nature, \dodoi{10.1038/s41586-024-07433-w}

\bibitem[{Pastorello {et~al.}(2005)Pastorello, Baron, Branch, Zampieri, Turatto, Ramina, Benetti, Cappellaro, Salvo, Patat, Piemonte, Sollerman, Leibundgut, \& Altavilla}]{pas05}
Pastorello, A., Baron, E., Branch, D., {et~al.} 2005, Monthly Notices of the Royal Astronomical Society, 360, 950, \dodoi{10.1111/j.1365-2966.2005.09079.x}

\bibitem[{{Peters} \& {Mathews}(1963)}]{pet63}
{Peters}, P.~C., \& {Mathews}, J. 1963, Physical Review, 131, 435, \dodoi{10.1103/PhysRev.131.435}

\bibitem[{Pietrzyński {et~al.}(2019)Pietrzyński, Graczyk, Gallenne, Gieren, Thompson, Pilecki, Karczmarek, Górski, Suchomska, Taormina, Zgirski, Wielgórski, Kołaczkowski, Konorski, Villanova, Nardetto, Kervella, Bresolin, Kudritzki, Storm, Smolec, \& Narloch}]{pie19}
Pietrzyński, G., Graczyk, D., Gallenne, A., {et~al.} 2019, Nature, 567, 200, \dodoi{10.1038/s41586-019-0999-4}

\bibitem[{Piro \& Pfahl(2007)}]{pir07}
Piro, A.~L., \& Pfahl, E. 2007, The Astrophysical Journal, 658, 1173, \dodoi{10.1086/511672}

\bibitem[{Podsiadlowski {et~al.}(2004)Podsiadlowski, Mazzali, Nomoto, Lazzati, \& Cappellaro}]{pod04}
Podsiadlowski, P., Mazzali, P.~A., Nomoto, K., Lazzati, D., \& Cappellaro, E. 2004, The Astrophysical Journal, 607, L17, \dodoi{10.1086/421347}

\bibitem[{Pozanenko {et~al.}(2018)Pozanenko, Barkov, Minaev, Volnova, Mazaeva, Moskvitin, Krugov, Samodurov, Loznikov, \& Lyutikov}]{poz18}
Pozanenko, A.~S., Barkov, M.~V., Minaev, P.~Y., {et~al.} 2018, The Astrophysical Journal Letters, 852, L30, \dodoi{10.3847/2041-8213/aaa2f6}

\bibitem[{Raskin {et~al.}(2008)Raskin, Scannapieco, Rhoads, \& Della~Valle}]{ras08}
Raskin, C., Scannapieco, E., Rhoads, J., \& Della~Valle, M. 2008, The Astrophysical Journal, 689, 358, \dodoi{10.1086/592495}

\bibitem[{{Reynolds}(2021)}]{rey21}
{Reynolds}, C.~S. 2021, \araa, 59, 117, \dodoi{10.1146/annurev-astro-112420-035022}

\bibitem[{{Rueda} {et~al.}(2021){Rueda}, {Ruffini}, {Moradi}, \& {Wang}}]{rue21}
{Rueda}, J.~A., {Ruffini}, R., {Moradi}, R., \& {Wang}, Y. 2021, International Journal of Modern Physics D, 30, 2130007, \dodoi{10.1142/S021827182130007X}

\bibitem[{Sahu {et~al.}(2009)Sahu, Tanaka, Anupama, Gurugubelli, \& Nomoto}]{sah09}
Sahu, D.~K., Tanaka, M., Anupama, G.~C., Gurugubelli, U.~K., \& Nomoto, K. 2009, The Astrophysical Journal, 697, 676, \dodoi{10.1088/0004-637x/697/1/676}

\bibitem[{Salpeter(1955)}]{sal55}
Salpeter, E.~E. 1955, The Astrophysical Journal, 121, 161, \dodoi{10.1086/145971}

\bibitem[{Sander {et~al.}(2019)Sander, Hamann, Todt, Hainich, Shenar, Ramachandran, \& Oskinova}]{san19}
Sander, A. A.~C., Hamann, W.-R., Todt, H., {et~al.} 2019, Astronomy \& Astrophysics, 621, A92, \dodoi{10.1051/0004-6361/201833712}

\bibitem[{{Shrestha} {et~al.}(2024){Shrestha}, {Bostroem}, {Sand}, {Hosseinzadeh}, {Andrews}, {Dong}, {Hoang}, {Janzen}, {Pearson}, {Jencson}, {Lundquist}, {Mehta}, {Ravi}, {Meza Retamal}, {Valenti}, {Brown}, {Jha}, {Macrie}, {Hsu}, {Farah}, {Howell}, {McCully}, {Newsome}, {Padilla Gonzalez}, {Pellegrino}, {Terreran}, {Kwok}, {Smith}, {Schwab}, {Martas}, {Munoz}, {Medina}, {Li}, {Diaz}, {Hiramatsu}, {Tucker}, {Wheeler}, {Wang}, {Zhai}, {Zhang}, {Gangopadhyay}, {Yang}, \& {Gutierez}}]{shr24}
{Shrestha}, M., {Bostroem}, K.~A., {Sand}, D.~J., {et~al.} 2024, arXiv e-prints, arXiv:2405.18490, \dodoi{10.48550/arXiv.2405.18490}

\bibitem[{Smartt(2009)}]{sma09}
Smartt, S.~J. 2009, Annual Review of Astronomy and Astrophysics, 47, 63, \dodoi{10.1146/annurev-astro-082708-101737}

\bibitem[{{Soderberg} {et~al.}(2012){Soderberg}, {Margutti}, {Zauderer}, {Krauss}, {Katz}, {Chomiuk}, {Dittmann}, {Nakar}, {Sakamoto}, {Kawai}, {Hurley}, {Barthelmy}, {Toizumi}, {Morii}, {Chevalier}, {Gurwell}, {Petitpas}, {Rupen}, {Alexander}, {Levesque}, {Fransson}, {Brunthaler}, {Bietenholz}, {Chugai}, {Grindlay}, {Copete}, {Connaughton}, {Briggs}, {Meegan}, {von Kienlin}, {Zhang}, {Rau}, {Golenetskii}, {Mazets}, \& {Cline}}]{sod12}
{Soderberg}, A.~M., {Margutti}, R., {Zauderer}, B.~A., {et~al.} 2012, \apj, 752, 78, \dodoi{10.1088/0004-637X/752/2/78}

\bibitem[{{Srivastav} {et~al.}(2024){Srivastav}, {Chen}, {Smartt}, {Nicholl}, {Smith}, {Young}, {Fulton}, {McCollum}, {Moore}, {Weston}, {Sheng}, {Aamer}, {Angus}, {Ramsden}, {Shingles}, {Gillanders}, {Rhodes}, {Andersson}, {Stevance}, {Denneau}, {Tonry}, {Weiland}, {Lawrence}, {Siverd}, {Erasmus}, {Koorts}, {Jordan}, {Suc}, {Rest}, {Stubbs}, \& {Sommer}}]{sri24}
{Srivastav}, S., {Chen}, T.~W., {Smartt}, S.~J., {et~al.} 2024, Transient Name Server AstroNote, 100, 1

\bibitem[{Srivastava {et~al.}(2019)Srivastava, Ballmer, Brown, Afle, Burrows, Radice, \& Vartanyan}]{sri19}
Srivastava, V., Ballmer, S., Brown, D.~A., {et~al.} 2019, Physical Review D, 100, 043026, \dodoi{10.1103/physrevd.100.043026}

\bibitem[{Stella \& Vietri(1998)}]{ste98}
Stella, L., \& Vietri, M. 1998, The Astrophysical Journal, 492, L59, \dodoi{10.1086/311075}

\bibitem[{Strolger {et~al.}(2015)Strolger, Dahlen, Rodney, Graur, Riess, McCully, Ravindranath, Mobasher, \& Shahady}]{str15}
Strolger, L.-G., Dahlen, T., Rodney, S.~A., {et~al.} 2015, The Astrophysical Journal, 813, 93, \dodoi{10.1088/0004-637x/813/2/93}

\bibitem[{Sukhbold {et~al.}(2018)Sukhbold, Woosley, \& Heger}]{suk18}
Sukhbold, T., Woosley, S.~E., \& Heger, A. 2018, The Astrophysical Journal, 860, 93, \dodoi{10.3847/1538-4357/aac2da}

\bibitem[{Sun {et~al.}(2022)Sun, Maund, Crowther, \& Liu}]{sun22}
Sun, N.-C., Maund, J.~R., Crowther, P.~A., \& Liu, L.-D. 2022, Monthly Notices of the Royal Astronomical Society: Letters, 512, L66, \dodoi{10.1093/mnrasl/slac023}

\bibitem[{Sun {et~al.}(2023)Sun, Maund, Shao, \& Janiak}]{sun23}
Sun, N.-C., Maund, J.~R., Shao, Y., \& Janiak, I.~A. 2023, Monthly Notices of the Royal Astronomical Society, 519, 3785, \dodoi{10.1093/mnras/stac3773}

\bibitem[{{Szczepa{\'n}czyk} {et~al.}(2021){Szczepa{\'n}czyk}, {Antelis}, {Benjamin}, {Cavagli{\`a}}, {Gondek-Rosi{\'n}ska}, {Hansen}, {Klimenko}, {Morales}, {Moreno}, {Mukherjee}, {Nurbek}, {Powell}, {Singh}, {Sitmukhambetov}, {Szewczyk}, {Valdez}, {Vedovato}, {Westhouse}, {Zanolin}, \& {Zheng}}]{szc21}
{Szczepa{\'n}czyk}, M.~J., {Antelis}, J.~M., {Benjamin}, M., {et~al.} 2021, \prd, 104, 102002, \dodoi{10.1103/PhysRevD.104.102002}

\bibitem[{Taddia {et~al.}(2019)Taddia, Sollerman, Fremling, Barbarino, Karamehmetoglu, Arcavi, Cenko, Filippenko, Gal-Yam, Hiramatsu, Hosseinzadeh, Howell, Kulkarni, Laher, Lunnan, Masci, Nugent, Nyholm, Perley, Quimby, \& Silverman}]{tad19}
Taddia, F., Sollerman, J., Fremling, C., {et~al.} 2019, Astronomy \& Astrophysics, 621, A71, \dodoi{10.1051/0004-6361/201834429}

\bibitem[{{The LIGO Scientific Collaboration} {et~al.}(2024){The LIGO Scientific Collaboration}, {The Virgo Collaboration}, \& {The KAGRA Collaboration}}]{LVK24}
{The LIGO Scientific Collaboration}, {The Virgo Collaboration}, \& {The KAGRA Collaboration}. 2024, Observation of Gravitational Waves from the Coalescence of a $2.5-4.5~M_\odot$ Compact Object and a Neutron Star,  arXiv, \dodoi{10.48550/ARXIV.2404.04248}

\bibitem[{{Thorne} {et~al.}(1986){Thorne}, {Price}, \& {MacDonald}}]{tho86}
{Thorne}, K.~S., {Price}, R.~H., \& {MacDonald}, D.~A. 1986, {Black holes: The membrane paradigm}

\bibitem[{{Tinney} {et~al.}(1998){Tinney}, {Stathakis}, {Cannon}, {Galama}, {Wieringa}, {Frail}, {Kulkarni}, \& {Higdon}}]{tin98}
{Tinney}, C., {Stathakis}, R., {Cannon}, R., {et~al.} 1998, IAU Circ., 6896, 1

\bibitem[{{Toscani} {et~al.}(2019){Toscani}, {Lodato}, \& {Nealon}}]{tos19}
{Toscani}, M., {Lodato}, G., \& {Nealon}, R. 2019, \mnras, 489, 699, \dodoi{10.1093/mnras/stz2201}

\bibitem[{{Toscani} {et~al.}(2020){Toscani}, {Rossi}, \& {Lodato}}]{tos20}
{Toscani}, M., {Rossi}, E.~M., \& {Lodato}, G. 2020, \mnras, 498, 507, \dodoi{10.1093/mnras/staa2290}

\bibitem[{Utrobin \& Chugai(2011)}]{utr11}
Utrobin, V.~P., \& Chugai, N.~N. 2011, Astronomy \& Astrophysics, 532, A100, \dodoi{10.1051/0004-6361/201117137}

\bibitem[{Van~Putten(2013)}]{van13}
Van~Putten, M.~H. 2013, Acta Polytechnica, 53, 736, \dodoi{10.14311/ap.2013.53.0736}

\bibitem[{{van Putten}(1999)}]{van99}
{van Putten}, M. H.~P.~M. 1999, Science, 284, 115, \dodoi{10.1126/science.284.5411.115}

\bibitem[{van Putten(2001)}]{van01}
van Putten, M. H. P.~M. 2001, Physical Review Letters, 87, 091101, \dodoi{10.1103/physrevlett.87.091101}

\bibitem[{van Putten(2002)}]{van02}
---. 2002, The Astrophysical Journal, 575, L71, \dodoi{10.1086/342781}

\bibitem[{van Putten(2004)}]{van04}
---. 2004, The Astrophysical Journal, 611, L81, \dodoi{10.1086/423934}

\bibitem[{{van Putten}(2008)}]{van08}
{van Putten}, M. H.~P.~M. 2008, \apjl, 684, L91, \dodoi{10.1086/592216}

\bibitem[{{van Putten}(2012)}]{van12}
{van Putten}, M.~H.~P.~M. 2012, Progress of Theoretical Physics, 127, 331, \dodoi{10.1143/PTP.127.331}

\bibitem[{van Putten(2015)}]{van15}
van Putten, M. H. P.~M. 2015, The Astrophysical Journal, 810, 7, \dodoi{10.1088/0004-637x/810/1/7}

\bibitem[{van Putten(2016)}]{van16}
---. 2016, The Astrophysical Journal, 819, 169, \dodoi{10.3847/0004-637x/819/2/169}

\bibitem[{{van Putten} \& {Della Valle}(2017)}]{van17}
{van Putten}, M. H.~P.~M., \& {Della Valle}, M. 2017, \mnras, 464, 3219, \dodoi{10.1093/mnras/stw2496}

\bibitem[{{van Putten} \& {Della Valle}(2019)}]{van19b}
---. 2019, \mnras, 482, L46, \dodoi{10.1093/mnrasl/sly166}

\bibitem[{van Putten \& Della~Valle(2023)}]{van23}
van Putten, M. H. P.~M., \& Della~Valle, M. 2023, Astronomy \& Astrophysics, 669, A36, \dodoi{10.1051/0004-6361/202142974}

\bibitem[{van Putten {et~al.}(2011)van Putten, Della~Valle, \& Levinson}]{van11}
van Putten, M. H. P.~M., Della~Valle, M., \& Levinson, A. 2011, Astronomy \& Astrophysics, 535, L6, \dodoi{10.1051/0004-6361/201118080}

\bibitem[{{van Putten} {et~al.}(2019){van Putten}, {Della Valle}, \& {Levinson}}]{van19a}
{van Putten}, M. H.~P.~M., {Della Valle}, M., \& {Levinson}, A. 2019, \apjl, 876, L2, \dodoi{10.3847/2041-8213/ab18a2}

\bibitem[{van Putten {et~al.}(2014)van Putten, Lee, Della~Valle, Amati, \& Levinson}]{van14}
van Putten, M. H. P.~M., Lee, G.~M., Della~Valle, M., Amati, L., \& Levinson, A. 2014, Monthly Notices of the Royal Astronomical Society: Letters, 444, L58, \dodoi{10.1093/mnrasl/slu113}

\bibitem[{van Putten {et~al.}(2004{\natexlab{a}})van Putten, Lee, Lee, \& Kim}]{van04b}
van Putten, M. H. P.~M., Lee, H.~K., Lee, C.~H., \& Kim, H. 2004{\natexlab{a}}, Physical Review D, 69, 104026, \dodoi{10.1103/physrevd.69.104026}

\bibitem[{{van Putten} \& {Levinson}(2003)}]{van03b}
{van Putten}, M. H.~P.~M., \& {Levinson}, A. 2003, \apj, 584, 937, \dodoi{10.1086/345900}

\bibitem[{van Putten {et~al.}(2019)van Putten, Levinson, Frontera, Guidorzi, Amati, \& Della~Valle}]{van19}
van Putten, M. H. P.~M., Levinson, A., Frontera, F., {et~al.} 2019, The European Physical Journal Plus, 134, \dodoi{10.1140/epjp/i2019-12932-3}

\bibitem[{van Putten {et~al.}(2004{\natexlab{b}})van Putten, Levinson, Lee, Regimbau, Punturo, \& Harry}]{van04c}
van Putten, M. H. P.~M., Levinson, A., Lee, H.~K., {et~al.} 2004{\natexlab{b}}, Physical Review D, 69, 044007, \dodoi{10.1103/physrevd.69.044007}

\bibitem[{{van Putten} \& {Regimbau}(2003)}]{van03}
{van Putten}, M. H.~P.~M., \& {Regimbau}, T. 2003, \apjl, 593, L15, \dodoi{10.1086/378146}

\bibitem[{{van Putten} \& {Sarkar}(2000)}]{van2000}
{van Putten}, M. H.~P.~M., \& {Sarkar}, A. 2000, \prd, 62, 041502, \dodoi{10.1103/PhysRevD.62.041502}

\bibitem[{Vartanyan {et~al.}(2023)Vartanyan, Burrows, Wang, Coleman, \& White}]{var23}
Vartanyan, D., Burrows, A., Wang, T., Coleman, M. S.~B., \& White, C.~J. 2023, The Gravitational-Wave Signature of Core-Collapse Supernovae,  arXiv, \dodoi{10.48550/ARXIV.2302.07092}

\bibitem[{{Wessel} {et~al.}(2021){Wessel}, {Paschalidis}, {Tsokaros}, {Ruiz}, \& {Shapiro}}]{wes21}
{Wessel}, E., {Paschalidis}, V., {Tsokaros}, A., {Ruiz}, M., \& {Shapiro}, S.~L. 2021, \prd, 103, 043013, \dodoi{10.1103/PhysRevD.103.043013}

\bibitem[{{West} {et~al.}(1987){West}, {Lauberts}, {Jorgensen}, \& {Schuster}}]{wes87}
{West}, R.~M., {Lauberts}, A., {Jorgensen}, H.~E., \& {Schuster}, H.~E. 1987, \aap, 177, L1

\bibitem[{Wilkins(1972)}]{wil72}
Wilkins, D.~C. 1972, Physical Review D, 5, 814, \dodoi{10.1103/physrevd.5.814}

\bibitem[{Woosley(1993)}]{woo93}
Woosley, S.~E. 1993, The Astrophysical Journal, 405, 273, \dodoi{10.1086/172359}

\bibitem[{{Woosley} \& {Bloom}(2006)}]{woo06}
{Woosley}, S.~E., \& {Bloom}, J.~S. 2006, \araa, 44, 507, \dodoi{10.1146/annurev.astro.43.072103.150558}

\bibitem[{Woosley {et~al.}(1999)Woosley, Eastman, \& Schmidt}]{woo99}
Woosley, S.~E., Eastman, R.~G., \& Schmidt, B.~P. 1999, The Astrophysical Journal, 516, 788, \dodoi{10.1086/307131}

\bibitem[{Xiang {et~al.}(2021)Xiang, Wang, Lin, Mo, Lin, Burke, Hiramatsu, Hosseinzadeh, Howell, McCully, Valenti, Vinkó, Wheeler, Ehgamberdiev, Mirzaqulov, Bódi, Bognár, Cseh, Hanyecz, Ignácz, Kalup, Könyves-Tóth, Kriskovics, Ordasi, Pál, Sárneczky, Seli, Szakáts, Arranz-Heras, Benavides-Palencia, Cejudo-Martínez, De~la Fuente-Fernández, Escartín-Pérez, la~Cuesta, González-Carballo, González-Farfán, Limón-Martínez, Mantero, Naves-Nogués, Morales-Aimar, Ruíz-Ruíz, Soldán-Alfaro, Valero-Pérez, Violat-Bordonau, Zhang, Zhang, Li, Chen, Sai, \& Li}]{xia21}
Xiang, D., Wang, X., Lin, W., {et~al.} 2021, The Astrophysical Journal, 910, 42, \dodoi{10.3847/1538-4357/abdeba}

\end{thebibliography}

\end{document}